\documentclass[preprint]{ptephy}

\preprintnumber{KYUSHU-HET-184}

\usepackage{amssymb}
\usepackage{amsthm}
\usepackage{amsmath}
\usepackage{booktabs}
\usepackage{bbm}
\usepackage{mathtools}
\usepackage{subcaption}

\DeclareMathOperator{\re}{Re}
\DeclareMathOperator{\im}{Im}

\DeclareMathOperator{\sign}{sign}

\numberwithin{equation}{section}

\usepackage{url}

\begin{document}

\title{Numerical study of the $\mathcal{N}=2$ Landau--Ginzburg model}

\author{%
\name{\fname{Okuto} \surname{Morikawa}}{1}
and
\name{\fname{Hiroshi} \surname{Suzuki}}{1,\ast}
}

\address{%
\affil{1}{Department of Physics, Kyushu University
744 Motooka, Nishi-ku, Fukuoka, 819-0395, Japan}
\email{hsuzuki@phys.kyushu-u.ac.jp}
}

\date{\today}

\begin{abstract}
It is believed that the two-dimensional massless $\mathcal{N}=2$ Wess--Zumino
model becomes the $\mathcal{N}=2$ superconformal field theory (SCFT) in the
infrared (IR) limit. We examine this theoretical conjecture of the
Landau--Ginzburg (LG) description of the $\mathcal{N}=2$ SCFT by numerical
simulations on the basis of a supersymmetric-invariant momentum-cutoff
regularization. We study a single supermultiplet with cubic and quartic
superpotentials. From two-point correlation functions in the IR region, we
measure the scaling dimension and the central charge, which are consistent with
the conjectured LG description of the $A_2$ and $A_3$ minimal models,
respectively. Our result supports the theoretical conjecture and, at the same
time, indicates a possible computational method of correlation functions in
the $\mathcal{N}=2$ SCFT from the LG description.
\end{abstract}
\subjectindex{B16, B24, B34}

\maketitle

\tableofcontents

\section{Introduction}
\label{sec:1}
In sufficiently low energies, any quantum field theory is expected to become
scale invariant, all massive modes being decoupled. Such a scale-invariant
theory would be described by a conformal field theory (CFT). If this low-energy
theory gives rise to a nontrivial CFT, the original field theory is called the
Landau--Ginzburg (LG) model or the LG description of the
CFT~\cite{Zamolodchikov:1986db}. The LG description thus provides a
Lagrangian-level realization of CFT, although the existence of the Lagrangian
of the latter is not always obvious.

As an example of the LG model, the two-dimensional (2D) $\mathcal{N}=2$
massless Wess--Zumino (WZ) model (which can be obtained by the dimensional
reduction of the four-dimensional WZ model~\cite{Wess:1974tw}) with a
quasi-homogeneous superpotential is considered to give an LG description of the
$\mathcal{N}=2$ superconformal field theory (SCFT)~\cite{DiVecchia:1985ief,%
DiVecchia:1986cdz,DiVecchia:1986fwg,Boucher:1986bh,Gepner:1986ip,%
Cappelli:1986hf,Cappelli:1986ed,Gepner:1986hr,Gepner:1987qi,Cappelli:1987xt,%
Kato:1987td,Gepner:1987vz}. There are various theoretical analyses which
support this correspondence~\cite{Kastor:1988ef,Vafa:1988uu,Martinec:1988zu,%
Lerche:1989uy,Howe:1989qr,Cecotti:1989jc,Howe:1989az,Cecotti:1989gv,%
Cecotti:1990kz,Witten:1993jg}. It is, however, still difficult to prove this
conjecture directly, because the 2D $\mathcal{N}=2$ massless WZ model is
strongly coupled at low energies and perturbation theory suffers from infrared
(IR) divergences; the LG description is truly a non-perturbative phenomenon.

A non-perturbative calculational method such as the lattice field theory may
provide an alternative approach to this issue. In~Ref.~\cite{Kawai:2010yj},
the scaling dimension of the scalar field in the IR limit of the 2D
$\mathcal{N}=2$ massless WZ model was measured by using a lattice formulation
from~Ref.~\cite{Kikukawa:2002as}. The case of a single supermultiplet with a
cubic superpotential~$W=\Phi^3$, which is considered to become the $A_2$
minimal model in the IR limit, is studied. In~Ref.~\cite{Kawai:2010yj}, good
agreement of the scaling dimension with that of the $A_2$ model was observed.
As is well-recognized, the lattice formulation is in general not compatible
with the supersymmetry (SUSY) that must be a crucial element of the above LG
correspondence. This is also the case for the lattice formulation
of~Ref.~\cite{Kikukawa:2002as}. However, the formulation
of~Ref.~\cite{Kikukawa:2002as} exactly preserves one nilpotent SUSY, utilizing
the existence of the the Nicolai or Nicolai--Parisi--Sourlas
map~\cite{Nicolai:1979nr,Nicolai:1980jc,Parisi:1982ud,Cecotti:1983up}.%
\footnote{This feature is common to the lattice formulation
of~Ref.~\cite{Sakai:1983dg}.} Because of this exactly preserved SUSY, and
because this 2D theory is super-renormalizable, it can be argued to all orders
of perturbation theory that the full SUSY is automatically restored in the
continuum limit.\footnote{For this issue, see also
Refs.~\cite{Giedt:2004qs,Kadoh:2010ca}. Ref.~\cite{Kadoh:2016eju} is a recent
review of SUSY on the lattice.} The study of~Ref.~\cite{Kawai:2010yj} thus
paved the way for the numerical investigation of the $\mathcal{N}=2$ LG model,
a triumph of the lattice field theory.%
\footnote{References~\cite{Beccaria:1998vi,Catterall:2001fr,Giedt:2005ae,%
Bergner:2007pu,Kastner:2008zc} are preceding studies on the 2D massive
$\mathcal{N}=2$ WZ model. It appears that the 2D massless $\mathcal{N}=2$ WZ
model is numerically studied in~Ref.~\cite{Nicolis:2017lqk}.}

Somewhat later, in~Ref.~\cite{Kamata:2011fr}, the same $W=\Phi^3$ model was
analyzed by using the formulation in~Ref.~\cite{Kadoh:2009sp}; a similar result
on the scaling dimension was obtained. A salient feature of the momentum cutoff
formulation of~Ref.~\cite{Kadoh:2009sp} is that it preserves the full set of
SUSY as well as the translational invariance even with a finite cutoff. The
formulation is (almost) identical to the dimensional reduction of the
lattice formulation~\cite{Bartels:1983wm} of the 4D WZ model on the basis of
the SLAC derivative~\cite{Drell:1976bq,Drell:1976mj}. Although this formulation
exactly preserves SUSY, it sacrifices the locality because of the SLAC
derivative. See Ref.~\cite{Bergner:2009vg} for an analysis of the issue of the
exact SUSY and the locality. Although the SLAC derivative generally suffers
from some pathology~\cite{Dondi:1976tx,Karsten:1979wh,Kato:2008sp}, for the 2D
$\mathcal{N}=2$ WZ model it can be argued~\cite{Kadoh:2009sp} to all orders of
perturbation theory that the locality is automatically restored in the
continuum limit. This is precisely because of the exactly preserved SUSY and
because this 2D theory is super-renormalizable. Since this formulation
preserves the full SUSY, the construction of the associated Noether current,
the supercurrent, is straightforward. Then, from the IR limit of the two-point
function of the supercurrent, the central charge being fairly consistent with
the $A_2$~model was observed. Thus, this study again supports the conjectured
LG correspondence.

In this paper, following on from the study of~Ref.~\cite{Kamata:2011fr}, we
carry out the numerical study of the $\mathcal{N}=2$ LG model on the basis of
the formulation of~Ref.~\cite{Kadoh:2009sp}. In several aspects we extend and
improve the analysis in~Ref.~\cite{Kamata:2011fr}. First, we study a higher
critical model~$W=\Phi^4$, which would correspond to the $A_3$ minimal model,
as well as~$W=\Phi^3$ to obtain further support for the LG correspondence and
the validity of the formulation. For the scaling dimension, in this paper we
use the two-point function in the momentum space instead of the susceptibility
of~Ref.~\cite{Kamata:2011fr}. Second, the numerical accuracy and the effective
number of configurations in the Monte Carlo simulation are quite improved.
Third, we also measure the central charge by using the two-point function of
the energy--momentum tensor, not only by that of the supercurrent.
In~Ref.~\cite{Kamata:2011fr}, it was reported that the former correlation
function was too noisy for extracting the central charge; in the present paper,
we avoid this problem by rewriting the correlation function of the
energy--momentum tensor by that of the supercurrent by using SUSY
Ward--Takahashi (WT) relations. It turns out that after this transformation,
the correlation function of the energy--momentum tensor is rather useful to
extract the central charge. We also repeat the calculation of the
``effective central charge'' in~Ref.~\cite{Kamata:2011fr} that is an analogue
of the Zamolodchikov $c$-function~\cite{Zamolodchikov:1986gt,Cappelli:1989yu}.
All our results below show a coherence picture being consistent with the
conjectured LG correspondence.

In view of the LG/Calabi--Yau
correspondence~\cite{Martinec:1988zu,Cecotti:1990wz,Greene:1988ut,%
Witten:1993yc}, we hope that this kind of numerical method will eventually
provide a computation method for scattering amplitudes in a superstring theory,
whose world sheet theory is given by an $\mathcal{N}=2$ SCFT but not
necessarily by the product of solvable minimal models.

\section{Formulation}
\label{sec:2}
\subsection{The classical action}
It is believed that the 2D $\mathcal{N}=2$ WZ model provides the LG description
of the 2D $\mathcal{N}=2$ SCFT.\footnote{Here, by $\mathcal{N}=2$, we mean
$\mathcal{N}=(2,2)$ and not $\mathcal{N}=(2,0)$.} The action of the 2D WZ model
can be obtained by the dimensional reduction of the 4D $\mathcal{N}=1$ WZ
model~\cite{Wess:1974tw} whose (Euclidean) action is given by
\begin{equation}
   S=\int d^4x\,d^4\theta\,\Bar{\Phi}\Phi
   -\int d^4x\,d^2\theta\,W(\Phi)-\int d^4x\,d^2\Bar{\theta}\,W(\Bar{\Phi}).
\label{eq:(2.1)}
\end{equation}
Here, $\theta$ and~$\Bar{\theta}$ are Grassmann coordinates and $\Phi$ is the
chiral superfield,
\begin{equation}
   \Phi(x,\theta)
   =A(y)+\sqrt{2}\,\sum_{\alpha=1}^2\theta^\alpha\psi_\alpha(y)
   +\sum_{\alpha=1}^2\theta^\alpha\theta_\alpha F(y),
\label{eq:(2.2)}
\end{equation}
consisting of a complex scalar~$A$, a left-handed spinor~$\psi$, and an
auxiliary field~$F$; the coordinate~$y$ is given by
\begin{equation}
   y_M=x_M
   +i\sum_{\alpha=1}^2\sum_{\Dot\alpha=\Dot{1}}^{\Dot{2}}
   \theta^\alpha\sigma_{M\alpha\Dot{\alpha}}\Bar{\theta}^{\Dot{\alpha}}\qquad
   \text{for $M=0$, $1$, $2$, $3$},
\label{eq:(2.3)}
\end{equation}
where $\sigma_0$ is the unit matrix and $\sigma_{1,2,3}$ the Pauli matrices. The
superpotential $W(\Phi)$ ($W(\Bar{\Phi})$) in~Eq.~\eqref{eq:(2.1)} is assumed
to be a polynomial of the superfield~$\Phi$ ($\Bar{\Phi}$).

Under the dimensional reduction, we eliminate the dependence on the coordinates
$x_2$ and~$x_3$. The coordinates $x_0$ and~$x_1$ are identified with the 2D
coordinates; in what follows, we use the complex coordinates quite often:
\begin{equation}
   z\equiv x_0+ix_1,\qquad\Bar{z}\equiv x_0-ix_1.
\label{eq:(2.4)}
\end{equation}
The corresponding derivatives are given by
\begin{equation}
   \partial\equiv\frac{\partial}{\partial z}
   =\frac{1}{2}\left(\partial_0-i\partial_1\right),\qquad
   \Bar{\partial}\equiv\frac{\partial}{\partial\Bar{z}}
   =\frac{1}{2}\left(\partial_0+i\partial_1\right).
\label{eq:(2.5)}
\end{equation}
With these notations,\footnote{Defining a two-component Dirac fermion by
$\psi\equiv\begin{pmatrix}\psi_1\\\Bar{\psi}_{\Dot{2}}\end{pmatrix}$
and~$\Bar{\psi}\gamma_0\equiv(\Bar{\psi}_{\Dot{1}},\psi_2)$, the 2D Dirac
matrices are given by
\begin{equation}
   \gamma_0=
   \begin{pmatrix}
   0 & 1 \\ 1 & 0
   \end{pmatrix},\qquad
   \gamma_1=
   \begin{pmatrix}
   0 & i \\ -i & 0
   \end{pmatrix},
\label{eq:(2.6)}
\end{equation}
that is,
\begin{equation}
   \gamma_z=
   \begin{pmatrix}
   0 & 1 \\ 0 & 0
   \end{pmatrix},\qquad
   \gamma_{\Bar{z}}=
   \begin{pmatrix}
   0 & 0 \\ 1 & 0
   \end{pmatrix}.
\label{eq:(2.7)}
\end{equation}}
the Euclidean action of the 2D $\mathcal{N}=2$ WZ model is given
by\footnote{The Euclidean action of the auxiliary field in the Wess--Zumino
model has the ``wrong sign'', i.e, the sign is opposite to the Gaussian one. In
this sense, the functional integral containing the Euclidean action of the
auxiliary field is merely a formal expression. We understand that the auxiliary
field is always expressed by using the equation of motion. The functional
integral then becomes perfectly well defined under this understanding. Our
computation below is based on such a well-defined functional integral.}
\begin{align}
   S&=\int d^2x\,\Biggl[
   4\partial A^*\Bar{\partial}A-F^*F-F^*W'(A)^*-FW'(A)
\notag\\
   &\qquad\qquad\qquad{}
   +\left(\Bar{\psi}_{\Dot{1}},\psi_2\right)
   \begin{pmatrix}
   2\partial&W''(A)^*\\W''(A)&2\Bar{\partial}
   \end{pmatrix}
   \begin{pmatrix}
   \psi_1\\\Bar{\psi}_{\Dot{2}}
   \end{pmatrix}
   \Biggr].
\label{eq:(2.8)}
\end{align}
The basic symmetries of this system, including SUSY, are summarized
in~Appendix~\ref{sec:A}.

\subsection{Momentum cutoff regularization}
We quantize the system of~Eq.~\eqref{eq:(2.8)} by employing a momentum cutoff
regularization; this approach is studied in~Ref.~\cite{Kadoh:2009sp}. As
emphasized in~Ref.~\cite{Kadoh:2009sp}, this regularization exactly preserves
important symmetries of the system, SUSY and the translational invariance. This
is the good news. The bad news is that the regularization breaks the locality.
In fact, this formulation is (when the integers~$L_\mu/a$ are odd implying a
spacetime lattice with periodic boundary conditions; see below) nothing but the
dimensional reduction of the SUSY-invariant lattice formulation of the 4D WZ
model of~Ref.~\cite{Bartels:1983wm} that is based on the SLAC
derivative~\cite{Drell:1976bq,Drell:1976mj}. It is well recognized that the
SLAC derivative generally suffers from some
pathology~\cite{Dondi:1976tx,Karsten:1979wh,Kato:2008sp}. On the other hand,
for the 2D $\mathcal{N}=2$ WZ model, one can argue to all orders of
perturbation theory that the locality is automatically restored when the UV
cutoff is removed, thanks to the exactly preserved SUSY~\cite{Kadoh:2009sp}.
However, since this argument is based on perturbation theory, whose validity
for the present \emph{massless\/} WZ model is not clear due to the IR
divergences, strictly speaking, the theoretical basis of our numerical
simulation is not quite obvious. Nevertheless, our numerical results below (and
those of~Ref.~\cite{Kamata:2011fr}) show a coherent picture which strongly
suggests the validity of the approach. We want to leave understanding the
observed validity of our formulation as a future problem.

Now, let us suppose that the system is defined in a box of physical
size~$L_0\times L_1$. The Fourier transformation of each field $\varphi(x)$
in~Eq.~\eqref{eq:(2.8)} is then defined by
\begin{equation}
   \varphi(x)=\frac{1}{L_0L_1}\sum_pe^{ipx}\varphi(p),\qquad
   \varphi(p)=\int d^2x\,e^{-ipx}\varphi(x),
\label{eq:(2.9)}
\end{equation}
where
\begin{equation}
   p_\mu=\frac{2\pi}{L_\mu}n_\mu,\qquad(n_\mu=0,\pm1,\pm2,\dotsc).
\label{eq:(2.10)}
\end{equation}
Note that
\begin{equation}
 \varphi^*(p)=\varphi(-p)^*.
\label{eq:(2.11)}
\end{equation}
After eliminating the auxiliary field~$F$ by the equation of motion, the
action in~Eq.~\eqref{eq:(2.8)} in terms of the Fourier modes yields
\begin{equation}
   S=S_B+\frac{1}{L_0L_1}\sum_p
   \left(\Bar{\psi}_{\Dot{1}},\psi_2\right)(-p)
   \begin{pmatrix}
   2ip_z&W''(A)^**\\W''(A)*&2ip_{\Bar{z}}
   \end{pmatrix}
   \begin{pmatrix}
   \psi_1\\\Bar{\psi}_{\Dot{2}}
   \end{pmatrix}(p),
\label{eq:(2.12)}
\end{equation}
where $p_z\equiv(1/2)(p_0-ip_1)$, $p_{\Bar{z}}\equiv(1/2)(p_0+ip_1)$,
$*$~denotes the convolution
\begin{equation}
   (\varphi_1*\varphi_2)(p)
   \equiv\frac{1}{L_0L_1}\sum_q\varphi_1(q)\varphi_2(p-q),
\label{eq:(2.13)}
\end{equation}
and $S_B$ is the boson part of the action,
\begin{equation}
   S_B\equiv\frac{1}{L_0L_1}\sum_p N^*(-p)N(p),\qquad
   N(p)\equiv2ip_zA(p)+W'(A)^*(p).
\label{eq:(2.14)}
\end{equation}
It is understood that the field product in~$W''(A)$ and~$W''(A)^*$ is
defined by the convolution of~Eq.~\eqref{eq:(2.13)}.

In order to define the functional integral, we then introduce the momentum
cutoff~$\Lambda$ and restrict the momentum as
\begin{equation}
   |p_\mu|\leq\Lambda\equiv\frac{\pi}{a}\qquad\text{for $\mu=0$ and~$1$}.
\label{eq:(2.15)}
\end{equation}
All dimensionful quantities are measured in units of~$a$. For notational
simplicity, we set~$a=1$. With this understanding,
\begin{equation}
   p_\mu=\frac{2\pi}{L_\mu}n_\mu,\qquad|n_\mu|\leq\frac{L_\mu}{2}.
\label{eq:(2.16)}
\end{equation}
We then define the partition function by
\begin{equation}
   \mathcal{Z}
   =\int\prod_{|p_\mu|\leq\pi}
   \left[dA(p)dA^*(p)\prod_{\alpha=1}^2d\psi_\alpha(p)
   \prod_{\Dot{\alpha}=\Dot{1}}^{\Dot{2}}d\Bar{\psi}_{\Dot{\alpha}}(p)\right]\,
   e^{-S}.
\label{eq:(2.17)}
\end{equation}

Equation~\eqref{eq:(2.12)} is the action in classical theory and thus is
invariant under the SUSY transformation and the translation. Since these
transformations act on field variables \emph{linearly\/} (see
Appendix~\ref{sec:A} for their explicit form) and do not change the momentum
label~$p$, these transformations preserve the restriction on the Fourier modes
in~Eq.~\eqref{eq:(2.16)}. As the consequence, our formulation
in~Eq.~\eqref{eq:(2.17)} manifestly preserves these
symmetries~\cite{Kadoh:2009sp}.

\subsection{Nicolai map}
Our definition of the partition function in the regularized level,
Eq.~\eqref{eq:(2.17)} of the 2D $\mathcal{N}=2$ WZ model allows the Nicolai
or Nicolai--Parisi--Sourlas
map~\cite{Nicolai:1979nr,Nicolai:1980jc,Parisi:1982ud,Cecotti:1983up}, which
renders the partition function Gaussian
integrals~\cite{Kadoh:2009sp}.\footnote{This feature is common to the lattice
formulation in~Refs.~\cite{Kikukawa:2002as,Kawai:2010yj}.} The point is that
the Dirac determinant in~Eq.~\eqref{eq:(2.17)} coincides with the Jacobian
associated with the change of integration variables from~$(A,A^*)$
to~$(N,N^*)$, the variables defined in~Eq.~\eqref{eq:(2.14)}, up to the sign:
\begin{equation}
   \det
   \begin{pmatrix}
   2ip_z&W''(A)^**\\W''(A)*&2ip_{\Bar{z}}
   \end{pmatrix}
   =\det\frac{\partial(N,N^*)}{\partial(A,A^*)}.
\label{eq:(2.18)}
\end{equation}
Hence, after the integration over the fermion fields, the partition function is
represented as
\begin{align}
   \mathcal{Z}
   &=\int\prod_{|p_\mu|\leq\pi}\left[dA(p)dA^*(p)\right]\,e^{- S_B}
   \det\frac{\partial(N,N^*)}{\partial(A,A^*)}
\notag\\
   &=\int\prod_{|p_\mu|\leq\pi}\left[dN(p)dN^*(p)\right]\,e^{-S_B}
   \sum_i\left.\sign\det
   \frac{\partial(N,N^*)}{\partial(A,A^*)}\right|_{A=A_i,A^*=A^*_i}.
\label{eq:(2.19)}
\end{align}
where $A_i$ ($i=1$, $2$, \dots) are solutions of the set of equations
\begin{equation}
   2ip_zA(p)+W'(A)^*(p)-N(p)=0,\qquad p_\mu=\frac{2\pi}{L_\mu}n_\nu,\qquad
   |n_\mu|\leq\frac{L_\mu}{2},
\label{eq:(2.20)}
\end{equation}
and $A^*_i$ are their complex conjugate. Note that, as Eq.~\eqref{eq:(2.14)}
shows, $e^{-S_B}$ is Gaussian in terms of the variables~$(N,N^*)$; this is,
thus, a drastic simplification.

The representation in~Eq.~\eqref{eq:(2.19)} thus presents the following
simulation algorithm~\cite{Beccaria:1998vi} (see
also~Ref.~\cite{Luscher:2009eq}):
\begin{enumerate}
\item Generate complex random numbers~$N(p)$ for each~$p_\mu$
in~Eq.~\eqref{eq:(2.16)} whose real and imaginary parts obey the Gaussian
distribution.
\item Solve the multi-variable algebraic equations of~Eq.~\eqref{eq:(2.20)}
numerically with respect to~$A$ and (ideally) find all the solutions $A_i$
($i=1$, $2$, \dots).
\item Calculate the following sums over solutions:
\begin{align}
   &\sum_i\left.\sign\det\frac{\partial(N,N^*)}{\partial(A,A^*)}
   \right|_{A=A_i,A^*=A^*_i},
\label{eq:(2.21)}
\\
   &\sum_i\left.\sign\det\frac{\partial(N,N^*)}{\partial(A,A^*)}\,
   \mathcal{O}(A, A^*)\right|_{A=A_i,A^*=A^*_i},
\label{eq:(2.22)}
\end{align}
where $\mathcal{O}$ is an observable of interest. In~Appendix~\ref{sec:B},
we present a fast algorithm for the computation
of~$\sign\det\frac{\partial(N,N^*)}{\partial(A,A^*)}$.

\item Repeat steps~(1)--(3) and compute the averages over configurations
of~$N$:
\begin{align}
   \Delta&\equiv
   \left\langle\sum_i\left.\sign\det\frac{\partial(N,N^*)}{\partial(A,A^*)}
   \right|_{A=A_i,A^*=A^*_i}\right\rangle,
\label{eq:(2.23)}
\\
   \langle\mathcal{O}\rangle
   &=\frac{1}{\Delta}
   \left\langle\sum_i\left.\sign\det\frac{\partial(N,N^*)}{\partial(A,A^*)}\,
   \mathcal{O}(A,A^*)
   \right|_{A=A_i,A^*=A^*_i}\right\rangle.
\label{eq:(2.24)}
\end{align}
\end{enumerate}
Here, $\Delta$ is the normalized partition function, i.e., the Witten
index~\cite{Witten:1982df,Cecotti:1981fu}.\footnote{In our numerical
simulations, we find that the statistical error of~$\Delta$ is much smaller
than that of the numerator in the ratio of~Eq.~\eqref{eq:(2.24)}. Hence, we
estimate the statistical error of~$\langle\mathcal{O}\rangle$ by a simple
error-propagation rule in the ratio.} If the superpotential~$W$ is a polynomial
of degree~$n$, we should have $\Delta=n-1$.\footnote{This can be seen by
counting the number of classical vacua.}

Since it is easy to generate Gaussian random numbers without any notable
autocorrelation, the above algorithm is completely free from any undesired
autocorrelation and the critical slowing down; this is a remarkable feature of
this algorithm.

Unfortunately, in step~(2) we cannot judge whether all the solutions
of~Eq.~\eqref{eq:(2.20)} are found or not because we cannot know a priori the
total number of solutions~$A_i$ for a given~$N$. The best thing we can do is to
collect as many solutions as possible. For this issue, the stability of the
number of solutions under the increase of initial trial solutions in the solver
algorithm, the agreement of~$\Delta$ with the expected Witten index and the
observation of expected SUSY WT relations provide some consistency checks. In
any case, the physical quantities we will compute in what follows, the scaling
dimension and the central charge, cannot be free from the systematic error
associated with the ``undiscovered solutions.'' It is difficult to estimate the
size of this systematic error at this time and the quoted values of the scaling
dimension and the central charge should be taken with this reservation.

\section{Simulation setup and classification of configurations}
\label{sec:3}
In this paper we consider the 2D WZ model of~Eq.~\eqref{eq:(2.8)} with the
superpotential
\begin{equation}
   W(\Phi)=\frac{\lambda}{n}\Phi^n
\label{eq:(3.1)}
\end{equation}
with~$n=3$ and~$4$, which will be written in the abbreviated forms
as~$W=\Phi^3$ and~$W=\Phi^4$, respectively. We set the coupling constant
\begin{equation}
   \lambda=0.3
\label{eq:(3.2)}
\end{equation}
in units of~$a=1$, as in~Refs.~\cite{Kawai:2010yj,Kamata:2011fr}.

To solve~Eq.~\eqref{eq:(2.20)} with respect to~$A$, we employ the
Newton--Raphson (NR) method.\footnote{For the generation of the configurations
of~$N$ and for the computation of~$A$ and
$\sign\det\frac{\partial(N,N^*)}{\partial(A,A^*)}$ we used a \texttt{C++}
library \texttt{Eigen}~\cite{eigen}. In particular, we extensively used the
class \texttt{PartialPivLU}.} The quality of the obtained configuration~$A$ is
estimated by the following norm of the residue:
\begin{equation}
   \sqrt{\frac{\sum_p\left|2ip_zA(p)+W'(A)(-p)^*-N(p)\right|^2}
   {\sum_q\left|N(q)\right|^2}}.
\label{eq:(3.3)}
\end{equation}
As we will see below, maximum values of this number are smaller
than~$10^{-14}$ for all obtained configurations, and this is much smaller than
the corresponding number in~Ref.~\cite{Kamata:2011fr}.

For a fixed configuration of~$N$, we randomly generate\footnote{The initial
value of the real and imaginary parts of~$A(p)$ is generated by the Gaussian
random number with unit variance as in~Ref.~\cite{Kamata:2011fr}.} initial
trial configurations of~$A$ so that we obtain $100$ solutions for~$A$ allowing
repetition of identical solutions; this is another improvement compared to the
setup of~Ref.~\cite{Kamata:2011fr}. A randomly generated initial configuration
does not necessarily converge to a solution along the iteration in the NR
method; sometimes it diverges and does not provide any
solution.\footnote{In~Ref.~\cite{Kamata:2011fr}, the number of \emph{initial
trial configurations\/} is fixed to~$100$ but we found that this choice
sometimes misses some solutions for~$A$, especially for~$W=\Phi^4$.} Two
solutions $A_1$ and~$A_2$ are regarded as identical if the norm of the
difference of the solutions,
\begin{equation}
   \sqrt{\frac{\sum_p\left|A_1(p)-A_2(p)\right|^2}{\sum_q\left|A_1(q)\right|^2}}
\end{equation}
is smaller than~$10^{-13}$.

For both cases~$W=\Phi^3$ and~$W=\Phi^4$, for each box size~$L$,
\begin{equation}
   L\equiv L_0=L_1,
\label{eq:(3.4)}
\end{equation}
we generate~$640$ configurations of~$N$ using the Gaussian random number. The
box size~$L$ is taken as even integers from~$8$ to~$36$ for~$W=\Phi^3$ and
from~$8$ to~$30$ for~$W=\Phi^4$.

We tabulate the classification of configurations we obtained
in~Tables~\ref{table:1}--\ref{table:3} for~$W=\Phi^3$ and
in~Tables~\ref{table:4}--\ref{table:6} for~$W=\Phi^4$. The symbols such
as~$({+}{+}{+}{-})_2$, for example, imply the following: For a certain
configuration of~$N$, we found four solutions~$A_i$ ($i=1$, \dots, $4$);
$\sign\det\frac{\partial(N,N^*)}{\partial(A,A^*)}$ at three of those solutions
is positive and negative at one solution. The subscript~$2(=1+1+1-1)$ stands
for the contribution of that $N$ configuration to~$\Delta$
in~Eq.~\eqref{eq:(2.23)}. Table~\ref{table:3}, for example, shows that
for~$L=36$ we had $13$ such configurations of~$N$ out of~$640$ configurations.

In the tables, to indicate the quality of the configurations obtained we
list~$\Delta$ from~Eq.~\eqref{eq:(2.23)}, which should reproduce $2$ and~$3$
for~$W=\Phi^3$ and~$W=\Phi^4$, respectively. For~$W=\Phi^3$, our simulation
gives $\Delta=2$ exactly for all box sizes. For~$W=\Phi^4$, $\Delta$ deviates
from~$3$ for~$L\geq26$ but only slightly; from this, it might be possible to
roughly estimate that the systematic error associated with the solution search
(i.e., the possibility that some solutions are missed) is less than~$0.5\%$
even for~$W=\Phi^4$.

For the same purpose, we also list the one-point function,
\begin{equation}
   \delta\equiv\frac{\left\langle S_B\right\rangle}{(L_0+1)(L_1+1)}-1,
\label{eq:(3.5)}
\end{equation}
where~$S_B$ is defined in~Eq.~\eqref{eq:(2.14)}, which should identically
vanish if the SUSY is exactly
preserved~\cite{Catterall:2001fr,Kamata:2011fr}.\footnote{For the calculation
of the one-point function~$\delta$ and succeeding numerical analyses, we used
the programming
language~\texttt{Julia}~\cite{Julia,Bezanson:xxxxa,Bezanson:xxxxb}.}

We also show the maximal value of the norm of the residue
in~Eq.~\eqref{eq:(3.3)} and the computation time in
$\text{core}\cdot\text{hour}$ on an Intel Xeon E5 $2.0\,\text{GHz}$.

\begin{table}[ht]
 \centering
 \caption{Classification of configurations for~$W=\Phi^3$.}
 \begin{tabular}{lrrrrr}\toprule
  $L$ & 8 & 10 & 12 & 14 & 16 \\\midrule
  $({+}{+})_2$ & 640 & 640 & 640 & 639 & 639 \\
  $({+}{+}{+}{-})_2$ & 0 & 0 & 0 & 1 & 1 \\
  $\Delta$ & 2 & 2 & 2 & 2 & 2 \\
  $\delta$ & 0.0070(44) & $-$0.0046(36) & 0.0019(30) & $-$0.0020(25) & $-$0.0003(23) \\
  \midrule
  $\text{core}\cdot\text{hour}$ [h] & 0.77 & 2.23 & 5.5 & 12.37 & 25.62 \\
  \bottomrule
 \end{tabular}
 \label{table:1}
\end{table}%
\begin{table}[ht]
 \centering
 \caption{Classification of configurations for~$W=\Phi^3$ (continued).}
 \begin{tabular}{lrrrrr}\toprule
  $L$ & 18 & 20 & 22 & 24 & 26 \\\midrule
  $({+}{+})_2$ & 634 & 636 & 634 & 637 & 635 \\
  $({+}{+}{+}{-})_2$ & 6 & 4 & 6 & 3 & 5 \\
  $\Delta$ & 2 & 2 & 2 & 2 & 2 \\
  $\delta$ & $-$0.0000(20) & $-$0.0015(19) & $-$0.0006(17) & 0.0001(16) & $-$0.0026(15) \\
  \midrule
  $\text{core}\cdot\text{hour}$ [h] & 48.97 & 87.03 & 143.83 & 236.62 & 405.28 \\
  \bottomrule
 \end{tabular}
\end{table}%
\begin{table}[ht]
 \centering
 \caption{Classification of configurations for~$W=\Phi^3$ (continued).}
 \begin{tabular}{lrrrrr}\toprule
  $L$ & 28 & 30 & 32 & 34 & 36 \\\midrule
  $({+}{+})_2$ & 634 & 626 & 633 & 628 & 627 \\
  $({+}{+}{+}{-})_2$ & 6 & 14 & 7 & 12 & 13 \\
  $\Delta$ & 2 & 2 & 2 & 2 & 2 \\
  $\delta$ & $-$0.0002(13) & 0.0000(13) & 0.0014(12) & 0.0008(11) & 0.0007(11) \\
  \midrule
  $\text{core}\cdot\text{hour}$ [h] & 649.78 & 963.93 & 1382.07 & 1936.52 & 2699.42 \\
  \bottomrule
 \end{tabular}
 \label{table:3}
\end{table}%

\begin{table}[ht]
 \centering
 \caption{Classification of configurations for~$W=\Phi^4$.}
\begin{tabular}{lrrrr}\toprule
  $L$ & 8 & 10 & 12 & 14 \\\midrule
  $({+}{+}{+})_3$ & 638 & 638 & 638 & 638 \\
  $({+}{+}{+}{+}{-})_3$ & 2 & 2 & 2 & 2 \\
  $({+}{+}{+}{+}{+}{-}{-})_3$ & 0 & 0 & 0 & 0 \\
  $({+}{+}{+}{+})_4$ & 0 & 0 & 0 & 0 \\
  $({+}{+}{+}{+}{+}{-})_4$ & 0 & 0 & 0 & 0 \\
  $({+}{+})_2$ & 0 & 0 & 0 & 0 \\
  $\Delta$ & 3 & 3 & 3 & 3 \\
  $\delta$ & 0.0003(45) & 0.0035(36) & 0.0001(30) & $-$0.0015(26) \\
  \midrule
  $\text{core}\cdot\text{hour}$ [h] & 3.73 & 12.8 & 36.1 & 89.55 \\
  \bottomrule
 \end{tabular}
  \label{table:4}
\end{table}%
\begin{table}[ht]
 \centering
 \caption{Classification of configurations for~$W=\Phi^4$ (continued).}
 \begin{tabular}{lrrrr}\toprule
  $L$ & 16 & 18 & 20 & 22 \\\midrule
  $({+}{+}{+})_3$ & 634 & 635 & 632 & 627 \\
  $({+}{+}{+}{+}{-})_3$ & 6 & 5 & 6 & 13 \\
  $({+}{+}{+}{+}{+}{-}{-})_3$ & 0 & 0 & 2 & 0 \\
  $({+}{+}{+}{+})_4$ & 0 & 0 & 0 & 0 \\
  $({+}{+}{+}{+}{+}{-})_4$ & 0 & 0 & 0 & 0 \\
  $({+}{+})_2$ & 0 & 0 & 0 & 0 \\
  $\Delta$ & 3 & 3 & 3 & 3 \\
  $\delta$ & 0.0006(25) & 0.0014(20) & 0.0024(20) & 0.0023(18) \\
  \midrule
  $\text{core}\cdot\text{hour}$ [h] & 202.65 & 425.23 & 872.03 & 1661.22 \\
  \bottomrule
 \end{tabular}
\end{table}%
\begin{table}[ht]
 \centering
 \caption{Classification of configurations for~$W=\Phi^4$ (continued).}
 \begin{tabular}{lrrrr}\toprule
  $L$ & 24 & 26 & 28 & 30 \\\midrule
  $({+}{+}{+})_3$ & 625 & 616 & 614 & 615 \\
  $({+}{+}{+}{+}{-})_3$ & 15 & 23 & 20 & 22 \\
  $({+}{+}{+}{+}{+}{-}{-})_3$ & 0 & 0 & 2 & 0 \\
  $({+}{+}{+}{+})_4$ & 0 & 1 & 3 & 2 \\
  $({+}{+}{+}{+}{+}{-})_4$ & 0 & 0 & 1 & 0 \\
  $({+}{+})_2$ & 0 & 0 & 0 & 1 \\
  $\Delta$ & 3 & 3.002(2) & 3.006(3) & 3.002(3) \\
  $\delta$ & 0.0000(16) & 0.0004(16) & 0.0023(17) & $-$0.0010(15) \\
  \midrule
  $\text{core}\cdot\text{hour}$ [h] & 2917.48 & 5004.37 & 8273.47 & 12905.13 \\
  \bottomrule
 \end{tabular}
 \label{table:6}
\end{table}%

\begin{figure}[ht]
 \begin{center}
  \begin{subfigure}{0.48\columnwidth}
   \begin{center}
    \includegraphics[width=\columnwidth]{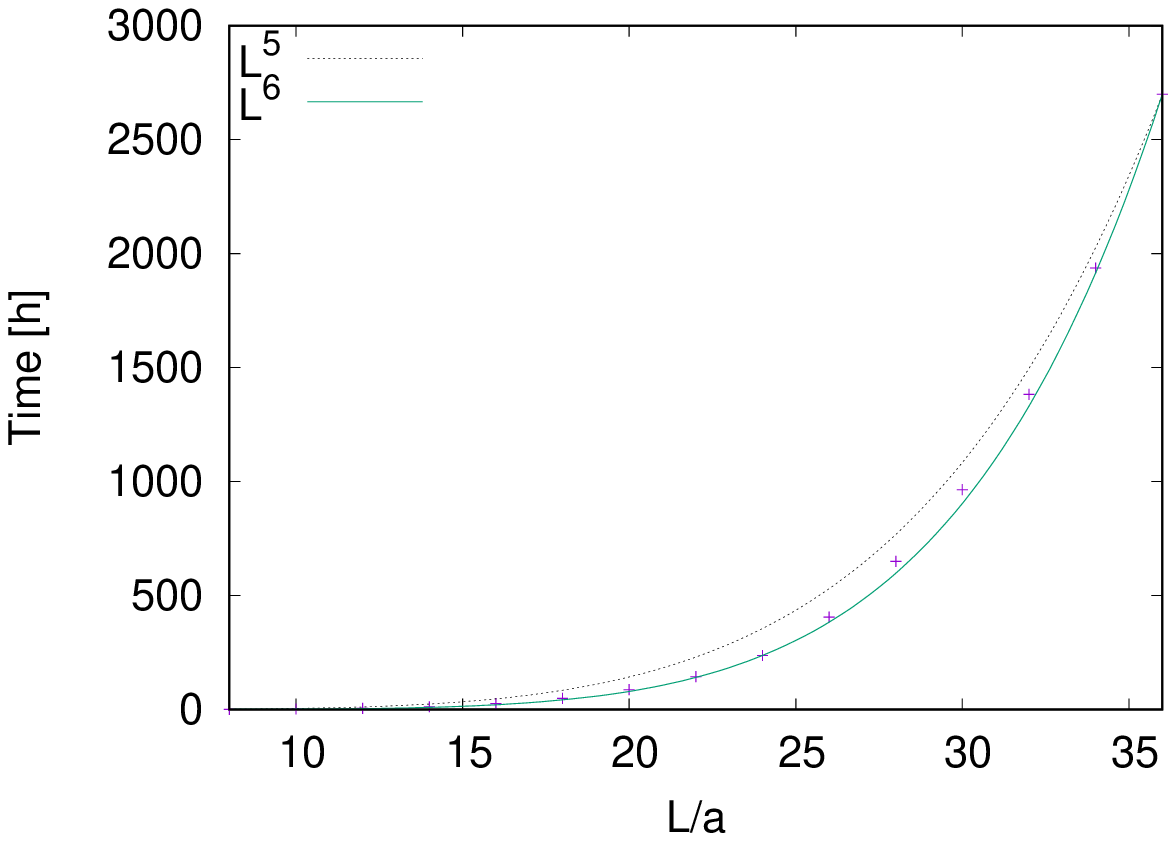}
    \caption{$W=\Phi^3$.}
   \end{center}
  \end{subfigure} \hspace*{1em}
  \begin{subfigure}{0.48\columnwidth}
   \begin{center}
    \includegraphics[width=\columnwidth]{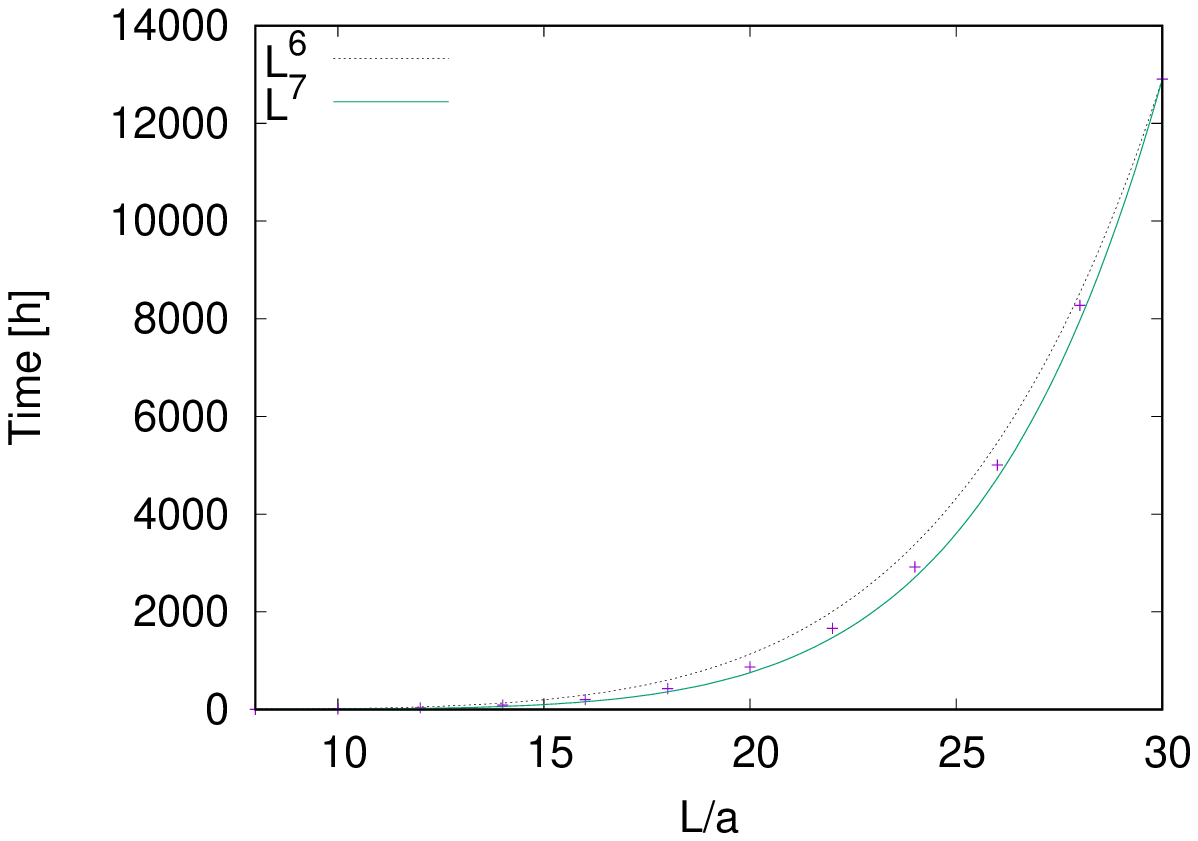}
    \caption{$W=\Phi^4$.}
   \end{center}
  \end{subfigure}
 \end{center}
 \caption{Computational time as a function of the lattice size.}
 \label{fig:time}
\end{figure}

The hot spot in our computation is the LU decomposition involved in the NR
method whose computational time scales as~$\propto N^3$ for a matrix of
size~$N$. Thus, we expect that the computational time scales as~$\propto L^6$
as a function of the lattice size~$L$. The actual computational time shown
in~Fig.~\ref{fig:time} is fairly well explained by this theoretical
expectation.

\section{SUSY Ward--Takahashi relation}
\label{sec:4}
As mentioned above, our formulation exactly preserves SUSY even with a finite
cutoff. Thus, barring the statistical error and the systematic error associated
with the solution search, SUSY WT relations should \emph{hold exactly} for
any parameter. The observation of these relations thus provides a useful check
of our simulation and gives a rough idea of the magnitude of the statistical
and systematic errors.

The simplest SUSY WT relation is $\delta=0$ for~$\delta$
in~Eq.~\eqref{eq:(3.5)}, and in~Tables~\ref{table:1}--\ref{table:6} we have
observed that this relation is reproduced quite well in our simulation. In this
section, we present results on two further SUSY WT relations on two-point
correlation functions which follow from the
identities~\cite{Kamata:2011fr}\footnote{In the present system, SUSY cannot be
spontaneously broken because of the non-zero Witten index.}
\begin{align}
   \left\langle Q_1(A(p)\Bar{\psi}_{\Dot{1}}(-p))\right\rangle&=0,
\label{eq:(4.1)}
\\
   \left\langle Q_2 (F^*(p) \psi_1(-p)) \right\rangle&=0,
\label{eq:(4.2)}
\end{align}
where the explicit form of the SUSY transformation is given
in~Appendix~\ref{sec:A}.

\begin{figure}[ht]
 \begin{center}
  \begin{subfigure}{0.45\columnwidth}
   \begin{center}
    \includegraphics[width=\columnwidth]{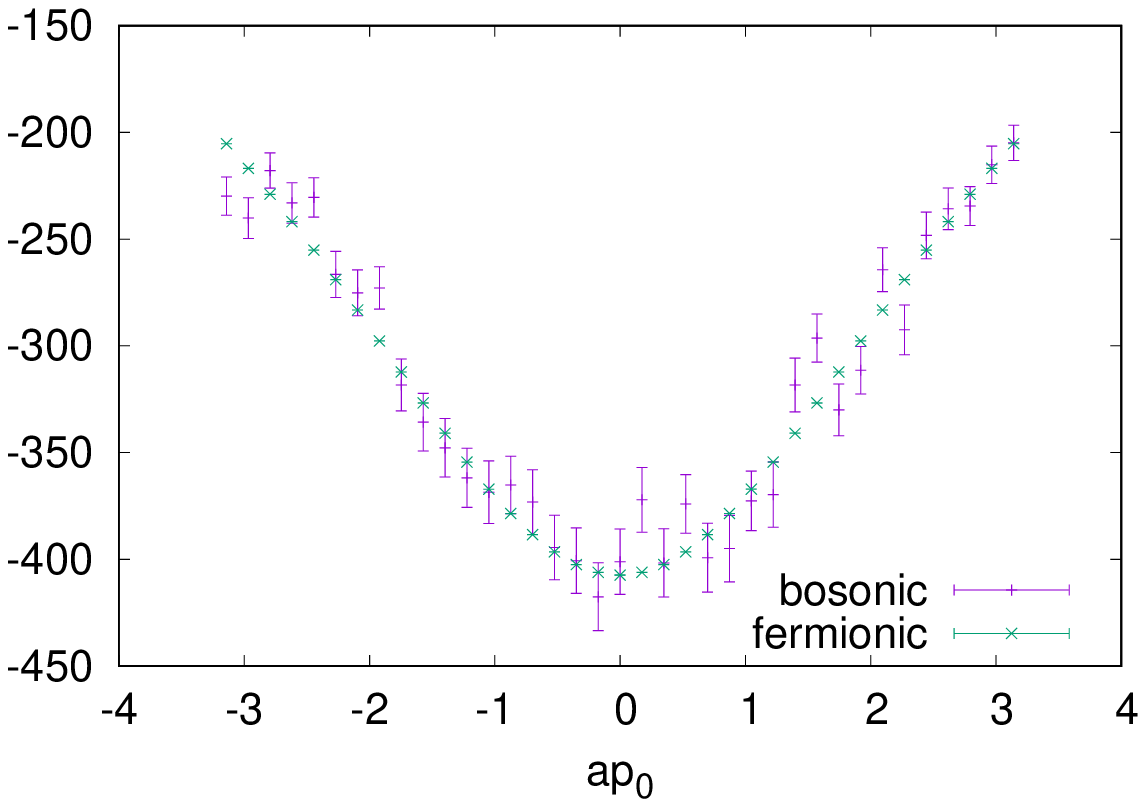}
    \caption{Real part~\eqref{eq:(4.4)}.}
   \end{center}
  \end{subfigure}\hspace{1em}
  \begin{subfigure}{0.45\columnwidth}
   \begin{center}
    \includegraphics[width=\columnwidth]{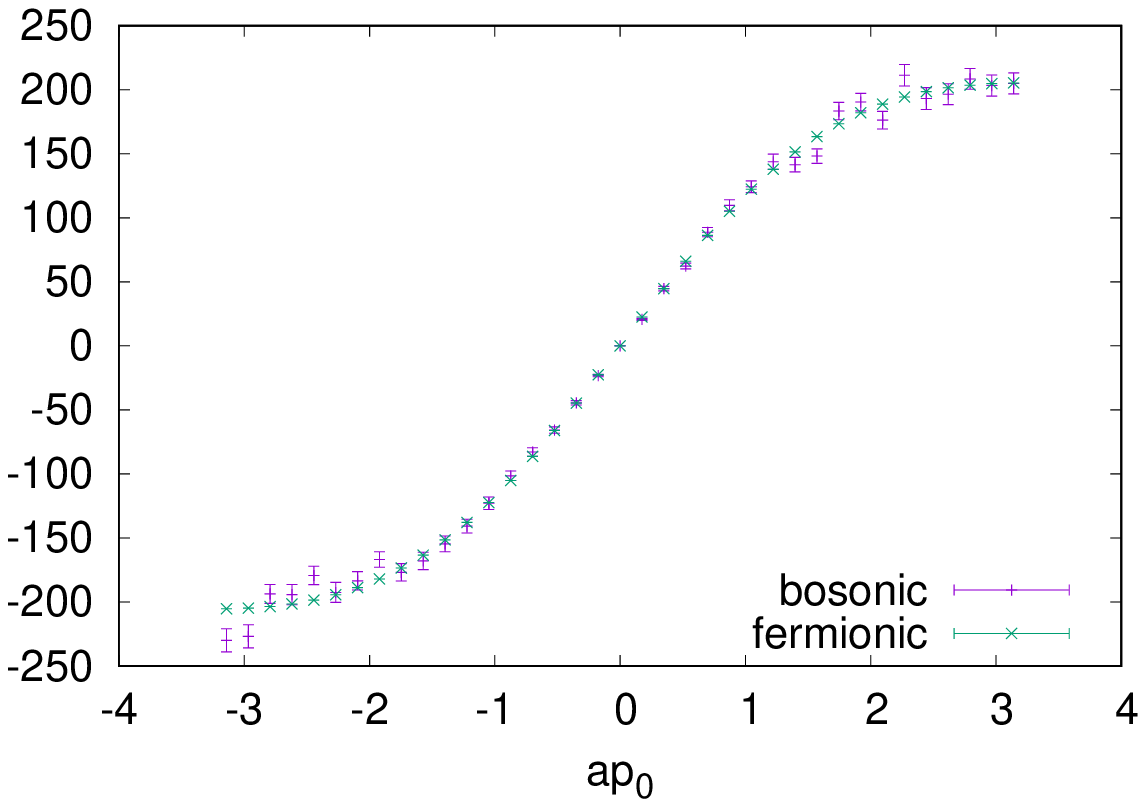}
    \caption{Imaginary part~\eqref{eq:(4.5)}.}
   \end{center}
  \end{subfigure}
 \end{center}
 \caption{SUSY WT relation of~Eq.~\eqref{eq:(4.3)} for~$W=\Phi^3$, $L=36$, and~$p_1=\pi$.}
 \label{fig:1}
\end{figure}

\begin{figure}[ht]
 \begin{center}
  \begin{subfigure}{0.45\columnwidth}
   \begin{center}
    \includegraphics[width=\columnwidth]{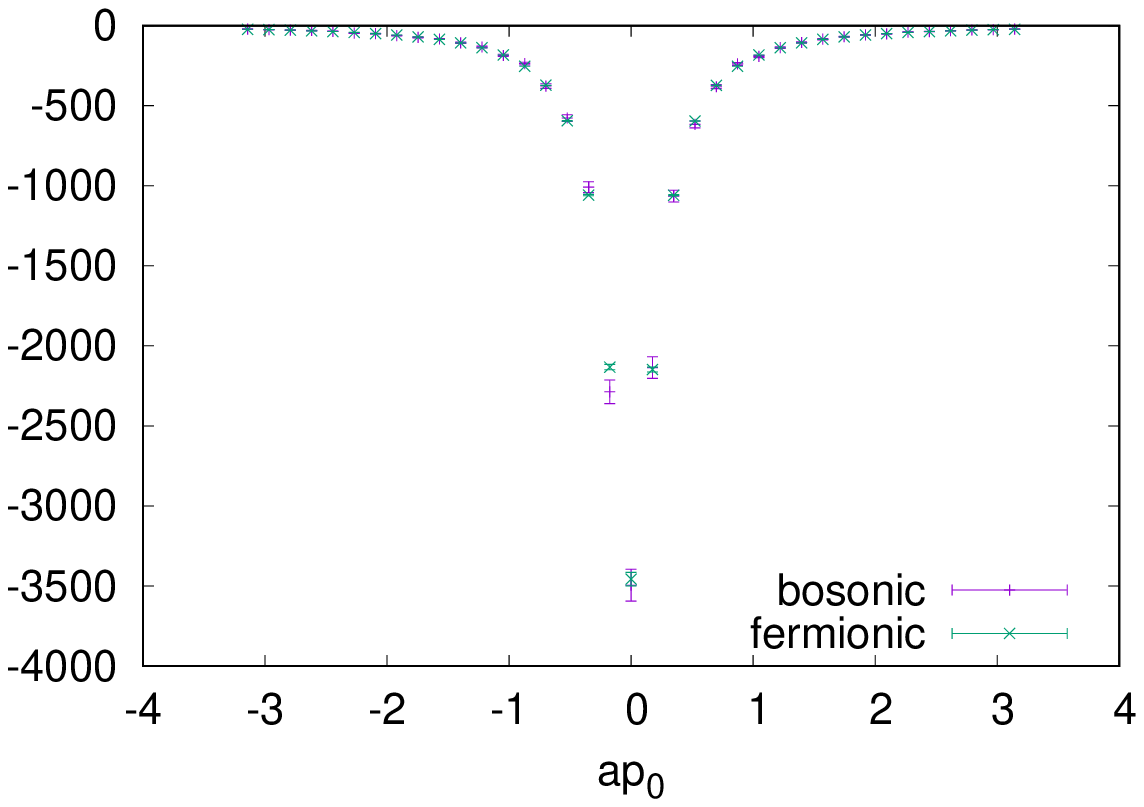}
    \caption{Real part~\eqref{eq:(4.4)}.}
   \end{center}
  \end{subfigure}\hspace{1em}
  \begin{subfigure}{0.45\columnwidth}
   \begin{center}
    \includegraphics[width=\columnwidth]{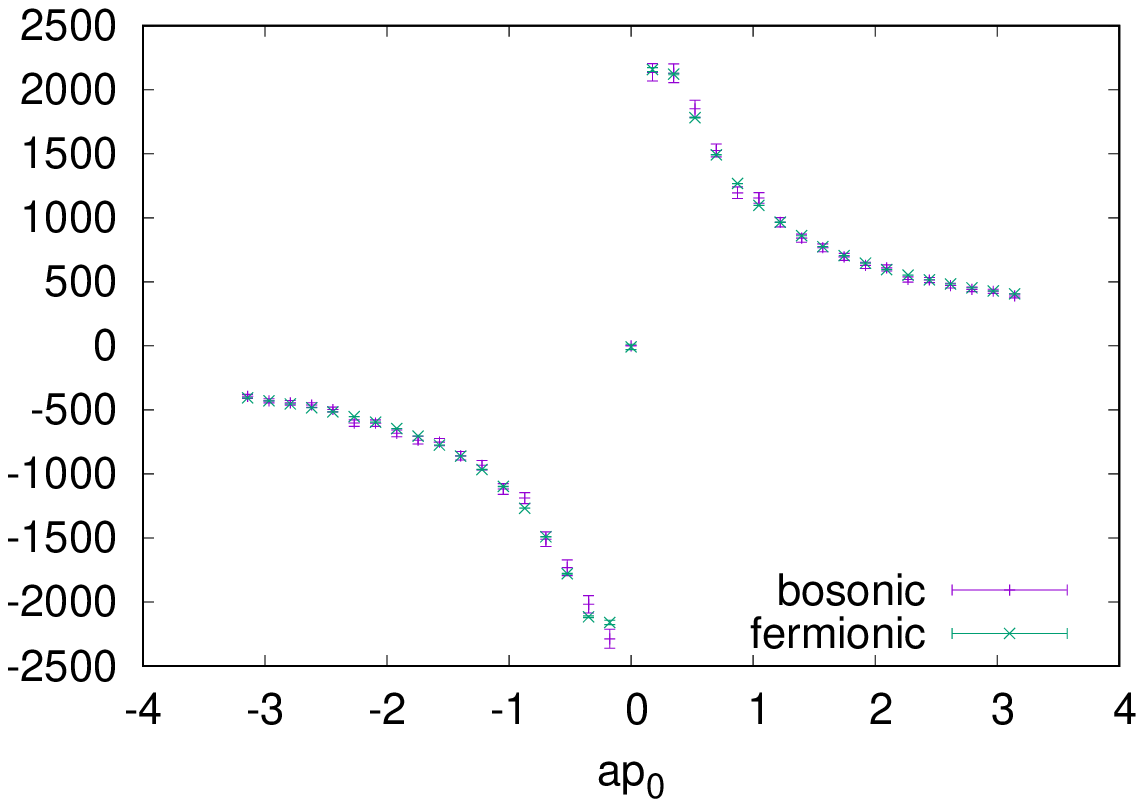}
    \caption{Imaginary part~\eqref{eq:(4.5)}.}
   \end{center}
  \end{subfigure} 
 \end{center}
 \caption{SUSY WT relation of~Eq.~\eqref{eq:(4.3)} for~$W=\Phi^3$, $L=36$, and~$p_1=\pi/18$.}
 \label{fig:2}
\end{figure}

\begin{figure}[ht]
 \begin{center}
  \begin{subfigure}{0.45\columnwidth}
   \begin{center}
    \includegraphics[width=\columnwidth]{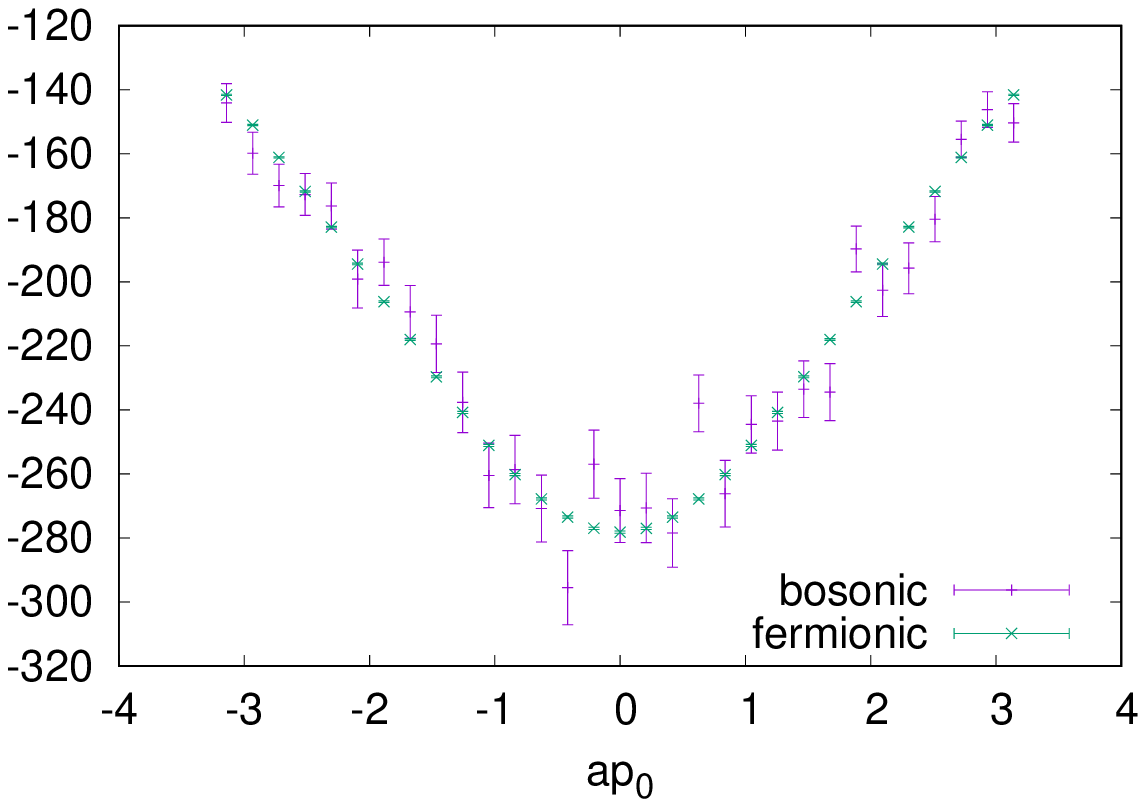}
    \caption{Real part~\eqref{eq:(4.4)}.}
   \end{center}
  \end{subfigure}\hspace{1em}
  \begin{subfigure}{0.45\columnwidth}
   \begin{center}
    \includegraphics[width=\columnwidth]{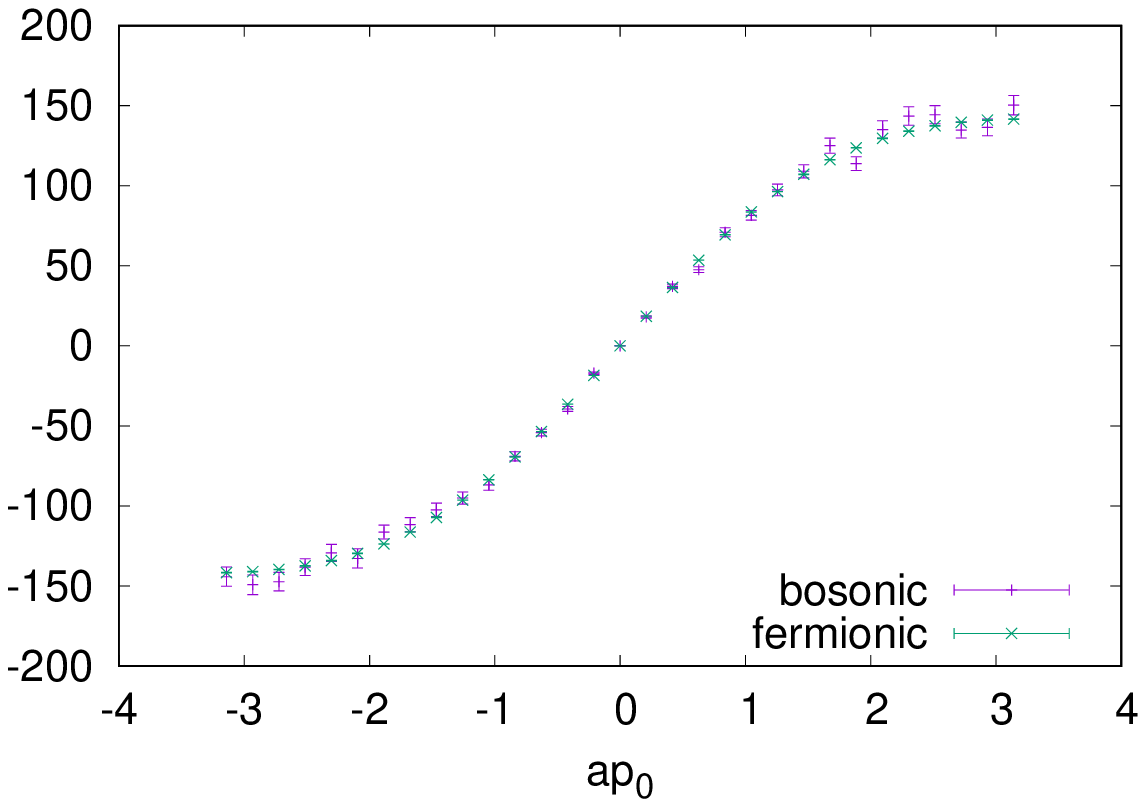}
    \caption{Imaginary part~\eqref{eq:(4.5)}.}
   \end{center}
  \end{subfigure}
 \end{center}
 \caption{SUSY WT relation of~Eq.~\eqref{eq:(4.3)} for~$W=\Phi^4$, $L=30$, and~$p_1=\pi$.}
 \label{fig:3}
\end{figure}

\begin{figure}[ht]
 \begin{center}
  \begin{subfigure}{0.45\columnwidth}
   \begin{center}
    \includegraphics[width=\columnwidth]{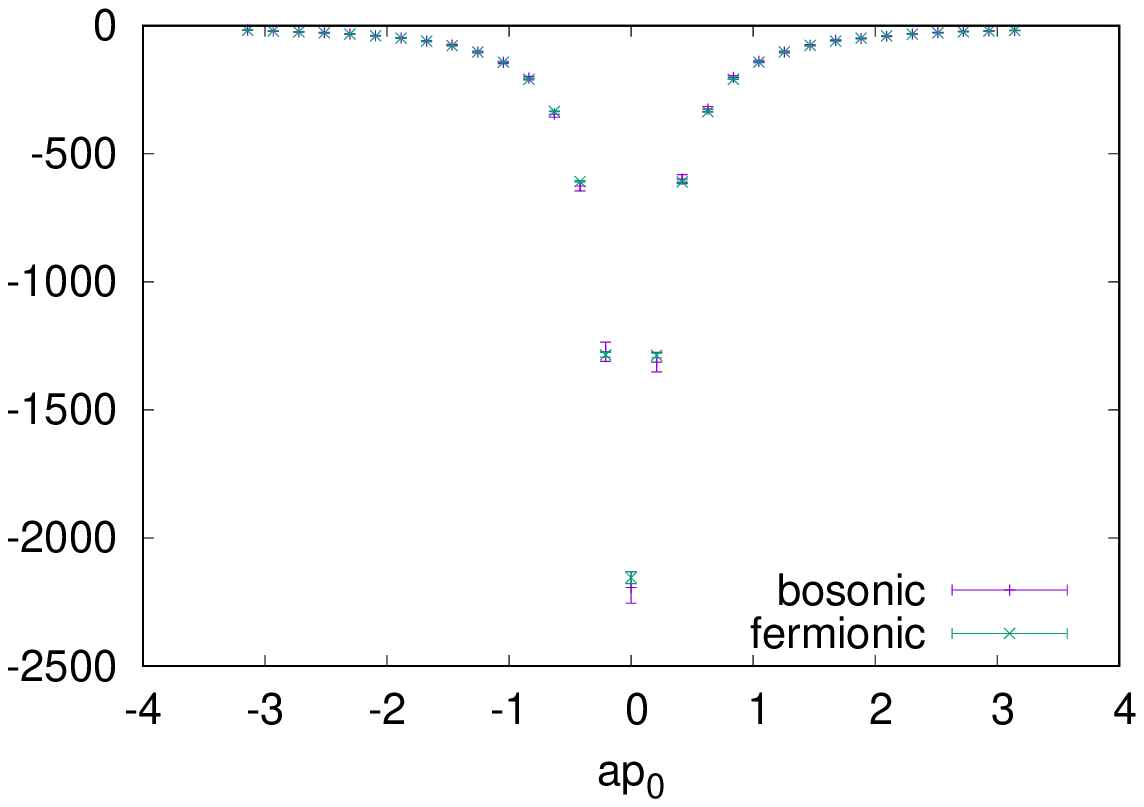}
    \caption{Real part~\eqref{eq:(4.4)}.}
   \end{center}
  \end{subfigure}\hspace{1em}
  \begin{subfigure}{0.45\columnwidth}
   \begin{center}
    \includegraphics[width=\columnwidth]{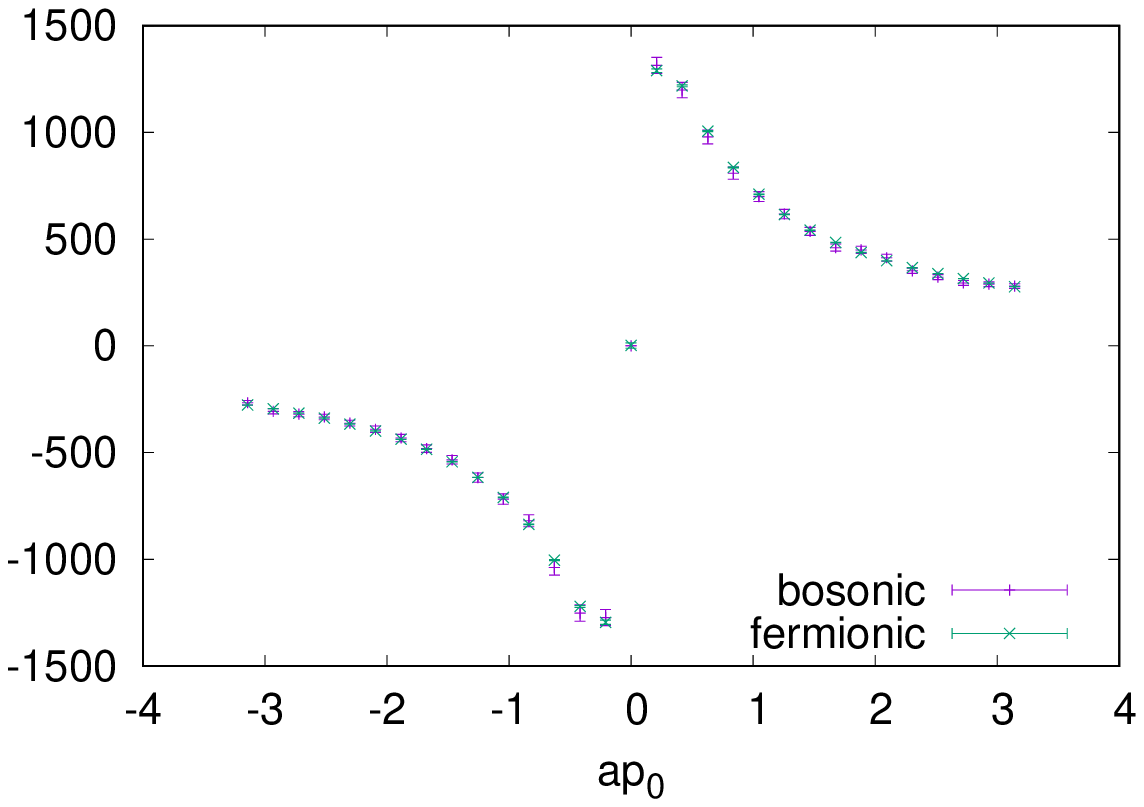}
    \caption{Imaginary part~\eqref{eq:(4.5)}.}
   \end{center}
  \end{subfigure}
 \end{center}
 \caption{SUSY WT relation of~Eq.~\eqref{eq:(4.3)} for~$W=\Phi^4$, $L=30$, and~$p_1=\pi/15$.}
 \label{fig:4}
\end{figure}

First, Eq.~\eqref{eq:(4.1)} yields
\begin{equation}
   2ip_{\Bar{z}}\left\langle A(p)A^*(-p)\right\rangle
   =-\left\langle\psi_1(p)\Bar{\psi}_{\Dot{1}}(-p)\right\rangle,
\label{eq:(4.3)}
\end{equation}
whose real and imaginary parts are
\begin{align}
   p_1\left\langle A(p)A^*(-p)\right\rangle
   &=\re\left\langle\psi_1(p)\Bar{\psi}_{\Dot{1}}(-p)\right\rangle,
\label{eq:(4.4)}
\\
   p_0\left\langle A(p)A^*(-p)\right\rangle
   &=-\im\left\langle\psi_1(p)\Bar{\psi}_{\Dot{1}}(-p)\right\rangle.
\label{eq:(4.5)}
\end{align}
In~Figs.~\ref{fig:1}--\ref{fig:4} we plot correlation functions in these
relations as functions of~$-\pi\leq p_0\leq\pi$. The box size is the maximal
one, i.e., $L=36$ for~$W=\Phi^3$ and $L=30$ for~$W=\Phi^4$. The spatial
momentum $p_1$ is fixed to be~$p_1=\pi$ (the largest positive value)
or~$p_1=2\pi/L$ (the smallest positive value). In the figures, the left panel
corresponds to the real part relation of~Eq.~\eqref{eq:(4.4)} and the right one
to the imaginary part of~Eq.~\eqref{eq:(4.5)}. In the plots, ``bosonic''
implies the correlation function on the left-hand side of the WT relation and
``fermionic'' implies the correlation function on the right-hand side. Errors
are statistical only.


\begin{figure}[ht]
 \begin{center}
  \begin{minipage}{0.45\columnwidth}
   \begin{center}
    \includegraphics[width=\columnwidth]{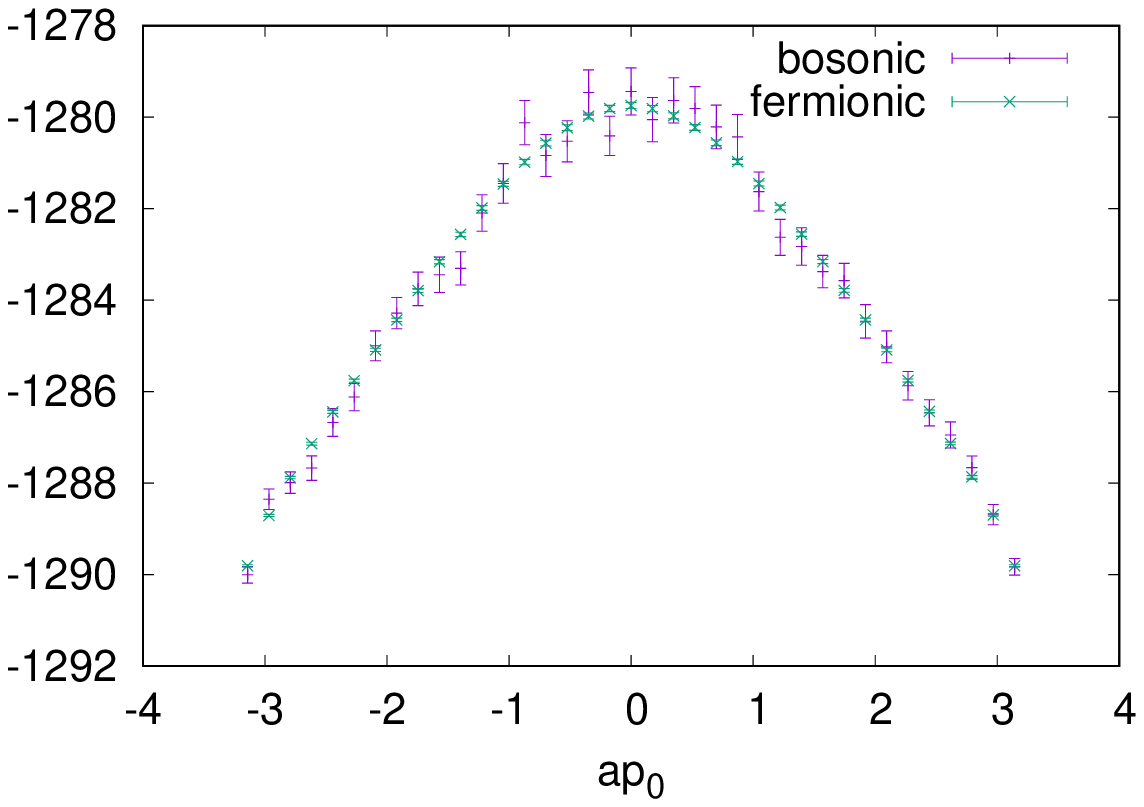}
    \caption{SUSY WT relation of~Eq.~\eqref{eq:(4.7)} for~$W=\Phi^3$, $L=36$, and~$p_1=\pi$.}
    \label{fig:5}
   \end{center}
  \end{minipage}\hspace{1em}
  \begin{minipage}{0.45\columnwidth}
   \begin{center}
    \includegraphics[width=\columnwidth]{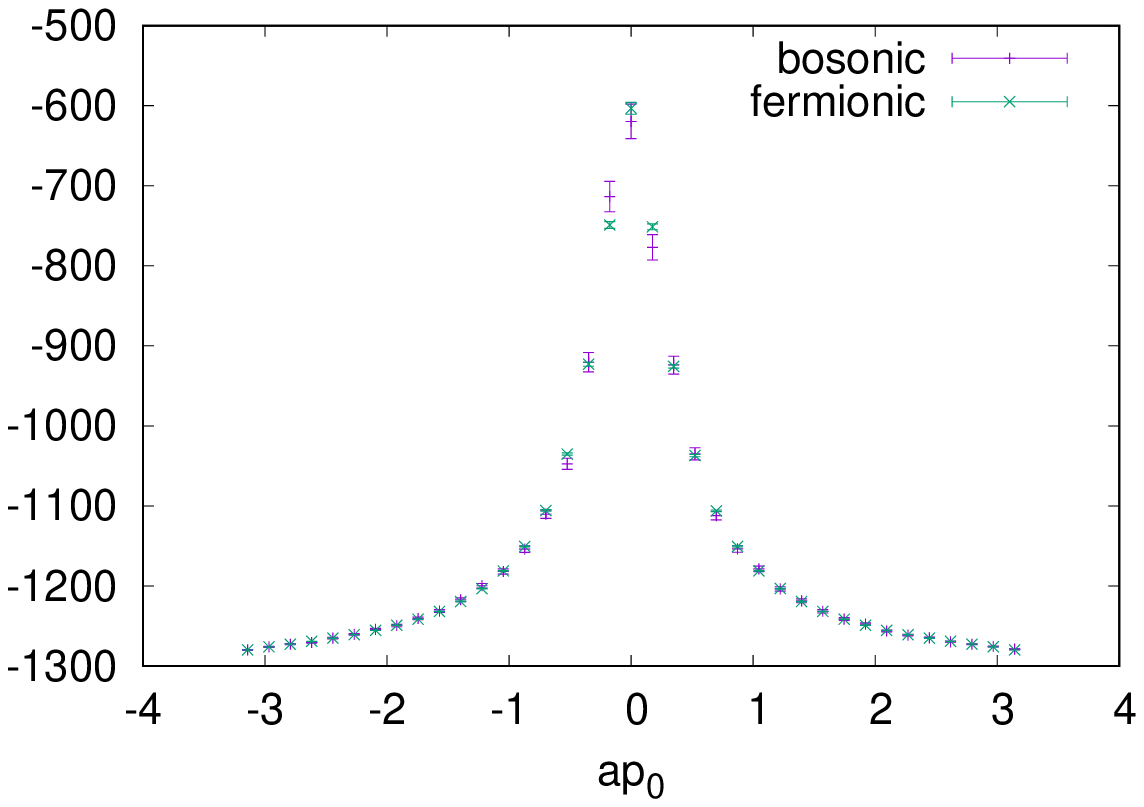}
    \caption{SUSY WT relation of~Eq.~\eqref{eq:(4.7)} for~$W=\Phi^3$, $L=36$, and~$p_1=\pi/18$.}
    \label{fig:6}
   \end{center}
  \end{minipage}
 \end{center}
\end{figure}

\begin{figure}[ht]
 \begin{center}
  \begin{minipage}{0.45\columnwidth}
   \begin{center}
    \includegraphics[width=\columnwidth]{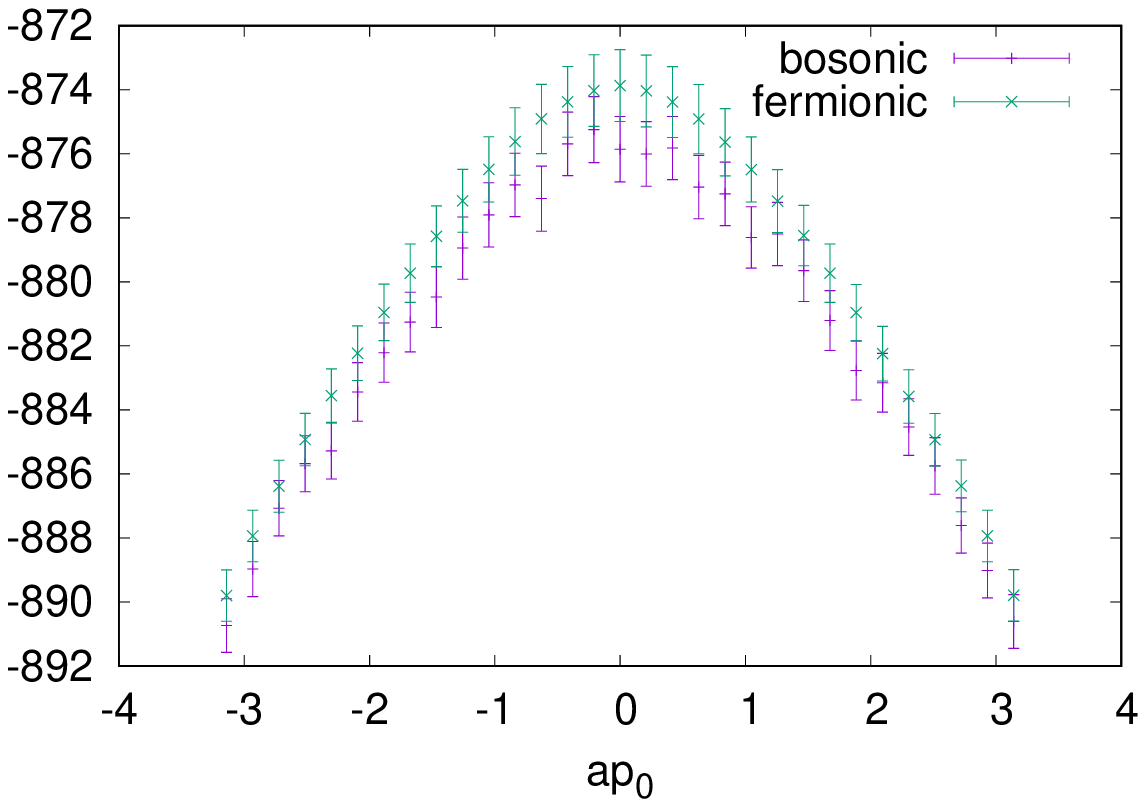}
    \caption{SUSY WT relation of~Eq.~\eqref{eq:(4.7)} for~$W=\Phi^4$, $L=30$, and~$p_1=\pi$.}
    \label{fig:7}
   \end{center}
  \end{minipage}\hspace{1em}
  \begin{minipage}{0.45\columnwidth}
   \begin{center}
    \includegraphics[width=\columnwidth]{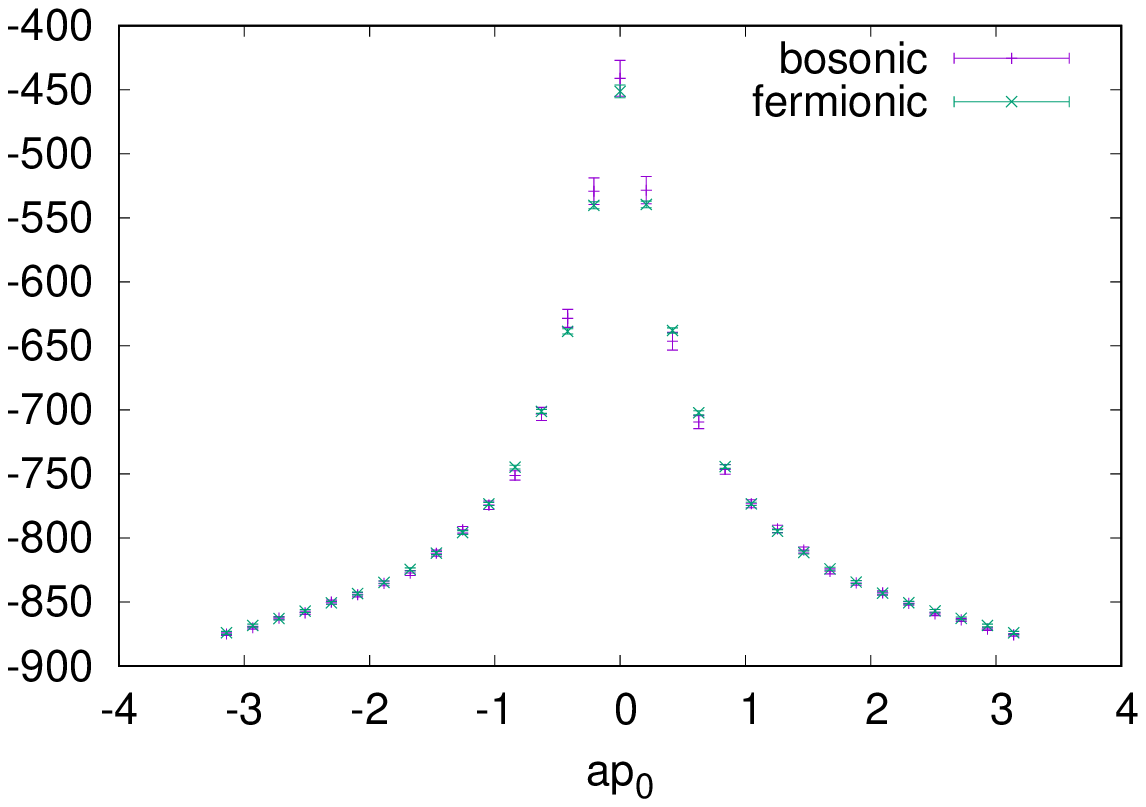}
    \caption{SUSY WT relation of~Eq.~\eqref{eq:(4.7)} for~$W=\Phi^4$, $L=30$, and~$p_1=\pi/15$.}
    \label{fig:8}
   \end{center}
  \end{minipage}
 \end{center}
\end{figure}

Next, Eq.~\eqref{eq:(4.2)} gives the relation
\begin{equation}
   \left\langle F(p)F^*(-p)\right\rangle
   =-2ip_z\left\langle\psi_1(p)\Bar{\psi}_{\Dot{1}}(-p)\right\rangle,
\label{eq:(4.6)}
\end{equation}
and the real and imaginary parts are given by
\begin{align}
   \left\langle F(p)F^*(-p)\right\rangle
   &=-p_1\re\left\langle\psi_1(p)\Bar{\psi}_{\Dot{1}}(-p)\right\rangle
   +p_0\im\left\langle\psi_1(p)\Bar{\psi}_{\Dot{1}}(-p)\right\rangle,
\label{eq:(4.7)}\\
   0&=-p_0\re\left\langle\psi_1(p)\Bar{\psi}_{\Dot{1}}(-p)\right\rangle
   -p_1\im\left\langle\psi_1(p)\Bar{\psi}_{\Dot{1}}(-p)\right\rangle.
\label{eq:(4.8)}
\end{align}
In~Figs.~\ref{fig:5}--\ref{fig:8} we plot correlation functions in the
real part relation of~Eq.~\eqref{eq:(4.7)}; the other conditions and
conventions are the same as above. For the computation of the left-hand side
of~Eq.~\eqref{eq:(4.7)} we have used the representation\footnote{A way to
derive this relation is to introduce the source term for the auxiliary field:
\begin{equation}
   S_J=\frac{1}{L_0L_1}\sum_p\left[F^*(-p)J(p)+J^*(-p)F(p)\right].
\end{equation}
Then, after a (formal) Gaussian integration over the auxiliary field, this term
changes to
\begin{equation}
   S_J\to\frac{1}{L_0L_1}\sum_p
   \left[-W'(A)(-p)J(p)-J^*(-p)W'(A)^*(p)+J^*(-p)J(p)\right].
\end{equation}
Therefore,
\begin{align}
   &\left\langle F^*(-p) F(p) \right\rangle
\notag\\
   &=(L_0L_1)^2\frac{\delta}{\delta J(p)}\frac{\delta}{\delta J^*(-p)}
\notag\\
   &\qquad{}
   \times\left.\left\langle\exp\left\{\frac{1}{L_0L_1}\sum_q
   \left[W'(A)(-q)J^*(q)+J(-q)W'(A)^*(q)-J(-q)J^*(q)\right]\right\}
   \right\rangle\right|_{J=0,J^*=0}
\notag\\
   &=\left\langle W'(A)^*(p)W'(A)(-p)\right\rangle
   -L_0L_1.
\end{align}
}
\begin{align}
   \left\langle F(p)F^*(-p)\right\rangle
   &=\left\langle W'(A)^*(p)W'(A)(-p)\right\rangle-L_0L_1
\notag\\
   &=\left\langle\left|N(p)-(ip_0+p_1)A(p)\right|^2\right\rangle-L_0L_1.
\label{eq:(4.9)}
\end{align}


\begin{figure}[ht]
 \begin{center}
  \begin{subfigure}{0.45\columnwidth}
   \begin{center}
    \includegraphics[width=\columnwidth]{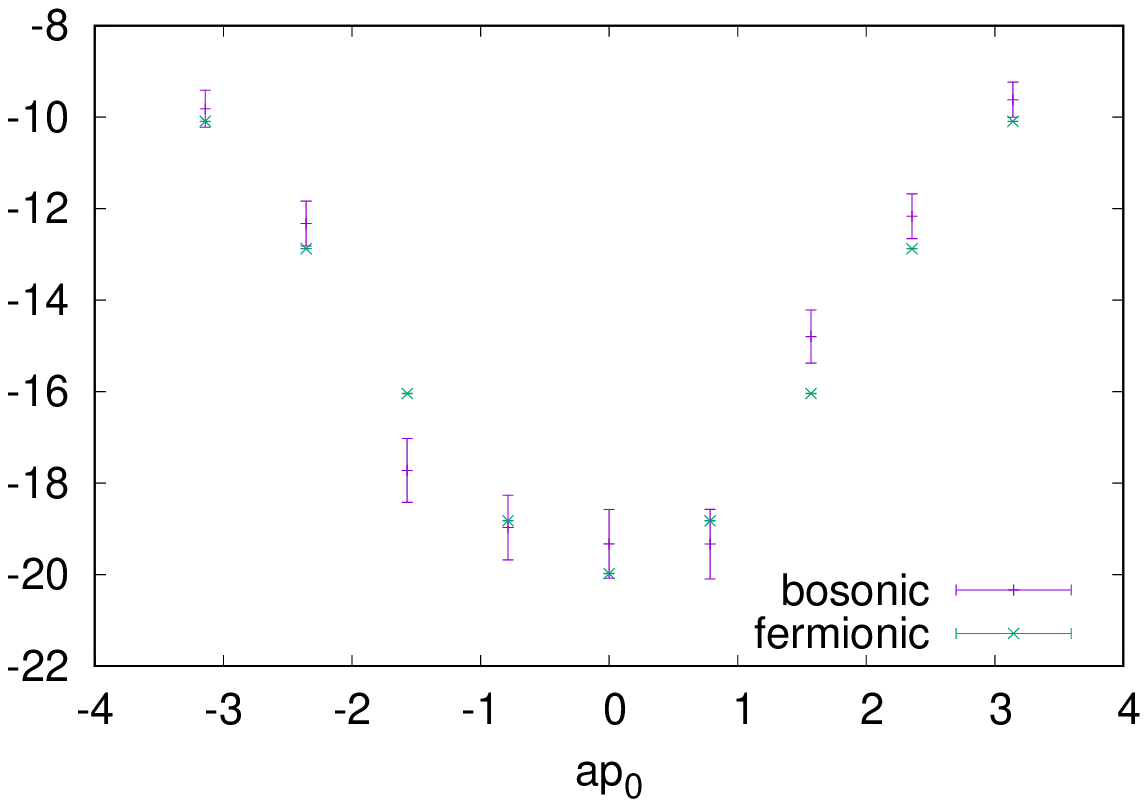}
    \caption{Number of configurations: $640$.}
   \end{center}
  \end{subfigure}\hspace{1em}
  \begin{subfigure}{0.45\columnwidth}
   \begin{center}
    \includegraphics[width=\columnwidth]{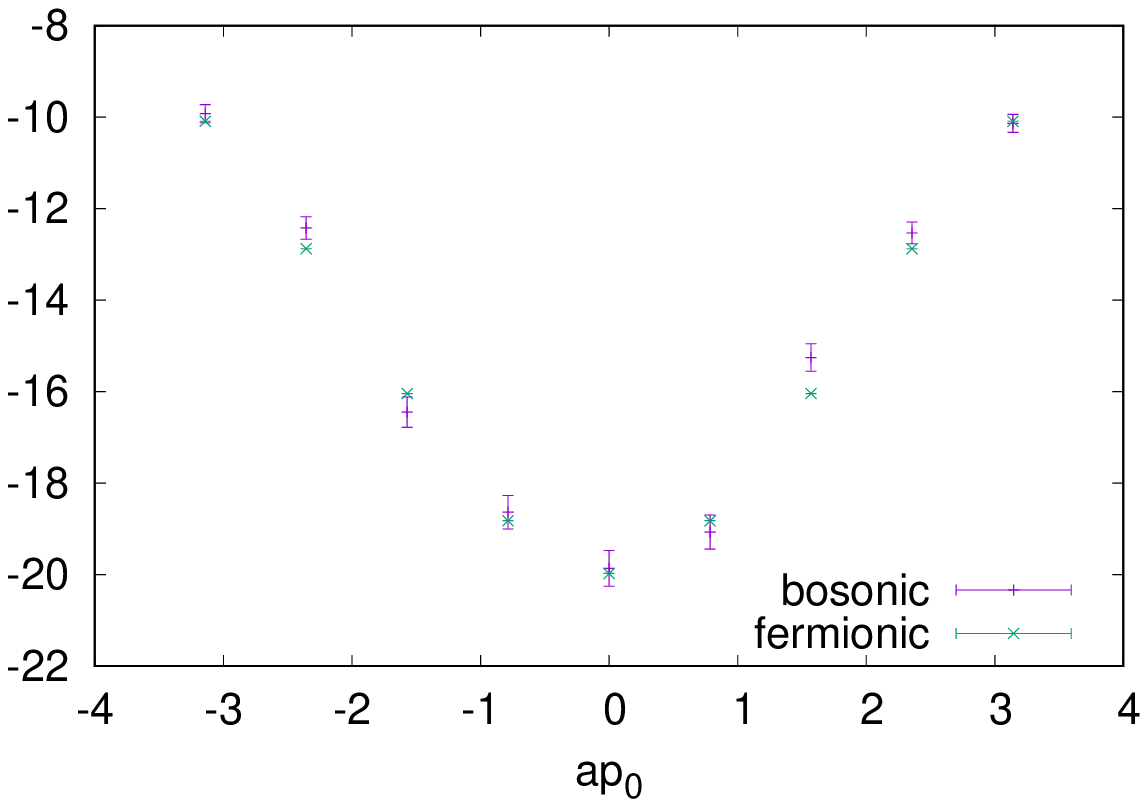}
    \caption{Number of configurations: $2560$.}
   \end{center}
  \end{subfigure}
 \end{center}
 \caption{SUSY WT relation of~Eq.~\eqref{eq:(4.4)} for~$W=\Phi^3$, $L=8$ and~$p_1=\pi$.}
 \label{fig:9}
\end{figure}

\begin{figure}[ht]
 \begin{center}
  \begin{subfigure}{0.45\columnwidth}
   \begin{center}
    \includegraphics[width=\columnwidth]{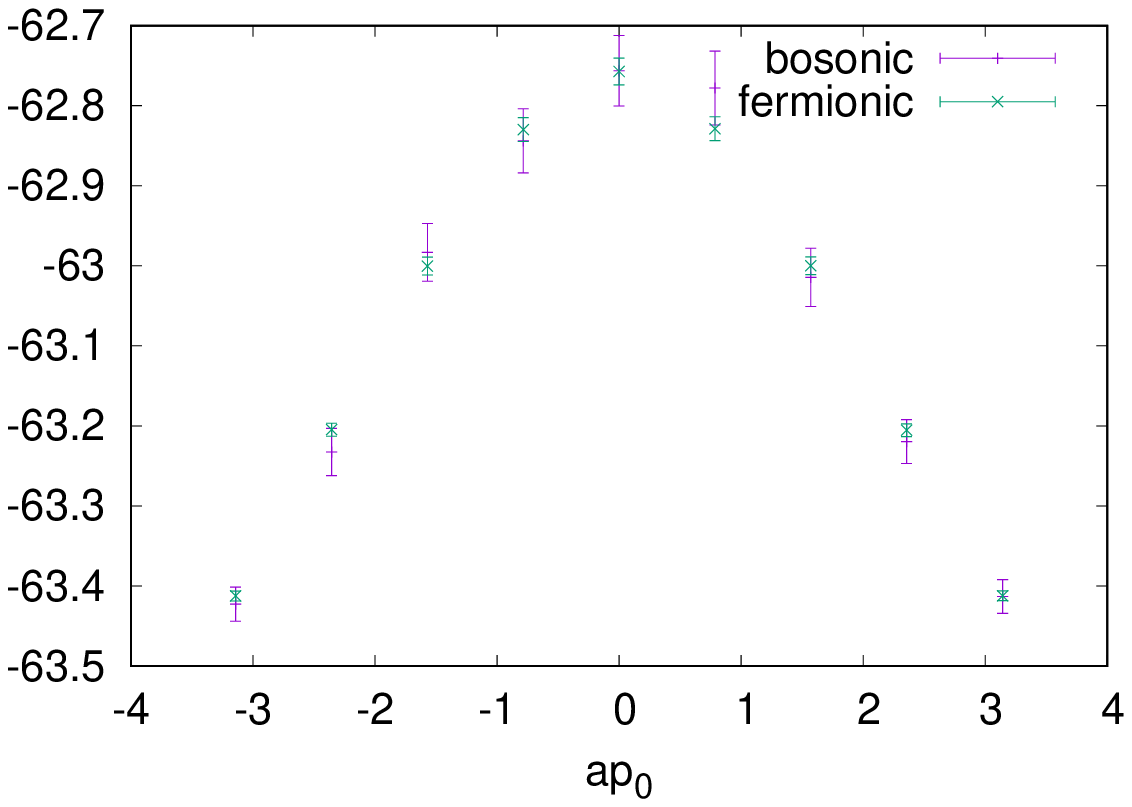}
    \caption{Number of configurations: $640$.}
   \end{center}
  \end{subfigure}\hspace{1em}
  \begin{subfigure}{0.45\columnwidth}
   \begin{center}
    \includegraphics[width=\columnwidth]{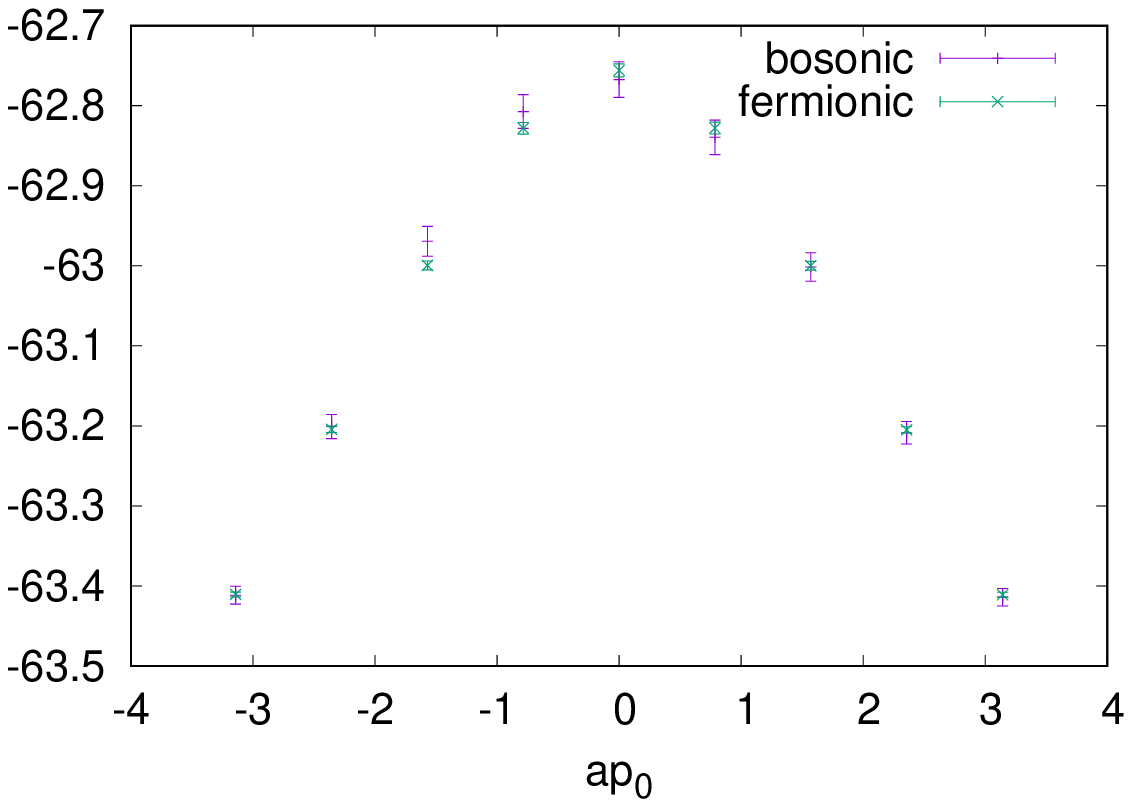}
    \caption{Number of configurations: $2560$.}
   \end{center}
  \end{subfigure}
 \end{center}
 \caption{SUSY WT relation of~Eq.~\eqref{eq:(4.7)} for~$W=\Phi^3$, $L=8$ and~$p_1=\pi$.}
 \label{fig:10}
\end{figure}

If the WT relations hold exactly, the ``bosonic'' points and the ``fermionic''
points in the plots should coincide with each other. Overall, we observe good
agreements within~$1\sigma$, as expected. However, there still exist some
deviations of order~$2\sigma$, especially in the real-part WT relations at
the largest spatial momentum~$p_1=\pi$. To argue that these deviations are a
result of statistical fluctuations and not due to the omission of some
solutions in our solution search, we carried out the measurements corresponding
to the left panels of~Figs.~\ref{fig:1} and~\ref{fig:5}, respectively but
for~$L=8$, by changing the number of configurations by four times, i.e., $640$
and~$2560$. The results are shown in~Figs.~\ref{fig:9} and~\ref{fig:10}. We see
that although for $640$ configurations there exist some discrepancies between
the ``bosonic'' and ``fermionic'' ones of order~$2\sigma$, when we increase the
number of configurations by four times, the statistical error is halved and the
discrepancies of the central values actually decrease. From this behavior, we
think that the observed discrepancies in the WT relations are due to
statistical fluctuations and they eventually disappear as the number of
configurations is increased sufficiently.


Finally, we mention a general tendency of the statistical error in the
correlation functions we found through the numerical simulation. Particularly
in the high momentum region, the correlation functions of the scalar field
suffer from larger statistical fluctuations than those of the fermion field (as
seen in the left panel of~Fig.~\ref{fig:1}). Actually, because of this problem
we could not directly examine four-point SUSY WT relations including a
four-point correlation function of~$A$ and~$A^*$. On the other hand, if we
assume the validity of SUSY WT relations, we can use them to rewrite some noisy
correlation functions into less noisy ones. This technique will be employed
frequently in the following sections.

\section{Scaling dimension}
\label{sec:5}
In this section, we measure the scaling dimension of the scalar field in the
IR limit from the two-point correlation function. If the expected LG
correspondence for the WZ model with~$W=\Phi^n$ holds, the chiral superfield is
identified with the chiral primary field in the $A_{n-1}$ minimal model with
the conformal dimension
\begin{equation}
   h=\Bar{h}=\frac{1}{2n}.
\label{eq:(5.1)}
\end{equation}
Thus the two-point function of the scalar field is expected to behave as
\begin{equation}
   \left\langle A(x)A^*(0)\right\rangle
   \propto\frac{1}{z^{2h}\Bar{z}^{2\Bar{h}}},
\label{eq:(5.2)}
\end{equation}
for large~$|z|$. To obtain the value of the scaling dimension~$h+\Bar{h}$,
in~Ref.~\cite{Kamata:2011fr}, the authors computed the susceptibility
\begin{equation}
   \chi_\phi
   \equiv\frac{1}{a^2}\int_{L_0L_1}d^2x\,
   \left\langle A(x)A^*(0)\right\rangle.
\label{eq:(5.3)}
\end{equation}
To avoid the UV ambiguity at the contact point~$x\sim0$, a small region
around~$x=0$ was excised~\cite{Kawai:2010yj}. Then, for the scaling dimension,
they obtained
\begin{equation}
   1-h-\Bar{h}=0.616(25)(13).
\label{eq:(5.4)}
\end{equation}
The expected value is~$1-h-\Bar{h}=2/3=0.666\dots$ for the~$A_2$ minimal model.
It turns out, however, that the susceptibility in~Eq.~\eqref{eq:(5.3)} is quite
sensitive to the size of the excised region with the formulation
of~Ref.~\cite{Kamata:2011fr}.

Here, we instead directly study the correlation function in the momentum space
$\langle A(p)A^*(-p)\rangle$. The Fourier transformation
of~Eq.~\eqref{eq:(5.2)} reads (assuming~$h=\Bar{h}$)
\begin{equation}
   \left\langle A(p)A^*(-p)\right\rangle
   \propto\frac{1}{(p^2)^{1-h-\Bar{h}}},
\label{eq:(5.5)}
\end{equation}
for $|p|$ small.

Also since the SUSY WT relation of~Eq.~\eqref{eq:(4.3)} shows that
\begin{equation}
   \left\langle\psi_1(p)\Bar{\psi}_{\Dot{1}}(-p)\right\rangle
   =-2ip_{\Bar{z}}\left\langle A(p)A^*(-p)\right\rangle
\label{eq:(5.6)}
\end{equation}
instead of the two-point function of the scalar field, we may use the
two-point function of the fermion field, which is less noisy, as already
mentioned.

\begin{figure}[ht]
 \begin{center}
  \begin{subfigure}{0.48\columnwidth}
   \begin{center}
    \includegraphics[width=\columnwidth]{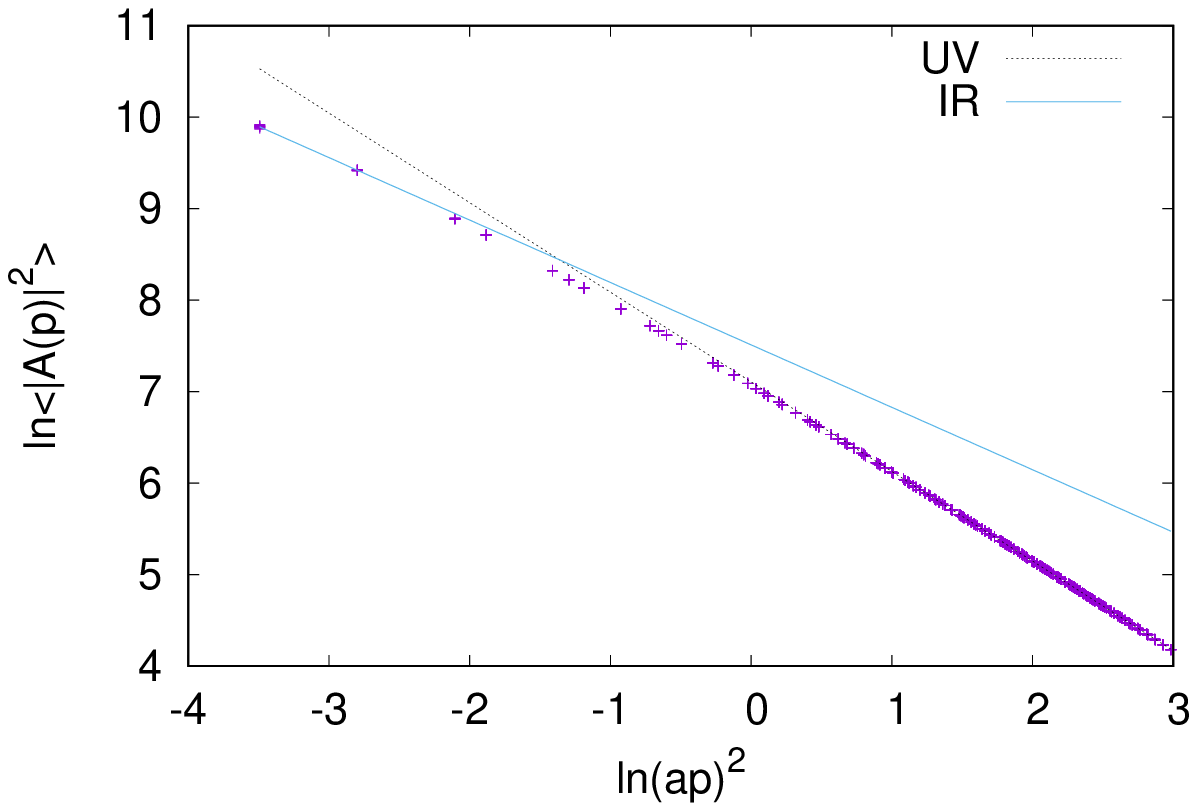}
    \caption{$W=\Phi^3$, $L=36$.}
    \label{fig:11(a)}
   \end{center}
  \end{subfigure} \hspace*{1em}
  \begin{subfigure}{0.48\columnwidth}
    \begin{center}
     \includegraphics[width=\columnwidth]{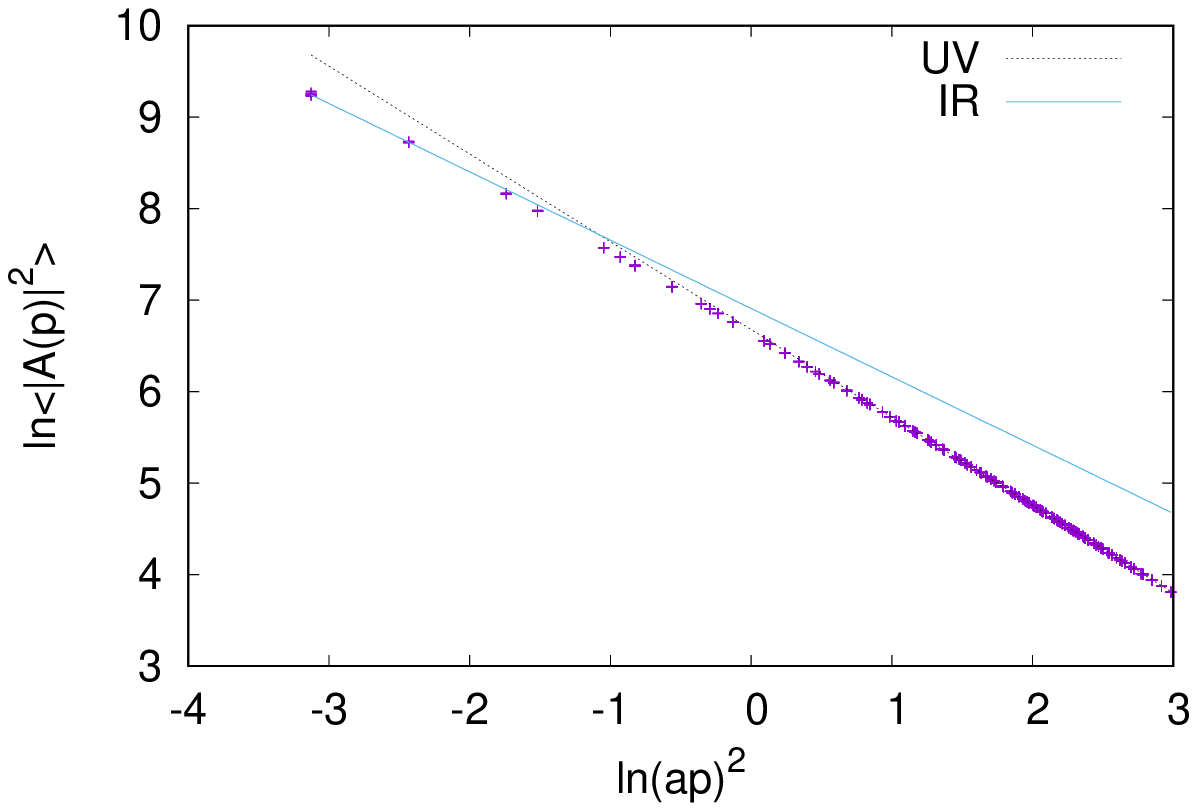}
     \caption{$W=\Phi^4$, $L=30$.}
     \label{fig:11(b)}
    \end{center}
  \end{subfigure}
 \end{center}
 \caption{$\ln\langle A(p)A^*(-p)\rangle$ as a function of~$\ln p^2$. The
 broken and solid lines are linear fits in the UV and IR regions,
 respectively.}
\label{fig:11}
\end{figure}

Figure~\ref{fig:11} shows~$\ln\langle A(p)A^*(-p)\rangle$ as a function
of~$\ln p^2$ in the case of the maximal box size, i.e., $L=36$ for~$W=\Phi^3$
and $L=30$ for~$W=\Phi^4$, respectively. We also show the fitting lines in
the UV region~$\frac{\pi}{\sqrt{2}}\leq|p|<\pi$ and in the IR
region~$\frac{2\pi}{L}\leq|p|<\frac{4\pi}{L}$. Table~\ref{table:7} summarizes
the scaling dimension obtained from the linear fit in the IR region, which is
one of our main results in this paper. Recall, however, that those numbers
may contain the systematic error associated with the solutions undiscovered
by the NR method.

It may be of some interest to see how the values are changed if we do not
include a few percent of ``strange solutions'' in~Tables
\ref{table:1}--\ref{table:6}, such as $({+}{+}{+}{-})_2$
in~Table~\ref{table:1}. Thus, we have computed the scaling dimension
$1-h-\Bar{h}$ by using only the $({+}{+})_2$-type solutions for~$W=\Phi^3$
($L=36$) and the $({+}{+}{+})_3$-type solutions for~$W=\Phi^4$ ($L=30$). The
result is:
\begin{align}
   1-h-\Bar{h}&=0.6716(82),\qquad W=\Phi^3,
\\
   1-h-\Bar{h}&=0.7364(83),\qquad W=\Phi^4.
\end{align}

\begin{table}[ht]
 \centering
 \caption{Scaling dimensions obtained from the linear fit in the IR
 region~$\frac{2\pi}{L}\leq|p|<\frac{4\pi}{L}$.}
 \begin{tabular}{lllll}\toprule
  $W$      & $L$  & $\chi^2/\text{d.o.f.}$ & $1-h-\Bar{h}$ & Expected value\\\midrule
  $\Phi^3$ & $36$ & 0.506           & 0.682(10)(7)     & 0.666\dots\\
  $\Phi^4$ & $30$ & 0.358           & 0.747(11)(12)     & 0.75\\
  \bottomrule
 \end{tabular}
\label{table:7}
\end{table}%

\begin{figure}[ht]
 \begin{center}
  \includegraphics[width=0.8\columnwidth]{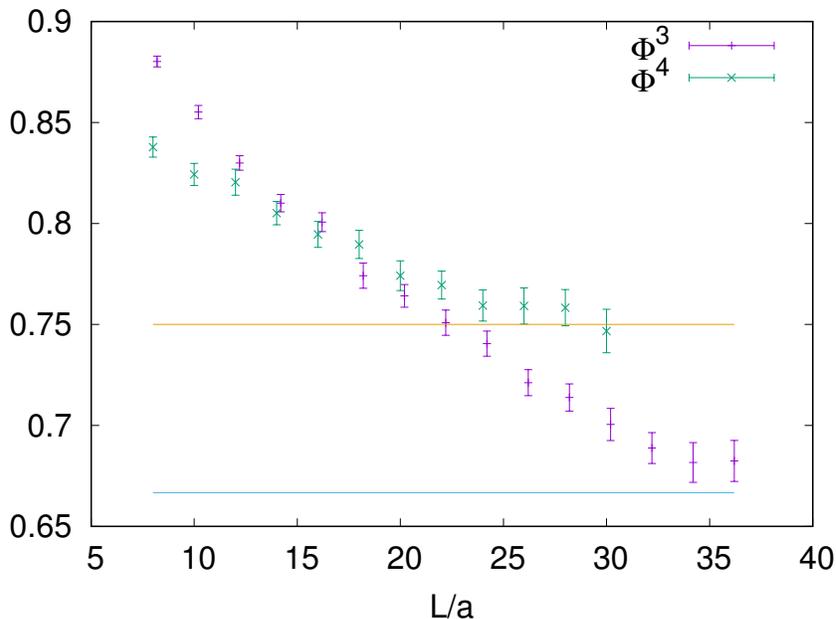}
  \caption{Scaling dimensions for~$W=\Phi^3$ and~$W=\Phi^4$ obtained with
  various box sizes.}
  \label{fig:12}
 \end{center}
\end{figure}

We also plotted in~Fig.~\ref{fig:12} the scaling dimension obtained by the
above method but with different box sizes~$L$. Two horizontal lines show the
expected values of~$1-h-\Bar{h}$ from the LG correspondence:
$1-h-\Bar{h} = 0.666\dots$ for~$W=\Phi^3$ and $1-h-\Bar{h}=0.75$
for~$W=\Phi^4$. We clearly see the tendency that the measured scaling dimension
approaches the expected value as $L$ increases. The approach appears not quite
smooth, however, so we do not try any fitting of this plot to extract the
$L\to\infty$ value; we suspect that this non-smoothness is due to statistical
fluctuations as we observed for the SUSY WT relation in the previous section.

From the $1-h-\Bar{h}$ case presented in~Fig.~\ref{fig:12}, we estimated the
systematic error associated with the finite-volume effect. We estimate it by
the maximum deviation of the central values at the three largest volumes; the
values obtained in this way are presented in the second parentheses
of~$1-h-\Bar{h}$ in~Table~\ref{table:7}.

\begin{figure}[ht]
 \begin{center}
  \begin{subfigure}{0.48\columnwidth}
   \begin{center}
    \includegraphics[width=\columnwidth]{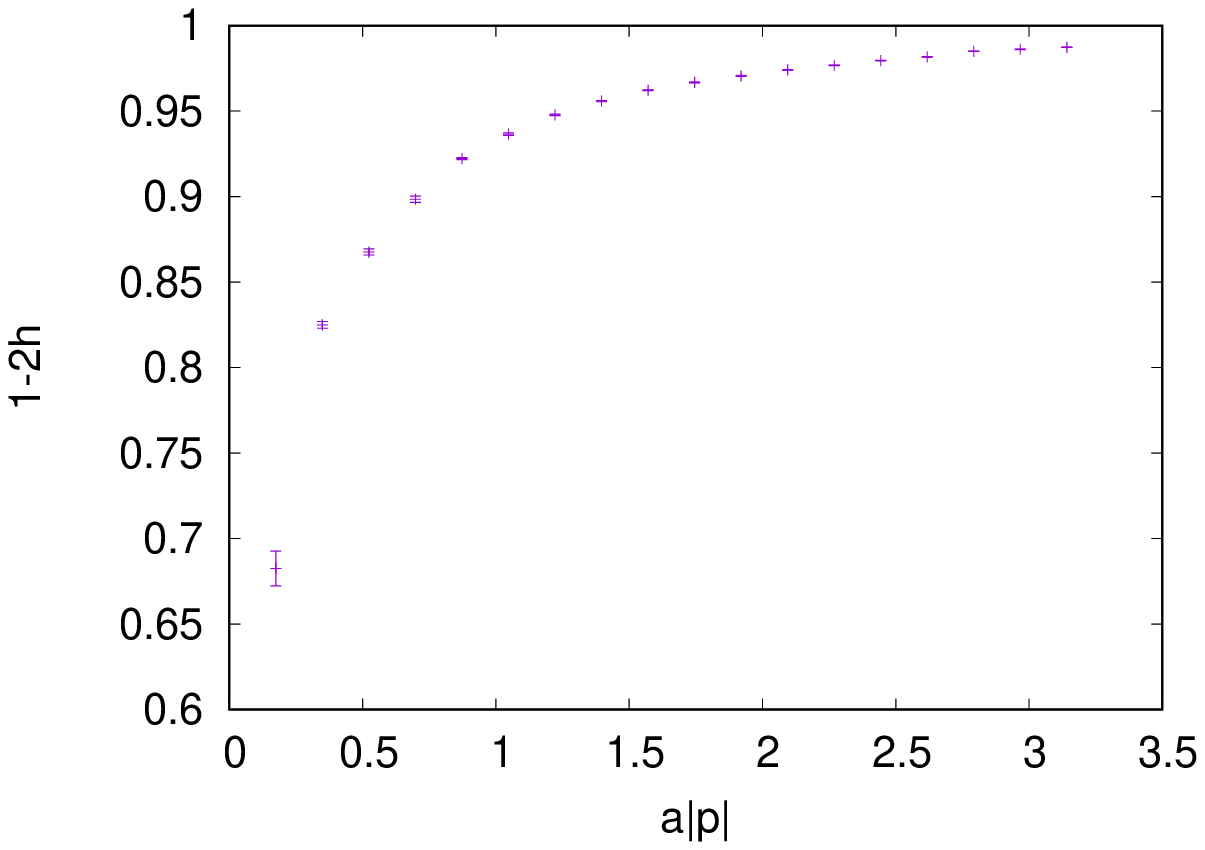}
    \caption{$W = \Phi^3$, $L=36$.}
    \label{fig:13(a)}
   \end{center}
  \end{subfigure} \hspace*{1em}
  \begin{subfigure}{0.48\columnwidth}
   \begin{center}
    \includegraphics[width=\columnwidth]{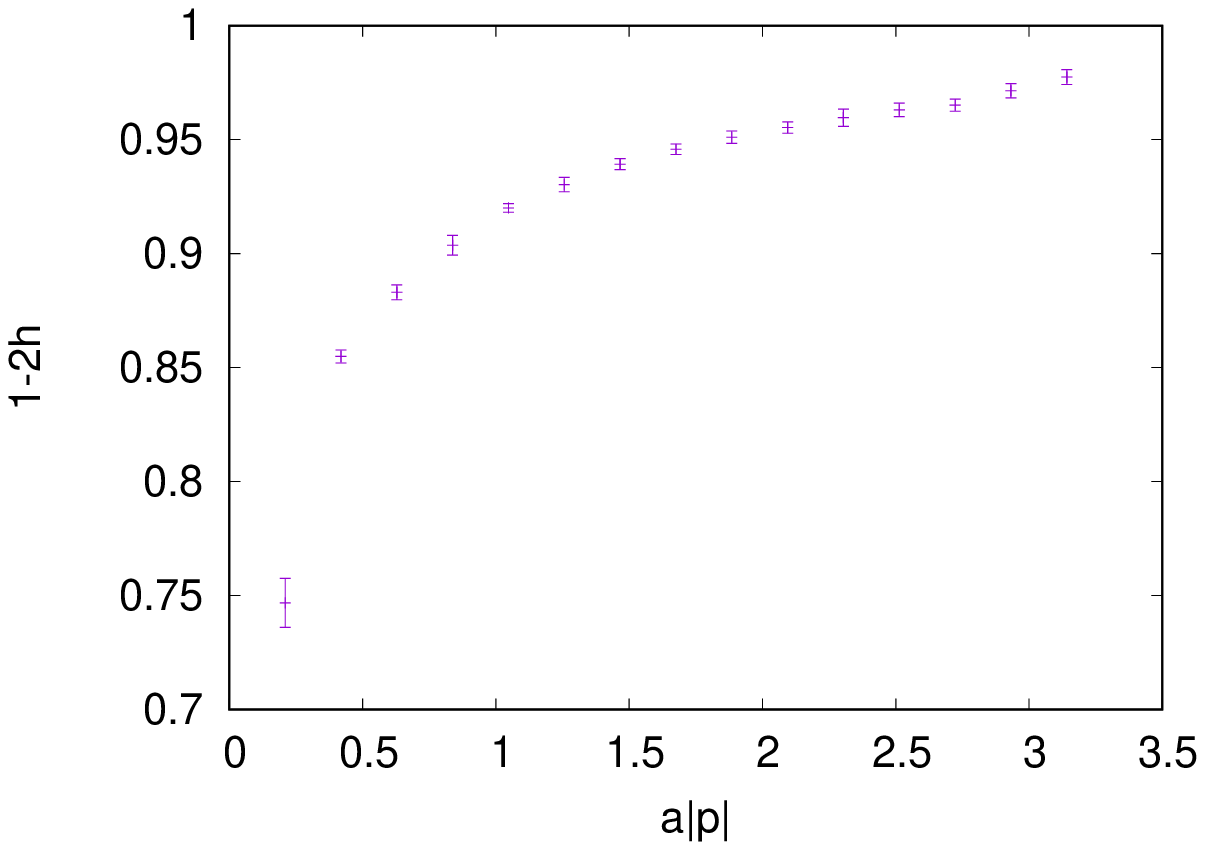}
    \caption{$W = \Phi^4$ $L=30$.}
    \label{fig:13(b)}
   \end{center}
  \end{subfigure}
 \end{center}
 \caption{Scaling dimensions obtained from the linear fitting in various
 momentum regions from IR to UV,
 $\frac{2\pi}{L}n\leq|p|<\frac{2\pi}{L}(n+1)$, where $n=1$, \dots, $L-1$.}
 \label{fig:13}
\end{figure}

It is also interesting to see the ``effective scaling dimension'' that is
obtained from the fitting in some restricted intermediate region of the
momentum norm~$|p|$. This is shown in~Fig.~\ref{fig:13}. In both panels, the
``effective scaling dimension'' smoothly changes from that in the IR region
(which is summarized in~Table~\ref{table:7}) and approaches $1-h-\Bar{h}\to1$
in the UV limit. This behavior is consistent with the expectation that the 2D
$\mathcal{N}=2$ WZ models become the free~$\mathcal{N}=2$ SCFT in the UV limit,
in which the chiral multiplet should have the scaling
dimension~$1-h-\Bar{h}=1$.

\section{Central charge}
\label{sec:6}
In this section we consider the measurement of the central charge~$c$, an
important quantity that characterizes CFT. This appears, in the first place, in
the operator product expansion (OPE) of the energy--momentum
tensor,\footnote{In this paper we follow the convention
of~Refs.~\cite{Polchinski:1998rq,Polchinski:1998rr}; this convention is
different from that of~Ref.~\cite{Kamata:2011fr}.}
\begin{equation}
   T(z)T(0)\sim\frac{c}{2z^4}+\frac{2}{z^2}T(0)+\frac{1}{z}\partial T(0),
\label{eq:(6.1)}
\end{equation}
where ``$\sim$'' implies ``$=$'' up to non-singular terms. The central charge
of the $A_n$ minimal model is
\begin{equation}
   c=\frac{3(n-2)}{n}=1,1.5,1.8,\dotsc,
\label{eq:(6.2)}
\end{equation}
for $n=3$, $4$, $5$, \dots.

From Eq.~\eqref{eq:(6.1)}, assuming rotational invariance,
\begin{equation}
   \langle T(z)T(0)\rangle=\frac{c}{2z^4}.
\label{eq:(6.3)}
\end{equation}
Similarly, in $\mathcal{N}=2$ SCFT, the two-point functions of the
supercurrent~$S^\pm$ and the~$U(1)$ current~$J$ are given by
\begin{align}
   \langle S^+(z)S^-(0)\rangle&=\frac{2c}{3z^3},
\label{eq:(6.4)}
\\
   \langle J(z)J(0)\rangle&=\frac{c}{3z^2}.
\label{eq:(6.5)}
\end{align}
Thus, the central charge may also be obtained by computing these two-point
functions.

To find the appropriate expression for the supercurrent, the energy--momentum
tensor, and the $U(1)$ current such that they form the superconformal multiplet
in $\mathcal{N}=2$ SCFT is itself an intriguing problem, because in our system
the $\mathcal{N}=2$ superconformal symmetry is expected to emerge only in the
IR limit. As explained in~Appendix~\ref{sec:A}, we adopt the expressions of the
former two which become (gamma-) traceless for the free massless WZ model,
$W'=0$. It appears that those expressions work as expected (see
also~Ref.~\cite{Kamata:2011fr}).

As in the previous section, we numerically compute the correlation function in
the momentum space. We consider the two-point functions of the
supercurrent, the energy--momentum tensor, and the~$U(1)$ current. As we will
explain, these two-point functions are related to each other by SUSY, which is
an exact symmetry of our formulation. Using this fact, the computation of the
whole correlation function can be reduced to that for the supercurrent
correlator.

\subsection{Central charge from the supercurrent correlator}
\label{sec:6.1}
The argument in~Appendix~\ref{sec:A} gives the supercurrent in the momentum
space,
\begin{align}
   S^+(p)&=S_z^+(p)
   =\frac{4\pi}{L_0L_1}\sum_qi(p-q)_zA(p-q)\Bar{\psi}_{\Dot{2}}(q),
\label{eq:(6.6)}
\\
   S^-(p)&=S_z^-(p)
   =-\frac{4\pi}{L_0L_1}\sum_qi(p-q)_zA^*(p-q)\psi_2(q).
\label{eq:(6.7)}
\end{align}
We thus compute the two-point function~$\langle S^+(p)S^-(-p)\rangle$. The
Fourier transformation of~Eq.~\eqref{eq:(6.4)} is, on the other hand,
\begin{align}
   \left\langle S^+(p)S^-(-p)\right\rangle
   &=L_0L_1\int_{L_0L_1}d^2x\,e^{-ipx}\left\langle S^+(x)S^-(0)\right\rangle
\notag\\
   &=L_0L_1 \int_{L_0L_1}d^2x\,e^{-ipx}
   \frac{2c\Bar{z}^3}{3(x^2+\delta^2)^3}
\notag\\
   &=L_0L_1\frac{-i\pi c}{6}\frac{\partial^3}{\partial p_{\Bar{z}}^3}
   \left(\frac{|p|}{\delta}\right)^2K_2(|p|\delta),
\label{eq:(6.8)}
\end{align}
where we have introduced a regulator~$\delta$ to tame the singularity at~$x=0$;
$K_2$~is the modified Bessel function of the second kind. Since we are
interested in the IR limit, taking the limit $|p|\delta\to0$, we have
\begin{equation}
   \left\langle S^+(p)S^-(-p)\right\rangle
   \to L_0L_1\frac{i\pi c}{3}\frac{p_z^2}{p_{\Bar{z}}}.
\label{eq:(6.9)}
\end{equation}
We fit the measured two-point function~$\langle S^+(p)S^-(-p)\rangle$ in the
IR region by this function.

\begin{figure}[ht]
 \begin{center}
  \begin{subfigure}{0.48\columnwidth}
   \begin{center}
    \includegraphics[width=\columnwidth]{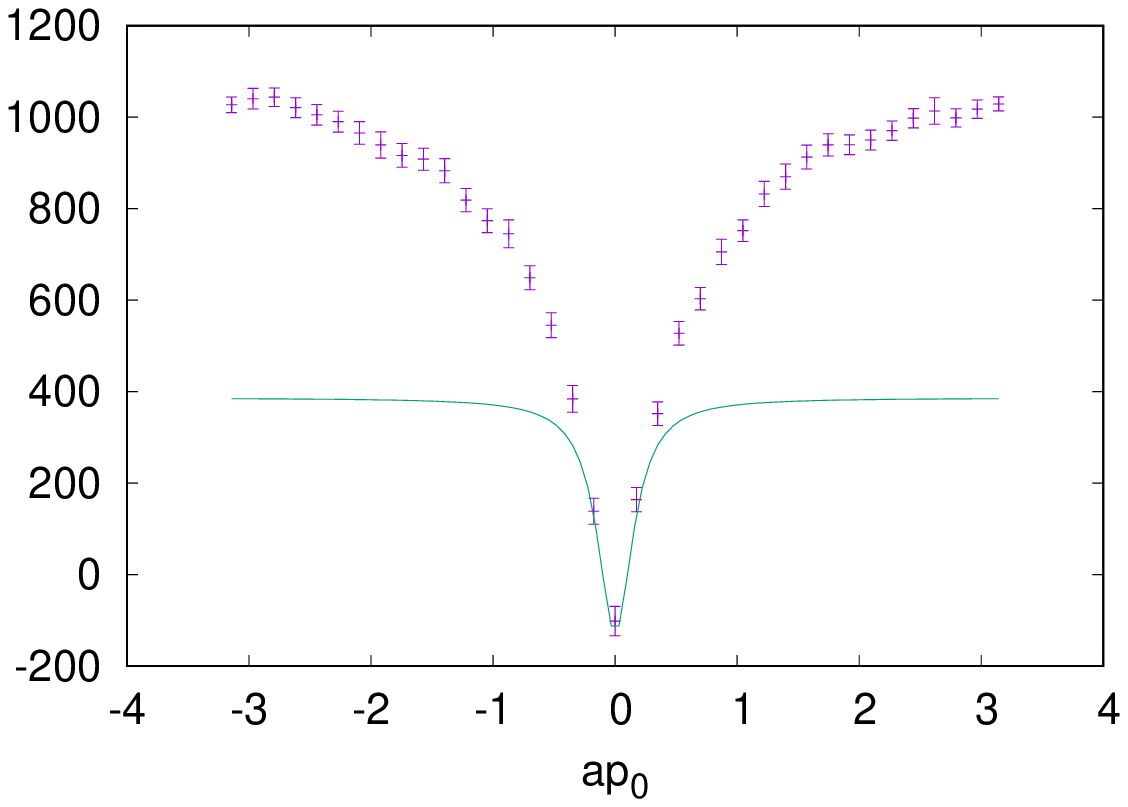}
    \caption{Real part}
   \end{center}
  \end{subfigure} 
  \begin{subfigure}{0.48\columnwidth}
   \begin{center}
    \includegraphics[width=\columnwidth]{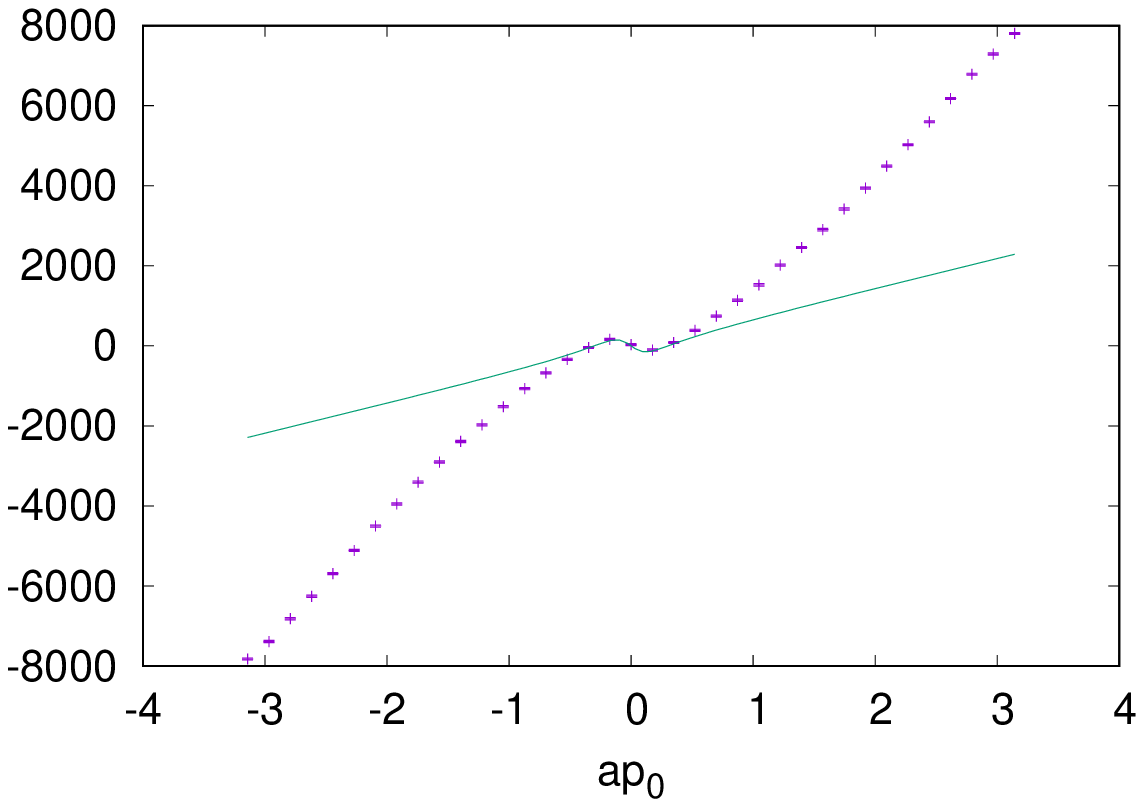}
    \caption{Imaginary part}
   \end{center}
  \end{subfigure}
 \end{center}
 \caption{$\langle S^+(p)S^-(-p)\rangle$ for~$W=\Phi^3$, $L=36$, and~$p_1=\pi/18$. The fitting curves from~Eq.~\eqref{eq:(6.9)} are also depicted.}
 \label{fig:14}
\end{figure}

\begin{figure}[ht]
 \begin{center}
  \begin{subfigure}{0.48\columnwidth}
    \begin{center}
     \includegraphics[width=\columnwidth]{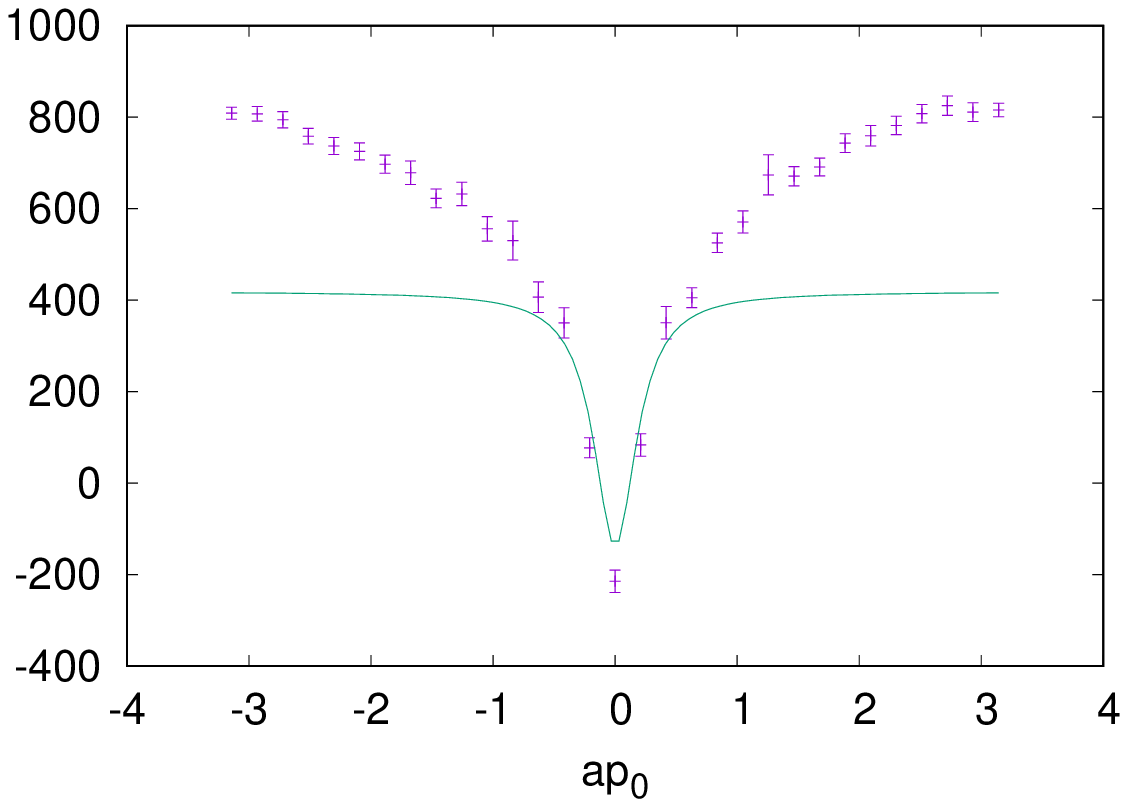}
     \caption{Real part}
    \end{center}
  \end{subfigure} \hspace{1em}
  \begin{subfigure}{0.48\columnwidth}
   \begin{center}
    \includegraphics[width=\columnwidth]{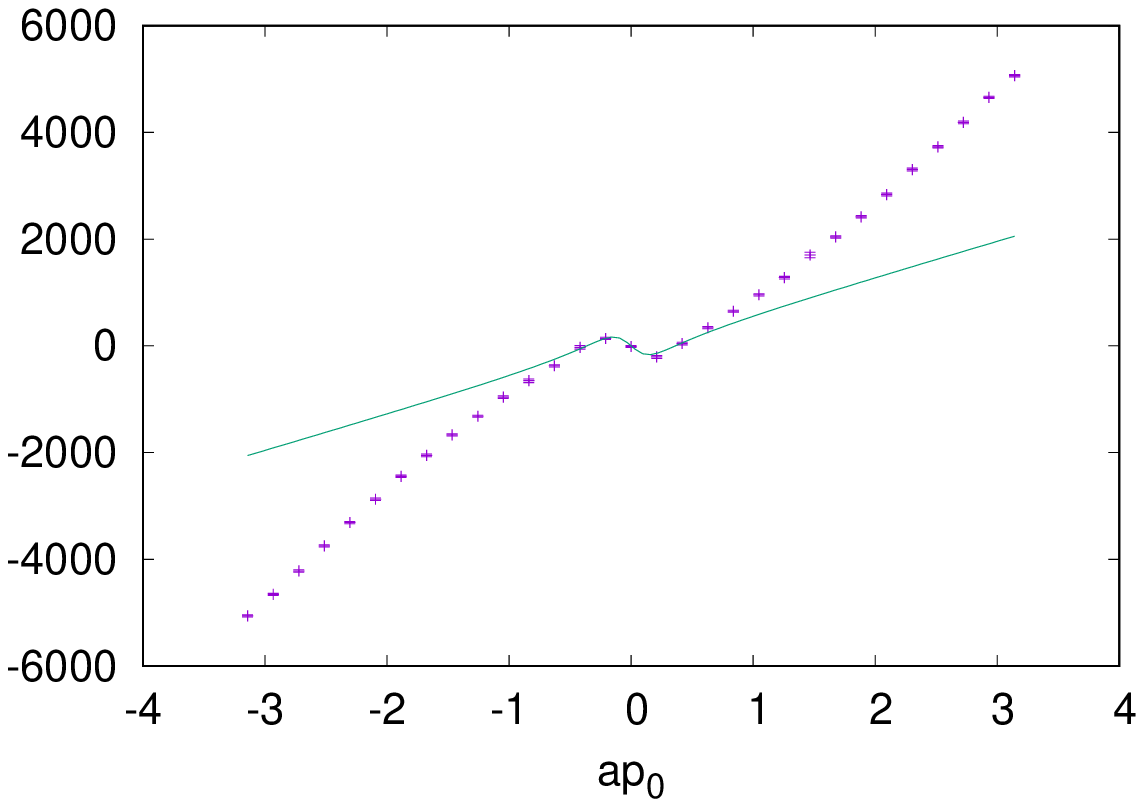}
    \caption{Imaginary part}
   \end{center}
  \end{subfigure}
 \end{center}
 \caption{$\langle S^+(p)S^-(-p)\rangle$ for~$W=\Phi^4$, $L=30$, and~$p_1=\pi/15$. The fitting curves from~Eq.~\eqref{eq:(6.9)} are also depicted.}
 \label{fig:15}
\end{figure}

We plot the two-point function~$\langle S^+(p)S^-(-p)\rangle$
in~Figs.~\ref{fig:14} and~\ref{fig:15} for the maximal box size, i.e.,
$L=36$ for~$W=\Phi^3$ and $L=30$ for~$W=\Phi^4$. In each figure, the left panel
is the real part of the correlation function and the right one is the imaginary
part. The spatial momentum~$p_1$ is fixed to the positive minimal value,
$p_1=2\pi/L$. In these figures we also show the function on the right-hand
side of~Eq.~\eqref{eq:(6.9)} with the central charge~$c$ obtained from the fit
in the IR region~$\frac{2\pi}{L}\leq|p|<\frac{4\pi}{L}$; the central charges
obtained in this way are tabulated in~Table~\ref{table:8}. Again, these numbers
may contain the systematic error associated with the solutions undiscovered
by the NR method.


\begin{table}[ht]
 \centering
 \caption{The central charges obtained from the fit of the supercurrent
 correlator. The fitting momentum region
 is~$\frac{2\pi}{L}\leq|p|<\frac{4\pi}{L}$.}
 \begin{tabular}{lllll}\toprule
  $W$      & $L$  & $\chi^2/\text{d.o.f.}$ & $c$ & Expected value\\\midrule
  $\Phi^3$ & $36$ & 0.928           & 1.087(68)(56) & 1             \\
  $\Phi^4$ & $30$ & 4.606           & 1.413(65)(31) & 1.5           \\
  \bottomrule
 \end{tabular}
 \label{table:8}
\end{table}%

Compared to the result of~Ref.~\cite{Kamata:2011fr} for~$W=\Phi^3$,
\begin{equation}
   c=1.09(14)(31),
\label{eq:(6.10)}
\end{equation}
the central charge we obtained is somewhat closer to the expected value with
the smaller statistical error.

\begin{figure}[ht]
 \begin{center}
  \includegraphics[width=0.8\columnwidth]{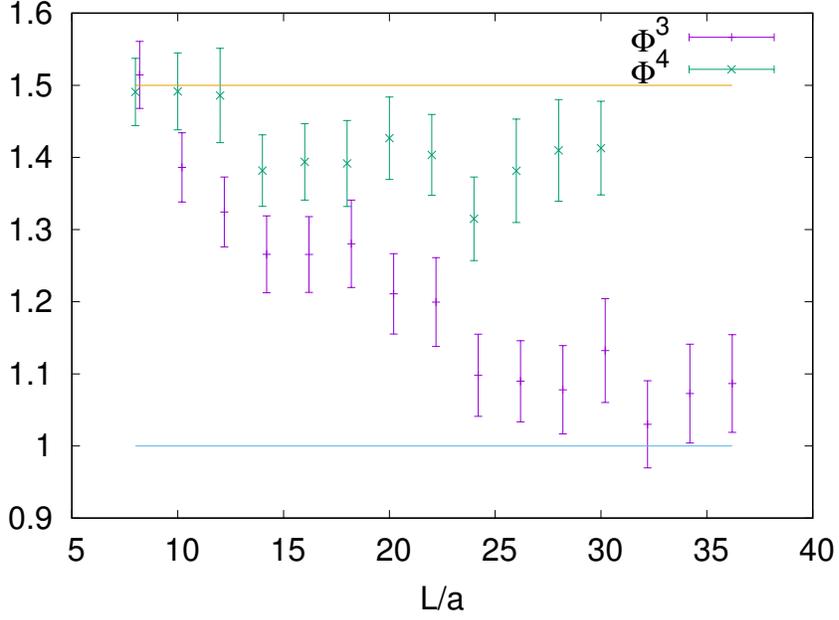}
  \caption{Central charges obtained by the fit for~$W=\Phi^3$ and~$W=\Phi^4$
  as a function of the box size~$L=8$--$36$.}
  \label{fig:16}
 \end{center}
\end{figure}

In Fig.~\ref{fig:16} we have plotted how the fitted central charge changes as
a function of the box size~$L$. From the $c$ presented in~Fig.~\ref{fig:16},
we estimated the systematic error associated with the finite-volume effect. We
estimate it by the maximum deviation of central values at the largest three
volumes; the values obtained in this way are presented in the second
parentheses for~$c$ in~Table~\ref{table:8}.

\begin{figure}[ht]
 \begin{center}
  \begin{subfigure}{0.48\columnwidth}
   \begin{center}
    \includegraphics[width=\columnwidth]{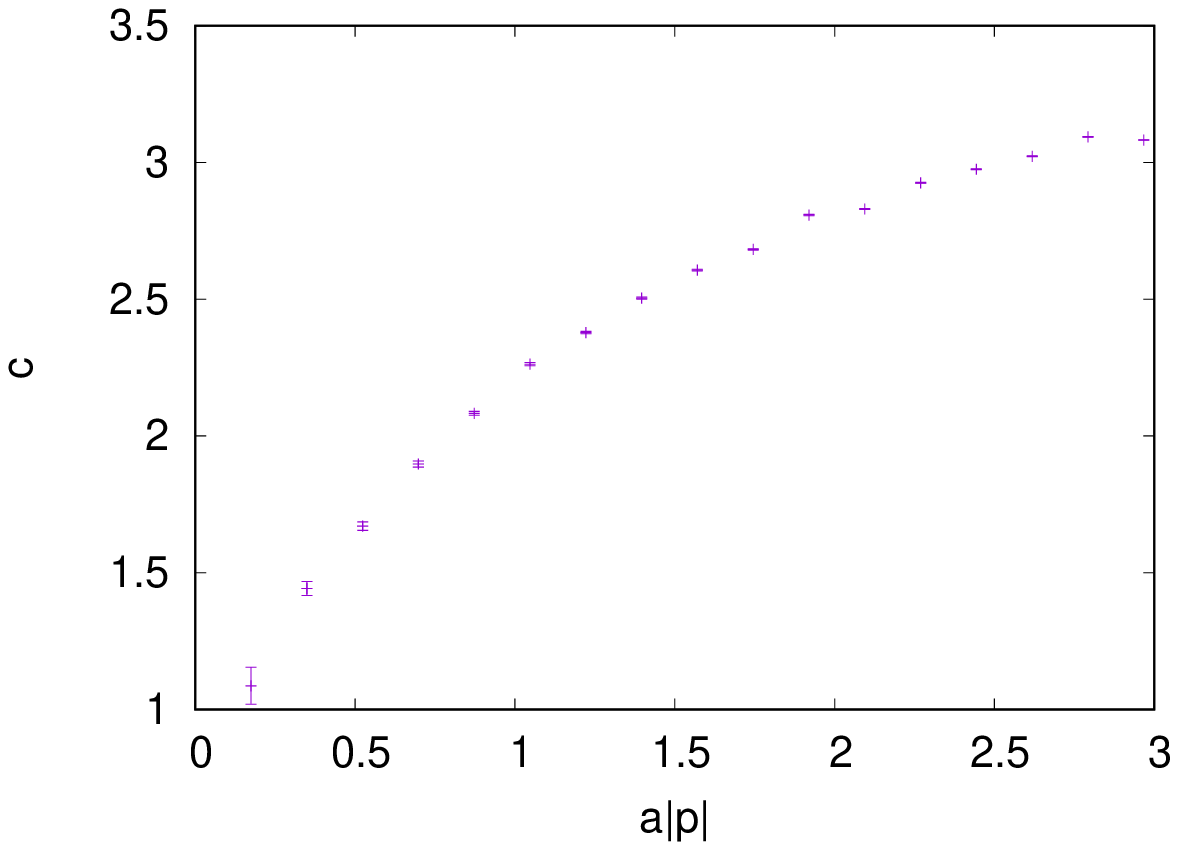}
    \caption{$W=\Phi^3$ and~$L=36$.}
   \end{center}
  \end{subfigure} \hspace*{1em}
  \begin{subfigure}{0.48\columnwidth}
   \begin{center}
    \includegraphics[width=\columnwidth]{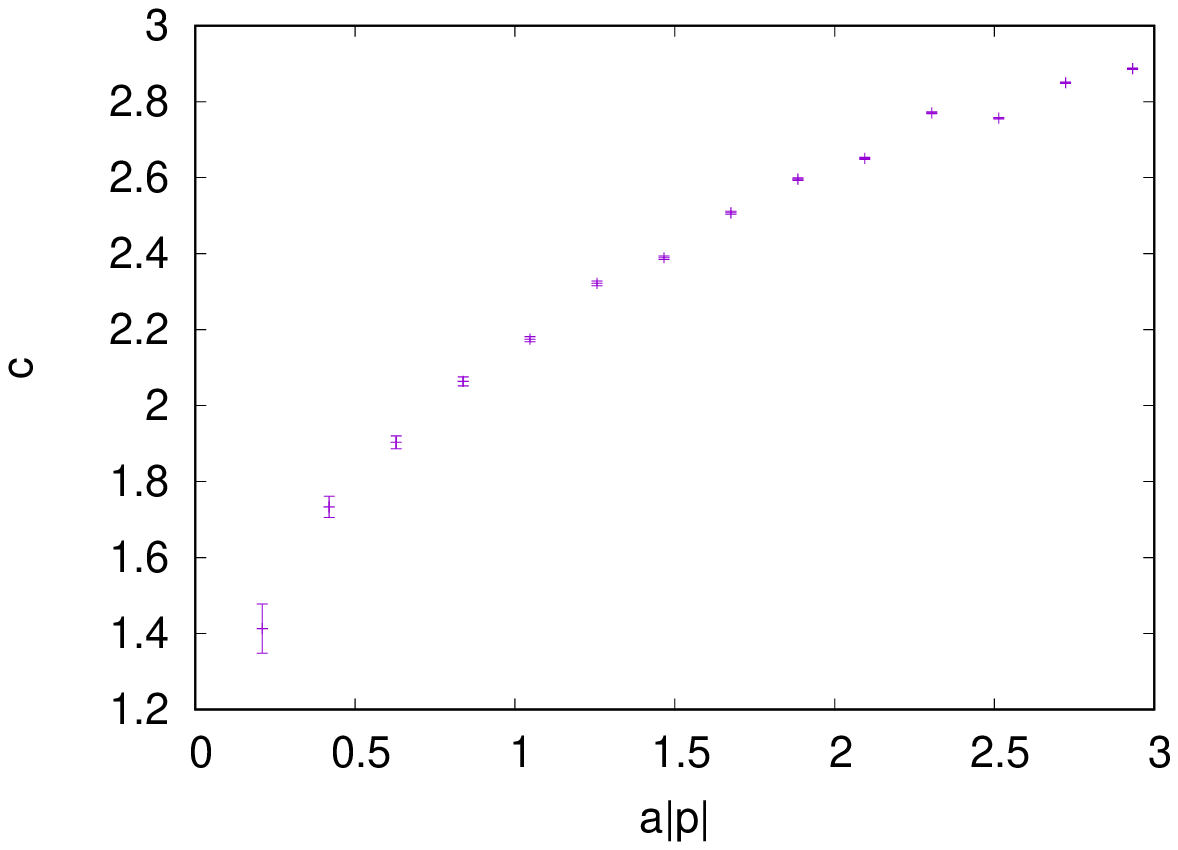}
    \caption{$W=\Phi^4$ and~$L=30$.}
   \end{center}
  \end{subfigure}
 \end{center}
 \caption{``Effective central charge'' obtained by the fit in various momentum
 regions, $\frac{2\pi}{L}n\leq|p|<\frac{2\pi}{L}(n+1)$ ($n=1$, \dots, $L-1$).}
 \label{fig:17}
\end{figure}

As for~Fig.~\ref{fig:13} in the previous section, it is interesting to see how
the central charge obtained by the fit changes as a function of the fitted
momentum region~\cite{Kamata:2011fr}. The result is shown
in~Fig.~\ref{fig:17}. This ``effective central charge'' depending on the
momentum region is analogous to the supersymmetric version of the
Zamolodchikov $c$-function~\cite{Zamolodchikov:1986gt,Cappelli:1989yu}. As
expected, the ``effective central charge'' changes from the IR value to~$c=3$
in the UV limit in which the system is expected to become a free
$\mathcal{N}=2$ SCFT.

\subsection{Central charge from the energy--momentum tensor correlator}
\label{sec:6.2}
As discussed in~Appendix~\ref{sec:A}, the energy--momentum tensor~$T=T_{zz}$,
which is expected to be consistent with the conformal symmetry, is given in the
momentum space by
\begin{align}
   T(p)&=\frac{\pi}{L_0L_1}\sum_q
   \Bigl[
   4(p-q)_zq_zA^*(p-q)A(q)
\notag\\
   &\qquad\qquad\qquad
   -iq_z\psi_2(p-q)\Bar{\psi}_{\Dot{2}}(q)
   +i(p-q)_z\psi_2(p-q)\Bar{\psi}_{\Dot{2}}(q)\Bigr].
\label{eq:(6.11)}
\end{align}
It turns out that this expression as it stands leads to a very noisy two-point
correlation function. Fortunately, noting the fact that the energy--momentum
tensor of~Eq.~\eqref{eq:(6.11)} is the SUSY transformation of the supercurrent
in~Eqs.~\eqref{eq:(6.6)} and~\eqref{eq:(6.7)},
\begin{equation}
   T(p)=\frac{1}{4}Q_2S^+(p)-\frac{1}{4}\Bar{Q}_{\Dot{2}}S^-(p),
\label{eq:(6.12)}
\end{equation}
where the SUSY transformation is given in~Appendix~\ref{sec:A}, we can express
the two-point function of the energy--momentum tensor by a linear combination
of two-point functions of the supercurrent which are less noisy:
\begin{equation}
   \left\langle T(p)T(-p)\right\rangle
   =-\frac{2ip_z}{16}\left\langle S^+(p)S^-(-p)+S^-(p)S^+(-p)\right\rangle.
\label{eq:(6.13)}
\end{equation}
Note that this relation holds exactly in our formulation that preserves SUSY.

The Fourier transformation of~Eq.~\eqref{eq:(6.3)} is, by the same procedure
as~Eqs.~\eqref{eq:(6.8)} and~\eqref{eq:(6.9)},
\begin{align}
   \langle T(p)T(-p)\rangle
   &=L_0L_1\frac{\pi c}{2\cdot4!}
   \frac{\partial^4}{\partial p_{\Bar{z}}^4}
   \left(\frac{|p|}{\delta}\right)^3K_3(|p|\delta)
\notag\\
   &\to L_0L_1\frac{\pi c}{12}\frac{p_z^3}{p_{\Bar{z}}}.
\label{eq:(6.14)}
\end{align}

\begin{figure}[ht]
 \begin{center}
  \begin{subfigure}{0.48\columnwidth}
   \begin{center}
    \includegraphics[width=\columnwidth]{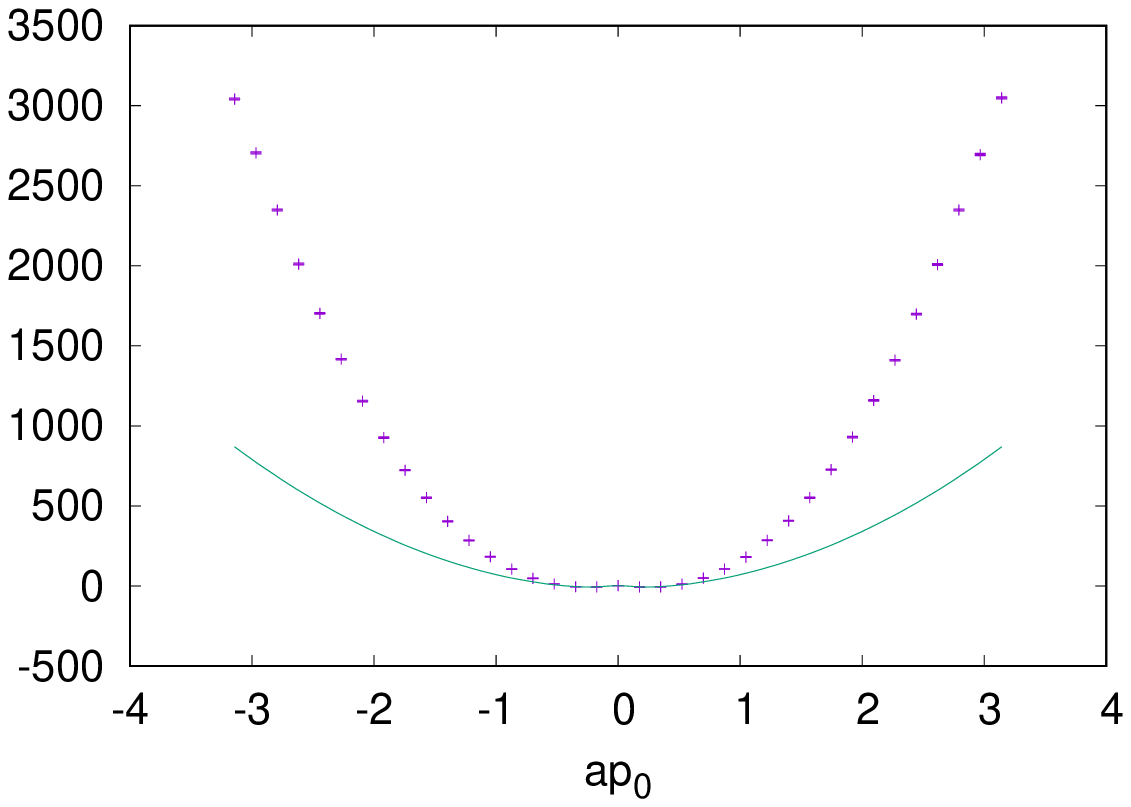}
    \caption{Real part}
   \end{center}
  \end{subfigure} \hspace{1em}
  \begin{subfigure}{0.48\columnwidth}
   \begin{center}
    \includegraphics[width=\columnwidth]{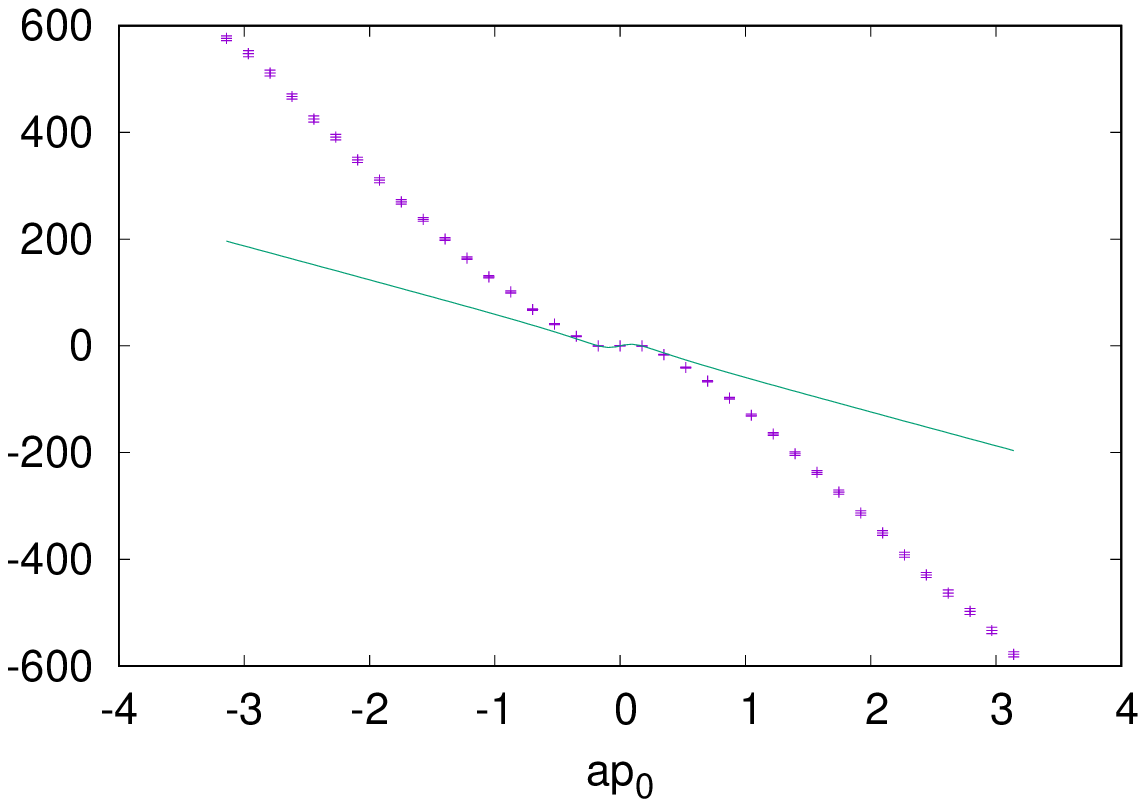}
    \caption{Imaginary part}
   \end{center}
  \end{subfigure}
 \end{center}
 \caption{$\langle T(p)T(-p)\rangle$ for~$W=\Phi^3$, $L=36$, and~$p_1=\pi/18$.
 The fitting curve of~Eq.~\eqref{eq:(6.14)} is also depicted.}
 \label{fig:18}
\end{figure}

\begin{figure}[ht]
 \begin{center}
  \begin{subfigure}{0.48\columnwidth}
   \begin{center}
    \includegraphics[width=\columnwidth]{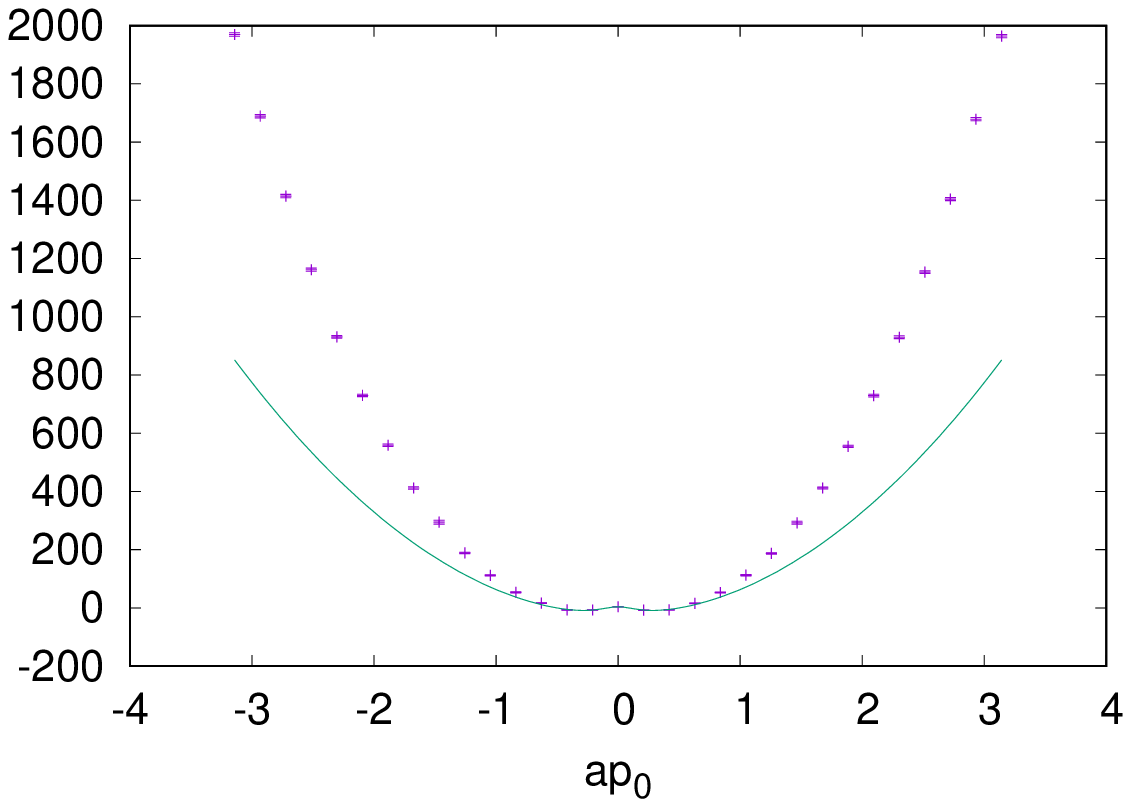}
    \caption{Real part}
   \end{center}
  \end{subfigure} \hspace{1em}
  \begin{subfigure}{0.48\columnwidth}
   \begin{center}
    \includegraphics[width=\columnwidth]{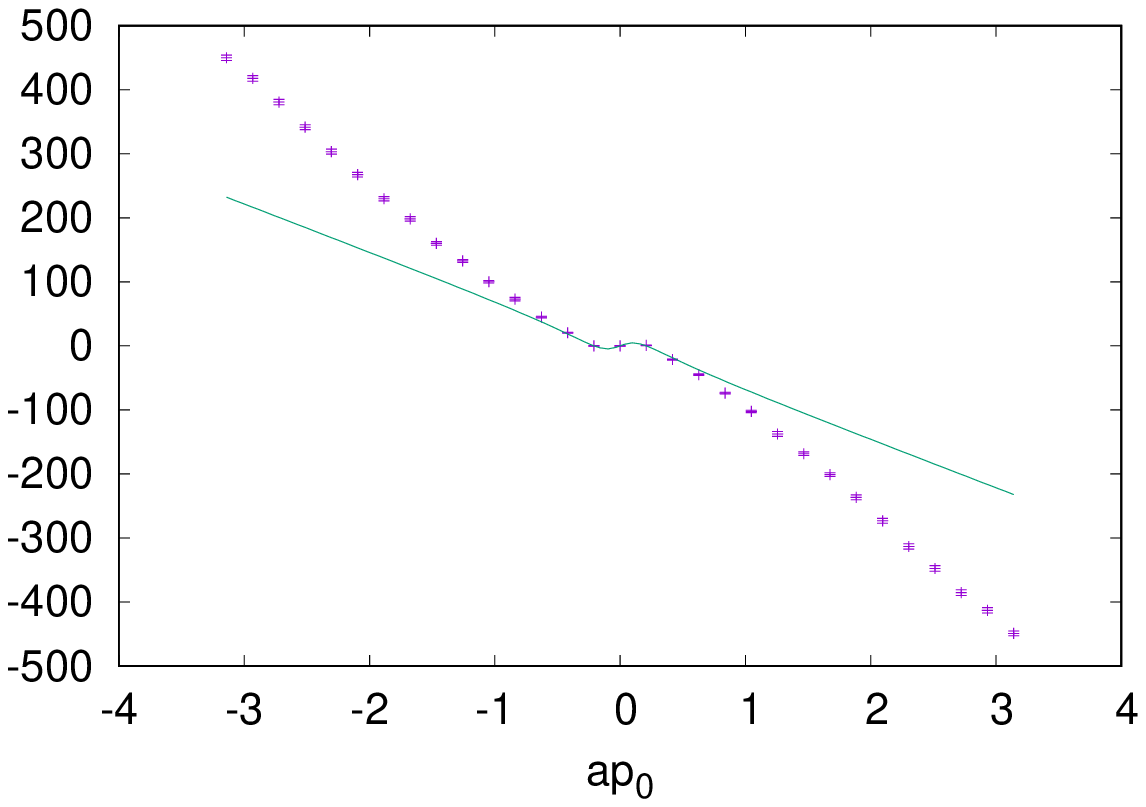}
    \caption{Imaginary part}
   \end{center}
  \end{subfigure}
 \end{center}
 \caption{$\langle T(p)T(-p)\rangle$ for~$W=\Phi^4$, $L=36$, and~$p_1=\pi/15$.
 The fitting curve by~Eq.~\eqref{eq:(6.14)} is also depicted.}
 \label{fig:19}
\end{figure}

We plot the two-point function~$\langle T(p)T(-p)\rangle$
of~Eq.~\eqref{eq:(6.13)} in~Figs.~\ref{fig:18} and~\ref{fig:19} for the maximal
box size, i.e., $L=36$ for~$W=\Phi^3$ and $L=30$ for~$W=\Phi^4$. In each
figure, the left panel is the real part of the correlation function and the
right one is the imaginary part. The spatial momentum~$p_1$ is fixed to the
positive minimal value, $p_1=2\pi/L$. In these figures we also show the
function in~Eq.~\eqref{eq:(6.14)} with the central charge~$c$ obtained from the
fit in the IR region~$\frac{2\pi}{L}\leq|p|<\frac{4\pi}{L}$. The central
charges obtained in this way are tabulated in~Table~\ref{table:9}; this is
another main result of this paper. Recall again, however, that these numbers
may contain the systematic error associated with the solutions undiscovered by
the NR method.


\begin{table}[ht]
 \centering
 \caption{The central charges obtained from the fit of the energy--momentum
 tensor correlator. The fitting momentum region
 is~$\frac{2\pi}{L}\leq|p|<\frac{4\pi}{L}$.}
 \begin{tabular}{lllll}\toprule
  $W$      & $L$  & $\chi^2/\text{d.o.f.}$ & $c$ & Expected value \\\midrule
  $\Phi^3$ & $36$ & 1.017           & 1.061(36)(34) & $1$            \\
  $\Phi^4$ & $30$ & 0.916           & 1.415(36)(36) & $1.5$          \\
  \bottomrule
 \end{tabular}
 \label{table:9}
\end{table}%

We repeated the computation of the central charge~$c$ by using only the
$({+}{+})_2$-type solutions for~$W=\Phi^3$ ($L=36$) and the
$({+}{+}{+})_3$-type solutions for~$W=\Phi^4$ ($L=30$), to see how the values
are changed if we do not include a few percent ``strange solutions.'' The
results are:
\begin{align}
   c&=1.057(34)\qquad(W=\Phi^3),
\\
   c&=1.288(28)\qquad(W=\Phi^4).
\end{align}

One may note that the fit in~Table~\ref{table:9} is better than that
in~Table~\ref{table:8}, in the sense that $\chi^2/\text{d.o.f.}$ is very close
to~$1$ in the former. This is due to the fact that the real and imaginary parts
of the two-point correlation function of~Eq.~\eqref{eq:(6.13)} are exactly
(anti-)symmetric under~$p\to-p$, while the numerical data
of~$\langle S^+(p)S^-(-p)\rangle$ itself does not possess this
property.\footnote{This (anti-)symmetry under $p\to-p$ is fulfilled within the
margin of the statistical error; one may also (anti-)symmetrize the two-point
function $\langle S^+(p)S^-(-p)\rangle$ by hand.} The number of data points is
thus effectively doubled.

\begin{figure}[ht]
 \begin{center}
  \includegraphics[width=0.8\columnwidth]{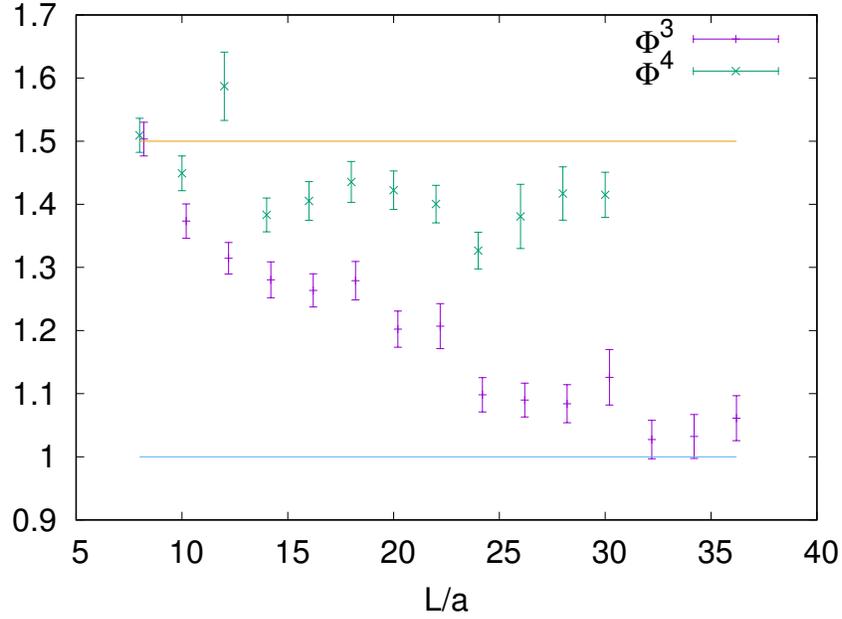}
  \caption{Central charges obtained by the fit for~$W=\Phi^3$ and~$W=\Phi^4$
  as a function of the box size~$L=8$--$36$.}
  \label{fig:20}
 \end{center}
\end{figure}

In Fig.~\ref{fig:20}, we plotted how the fitted central charge changes as a
function of the box size~$L$. From $c$ presented in~Fig.~\ref{fig:20}, we again
estimated the systematic error associated with the finite-volume effect. The
values obtained in this way are presented in the second parentheses for~$c$
in~Table~\ref{table:9}.

\begin{figure}[ht]
 \begin{center}
  \begin{subfigure}{0.48\columnwidth}
   \begin{center}
    \includegraphics[width=\columnwidth]{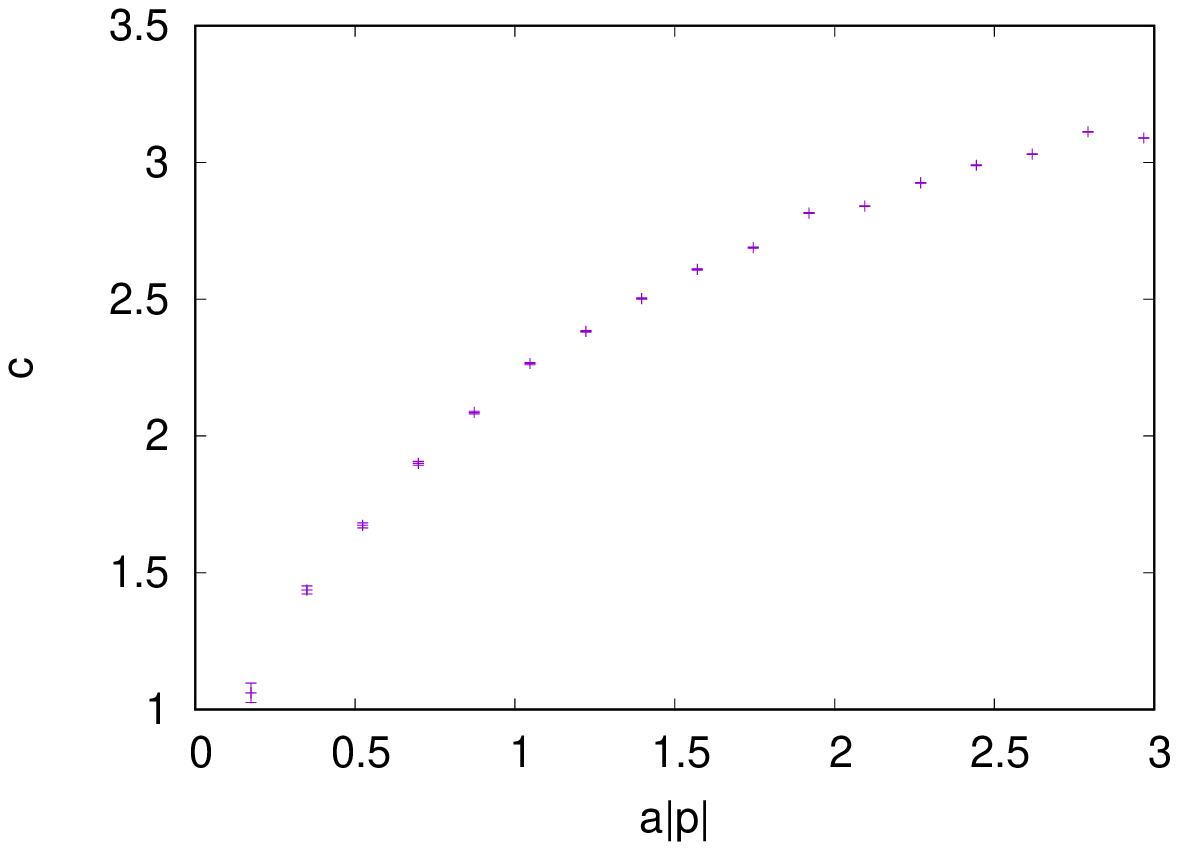}
    \caption{$W=\Phi^3$ and~$L=36$.}
   \end{center}
  \end{subfigure} \hspace*{1em}
  \begin{subfigure}{0.48\columnwidth}
   \begin{center}
    \includegraphics[width=\columnwidth]{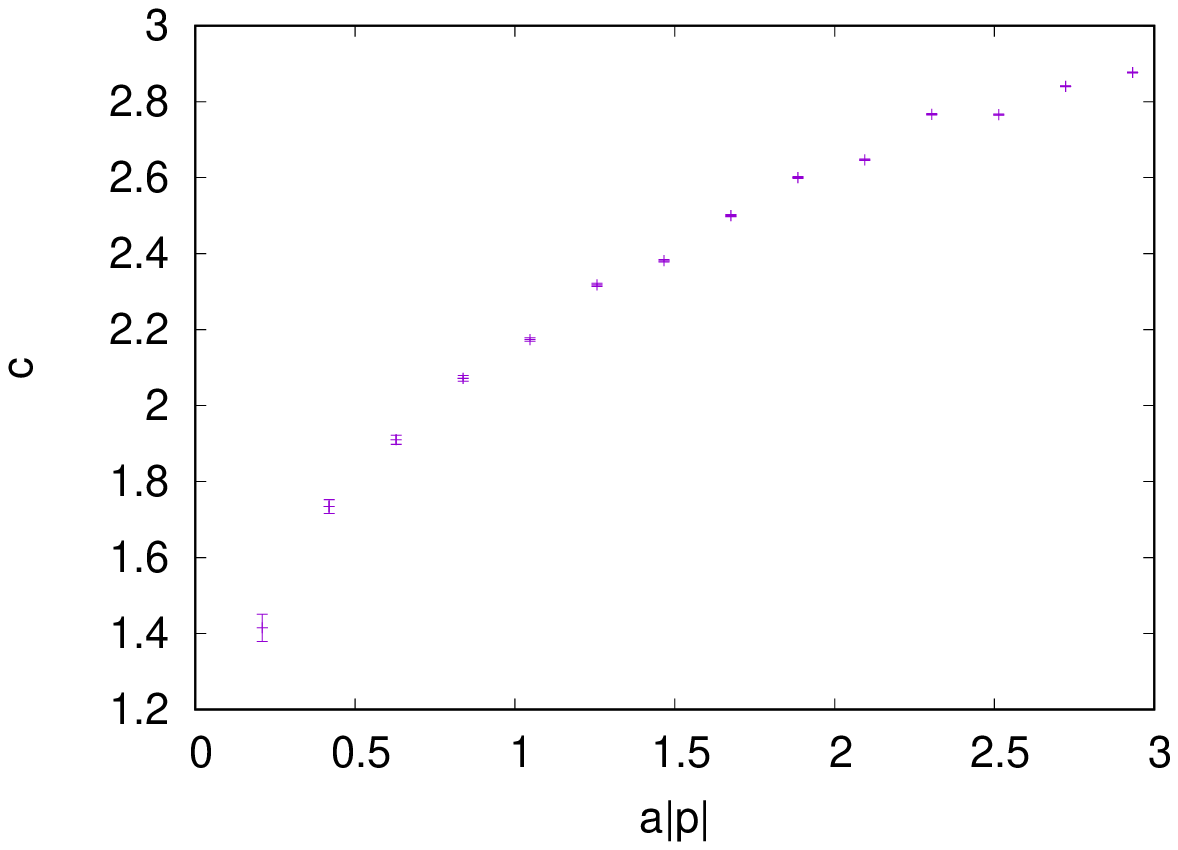}
    \caption{$W=\Phi^4$ and~$L=30$.}
   \end{center}
  \end{subfigure}
 \end{center}
 \caption{``Effective central charge'' obtained by the fit in various momentum
 regions, $\frac{2\pi}{L}n\leq|p|<\frac{2\pi}{L}(n+1)$ ($n=1$, \dots, $L-1$).}
 \label{fig:21}
\end{figure}

Also, in~Fig.~\ref{fig:21} the ``effective central charge'' obtained from the
fit in various momentum regions is depicted; from IR to UV, it again shows the
expected behavior analogously to the Zamolodchikov $c$-function.

\subsection{Central charge from the $U(1)$ current correlator}
\label{sec:6.3}
Finally, we consider the $U(1)$ current correlator. As discussed
in~Appendix~\ref{sec:A}, the~$U(1)$ current is given by
\begin{equation}
   J(p)=\frac{2\pi}{L_0L_1}\sum_q\Bar{\psi}_{\Dot{2}}(p-q)\psi_2(q).
\label{eq:(6.15)}
\end{equation}
The two-point function of this current is expected to behave in the IR limit as
\begin{align}
   \langle J(p)J(-p)\rangle
   &=L_0L_1\frac{-\pi c}{3}
   \frac{\partial^2}{\partial p_{\Bar{z}}^2}\frac{|p|}{\delta}K_1(|p|\delta)
\notag\\
   &\to L_0L_1\frac{-\pi c}{3}\frac{p_z}{p_{\Bar{z}}}.
\label{eq:(6.16)}
\end{align}
We note that the supercurrent~$S^\pm$ can be rewritten as the SUSY
transformation of~$J$,
\begin{equation}
   S^+(p)=\Bar{Q}_{\Dot{2}}J(p),\qquad S^-(p)=Q_2J(p).
\label{eq:(6.17)}
\end{equation}
Therefore,
\begin{equation}
   \left\langle S^+(p)S^-(-p)+S^-(p)S^+(-p)\right\rangle
   =-2ip_z\left\langle J(p)J(-p)\right\rangle.
\label{eq:(6.18)}
\end{equation}
This shows that the computation of the $U(1)$ current correlator is identical
to the energy--momentum tensor correlator of~Eq.~\eqref{eq:(6.13)} up to a
proportionality factor. We expect that we would obtain almost the same results
as the previous subsection, so we do not carry out the analysis on this
correlator.

\section{Conclusion}
\label{sec:7}
In this paper, following on from the study of~Ref.~\cite{Kamata:2011fr}, we
numerically studied the IR behavior of the 2D $\mathcal{N}=2$ WZ model with the
superpotentials $W=\Phi^3$ and~$W=\Phi^4$. We used the SUSY-invariant
momentum-cutoff formulation which allows, because of exact symmetries, a
straightforward construction of the Noether currents, i.e., the supercurrent
and the energy--momentum tensor. The simulation algorithm is free from
autocorrelation because it utilizes the Nicolai map. From two-point correlation
functions in the momentum space, we determined the scaling dimension of the
scalar field (Table~\ref{table:7}) and the central charge (Table~\ref{table:9})
in the IR region. It appears that these results, with the flow of the
``effective central charge'' in~Fig.~\ref{fig:21}, are consistent with the
conjectured LG correspondence to the $A_2$ and~$A_3$ minimal
SCFT.\footnote{Although those numbers may contain the systematic error
associated with the solutions undiscovered by the NR method.}

As future prospects, we may further extend the present study to WZ models with
multiple superfields and more complicated superpotentials such as the ADE-type
theories in~Table~\ref{table:10}~\cite{Vafa:1988uu} (the results of the
present paper apply to the $A_2$, $A_3$, and~$E_6$ models).

\begin{table}[h]
 \centering
 \caption{ADE-type theories}
 \begin{tabular}{lll}\toprule
  Algebra & Superpotential $W$ & Central charge $c$ \\\midrule
  $A_n$ & $\Phi^{n+1}$, $n\geqq1$ & $3-6/(n+1)$\\
  $D_n$ & $\Phi^{n-1}+\Phi\Phi^{\prime2}$, $n\geqq3$ & $3-6/2(n-1)$\\
  $E_6$ & $\Phi^3+\Phi^{\prime4}$ & $3-6/12$\\
  $E_7$ & $\Phi^3+\Phi\Phi^{\prime3}$ & $3-6/18$\\
  $E_8$ & $\Phi^3+\Phi^{\prime5}$ & $3-6/30$\\
  \bottomrule
 \end{tabular}
 \label{table:10}
\end{table}

For a possible application of the present calculational method to the
superstring compactification to the Calabi--Yau quintic threefold, the
simulation of the $W=\Phi^5$ model will be an important starting point.
We are now considering various possible extensions of the present study.

\section*{Acknowledgments}
We would like to thank Katsumasa Nakayama and Hisao Suzuki for helpful
suggestions and discussions. 
This work was supported by JSPS Grant-in-Aid for Scientific Research Grant
Numbers JP18J20935 (O.~M.) and JP16H03982 (H.~S.).

\appendix

\section{Symmetries and the Noether currents}
\label{sec:A}
In this Appendix we summarize basic symmetries of the 2D $\mathcal{N}=2$ WZ
model, i.e., SUSY, the translation, and the~$U(1)$ symmetry and the associated
Noether currents, the supercurrent, the energy--momentum tensor, and the $U(1)$
current. The explicit form of the former two Noether currents is ambiguous
because of freedom to add a divergence-free term and/or a term that is
proportional to the equation of motion. We remove such ambiguity by imposing
that they are (gamma-) traceless for the \emph{massless free WZ model}. This is
a natural requirement because the massless free WZ model itself is an
$\mathcal{N}=2$ SCFT which possesses the $\mathcal{N}=2$ superconformal
symmetry.

\subsection{SUSY and the supercurrent}
\label{sec:A.1}
The SUSY transformation in the 2D $\mathcal{N}=2$ WZ model consists of four
spinor components, $Q_\alpha$ ($\alpha=1$, $2$) and~$\Bar{Q}_{\Dot{\alpha}}$
($\Dot{\alpha}=\Dot{1}$, $\Dot{2}$). $Q_\alpha$ is defined by
\begin{align}
   Q_1\Bar{\psi}_{\Dot{1}}(x)&=-2\Bar{\partial}A^*(x),&
   Q_1A^*(x)&=0,
\label{eq:(A1)}
\\
   Q_1F^*(x)&=2\Bar{\partial}\Bar{\psi}_{\Dot{2}}(x),&
   Q_1\Bar{\psi}_{\Dot{2}}(x)&=0,
\label{eq:(A2)}
\\
   Q_1A(x)&=\psi_1(x),&
   Q_1\psi_1(x)&=0,
\label{eq:(A3)}
\\
   Q_1\psi_2(x)&=F(x),&
   Q_1F(x)&=0,
\label{eq:(A4)}
\end{align}
and
\begin{align}
   Q_2\Bar{\psi}_{\Dot{2}}(x)&=-2\partial A^*(x),&
   Q_2A^*(x)&=0,
\label{eq:(A5)}
\\
   Q_2F^*(x)&=-2\partial\Bar{\psi}_{\Dot{1}}(x),&
   Q_2\Bar{\psi}_{\Dot{1}}(x)&=0,
\label{eq:(A6)}
\\
   Q_2A(x)&=\psi_2(x),&
   Q_2\psi_2(x)&=0,
\label{eq:(A7)}
\\
   Q_2\psi_1(x)&=-F(x),&
   Q_2F(x)&=0.
\label{eq:(A8)}
\end{align}
$\Bar{Q}_{\Dot{\alpha}}$ is, on the other hand, defined by
\begin{align}
   \Bar{Q}_{\Dot{1}}\psi_1(x)&=-2\Bar{\partial}A(x),&
   \Bar{Q}_{\Dot{1}}A(x)&=0,
\label{eq:(A9)}
\\
   \Bar{Q}_{\Dot{1}}F(x)&=-2\Bar{\partial}\psi_2(x),&
   \Bar{Q}_{\Dot{1}}\psi_2(x)&=0,
\label{eq:(A10)}
\\
   \Bar{Q}_{\Dot{1}}A^*(x)&=\Bar{\psi}_{\Dot{1}}(x),&
   \Bar{Q}_{\Dot{1}}\Bar{\psi}_{\Dot{1}}(x)&=0,
\label{eq:(A11)}
\\
   \Bar{Q}_{\Dot{1}}\Bar{\psi}_{\Dot{2}}(x)&=-F^*(x),&
   \Bar{Q}_{\Dot{1}}F^*(x)&=0,
\label{eq:(A12)}
\end{align}
and
\begin{align}
   \Bar{Q}_{\Dot{2}}\psi_2(x)&=-2\partial A(x),&
   \Bar{Q}_{\Dot{2}}A(x)&=0,
\label{eq:(A13)}
\\
   \Bar{Q}_{\Dot{2}}F(x)&=2\partial\psi_1(x),&
   \Bar{Q}_{\Dot{2}}\psi_1(x)&=0,
\label{eq:(A14)}
\\
   \Bar{Q}_{\Dot{2}}A^*(x)&=\Bar{\psi}_{\Dot{2}}(x),&
   \Bar{Q}_{\Dot{2}}\Bar{\psi}_{\Dot{2}}(x)&=0,
\label{eq:(A15)}
\\
   \Bar{Q}_{\Dot{2}}\Bar{\psi}_{\Dot{1}}(x)&=F^*(x),&
   \Bar{Q}_{\Dot{2}}F^*(x)&=0.
\label{eq:(A16)}
\end{align}
We see that these transformations fulfill simple anti-commutation relations,
\begin{align}
   \{Q_1,\Bar{Q}_{\Dot{1}}\}&=-2\Bar{\partial},
\label{eq:(A17)}
\\
   \{Q_2,\Bar{Q}_{\Dot{2}}\}&=-2\partial,
\label{eq:(A18)}
\\
   \{Q_1,\Bar{Q}_{\Dot{2}}\}&=\{Q_2,\Bar{Q}_{\Dot{1}}\}=0,
\label{eq:(A19)}
\\
   \{Q_\alpha,Q_\beta\}&=\{\Bar{Q}_{\Dot{\alpha}},\Bar{Q}_{\Dot{\beta}}\}=0.
\label{eq:(A20)}
\end{align}

The supercurrent, the Noether current associated with SUSY can be read off by
considering the localized SUSY transformation in the action. That is, under
\begin{equation}
   \delta\varphi(x)
   =\sum_{\alpha=1}^2\xi^\alpha(x)Q_\alpha\varphi(x)
   -\sum_{\Dot{\alpha}=\Dot{1}}^{\Dot{2}}
   \Bar{\xi}^{\Dot{\alpha}}(x)\Bar{Q}_{\Dot{\alpha}}\varphi(x),
\label{eq:(A21)}
\end{equation}
where~$\varphi$ stands for a generic field and~$\xi^\alpha(x)$
and~$\Bar{\xi}^{\Dot{\alpha}}(x)$ are localized Grassmann parameters, the action
changes as
\begin{equation}
   \delta S
   =\frac{1}{2\pi}\int d^2x\,
   \sum_\mu\left[
   \xi^1(x)\partial_\mu\Bar{S}_\mu^+(x)+\xi^2(x)\partial_\mu S_\mu^-(x)
   +\Bar{\xi}^{\Dot{1}}(x)\partial_\mu\Bar{S}_\mu^-(x)
   +\Bar{\xi}^{\Dot{2}}(x)\partial_\mu S_\mu^+(x)\right].
\label{eq:(A22)}
\end{equation}
Here, superscripts~$\pm$ denote the $U(1)$ charge~$\pm1$, which will be defined
in~Sect.~\ref{sec:A.3} below. The definition of the supercurrent~$S_\mu^\pm$ is
still ambiguous because of the freedom to add a divergence-free term and/or a
term that is proportional to the equation of motion. We can remove the
ambiguity~\cite{Kamata:2011fr} by imposing the gamma-traceless condition,
\begin{equation}
   \sum_\mu\gamma_\mu
   \begin{pmatrix}
   \Bar{S}_\mu^\pm\\S_\mu^\pm
   \end{pmatrix}
   =0,
\label{eq:(A23)}
\end{equation}
that is,
\begin{equation}
   S_{\Bar{z}}^\pm=\Bar{S}_z^\pm=0,
\label{eq:(A24)}
\end{equation}
for the massless free WZ model, $W'=0$.

Calculating the above variation and imposing~Eq.~\eqref{eq:(A24)}, we
have~\cite{Kamata:2011fr}
\begin{align}
   S_z^+&=4\pi\Bar{\psi}_{\Dot{2}}\partial A,&
   S_{\Bar{z}}^+&=2\pi\psi_1W'(A),
\label{eq:(A25)}
\\
   S_z^-&=-4\pi\psi_2\partial A^*,&
   S_{\Bar{z}}^-&=2\pi\Bar{\psi}_{\Dot{1}}W'(A)^*,
\label{eq:(A26)}
\\
   \Bar{S}_z^+&=-2\pi\Bar{\psi}_{\Dot{2}}W'(A)^*,&
   \Bar{S}_{\Bar{z}}^+&=-4\pi\psi_1\Bar{\partial}A^*,
\label{eq:(A27)}
\\
   \Bar{S}_z^-&=-2\pi\psi_2W'(A),&
   \Bar{S}_{\Bar{z}}^-&=4\pi\Bar{\psi}_{\Dot{1}}\Bar{\partial}A.
\label{eq:(A28)}
\end{align}

\subsection{Translational invariance and the energy--momentum tensor}
\label{sec:A.2}
The energy--momentum tensor is the Noether current associated with the
translational invariance. To remove its ambiguity, we require the traceless
condition
\begin{equation}
   \sum_\mu T_{\mu\mu}=0,
\label{eq:(A29)}
\end{equation}
that is,
\begin{equation}
   T_{z\Bar{z}}=T_{\Bar{z}z}=0,
\label{eq:(A30)}
\end{equation}
for the massless free WZ model, $W'=0$.

The energy--momentum tensor, however, has wider ambiguity than the supercurrent
and, because of this, it is difficult to find the energy--momentum tensor which
fulfills the above requirement if we simply follow the above procedure, i.e.,
starting from the variation of the action under the localized translation and
then imposing the traceless condition.

A better strategy is the following: We consider the infinitesimal
transformation of the form
\begin{align}
   \delta A(x)&=-\sum_\mu v_\mu\partial_\mu A(x),
\label{eq:(A31)}
\\
   \delta\psi_1(x)&=-\sum_\mu v_\mu\partial_\mu\psi_1(x)
   -\frac{1}{2}(\Bar{\partial}v_{\Bar{z}})\psi_1(x),
\label{eq:(A32)}
\\
   \delta\Bar{\psi}_{\Dot{1}}(x)
   &=-\sum_\mu v_\mu\partial_\mu\Bar{\psi}_{\Dot{1}}(x)
   -\frac{1}{2}(\Bar{\partial}v_{\Bar{z}})\Bar{\psi}_{\Dot{1}}(x),
\label{eq:(A33)}
\\
   \delta\psi_2(x)
   &=-\sum_\mu v_\mu\partial_\mu\psi_2(x)-\frac{1}{2}(\partial v_z)\psi_2(x),
\label{eq:(A34)}
\\
   \delta\Bar{\psi}_{\Dot{2}}(x)
   &=-\sum_\mu v_\mu\partial_\mu\Bar{\psi}_{\Dot{2}}(x)
   -\frac{1}{2}(\partial v_z)\Bar{\psi}_{\Dot{2}}(x),
\label{eq:(A35)}
\\
   \delta F(x)
   &=-\sum_\mu v_\mu\partial_\mu F(x).
\label{eq:(A36)}
\end{align}
When the parameter~$v_\mu$ is constant, this is simply the translation that is
a symmetry of the WZ model. When $v_\mu\propto\epsilon_{\mu\nu}x_\nu$, this is
the infinitesimal Lorentz transformation that is also a symmetry of the WZ
model. Thus, localizing the parameter $v_\mu$ as~$v_\mu(x)$, the variation of
the action gives rise to a conserved current. By construction, this current is
a combination of the canonical energy--momentum tensor, the Lorentz current,
and the equation of motion. Moreover, when the parameters~$v_z$
and~$v_{\Bar{z}}$ are holomorphic and anti-holomorphic, respectively,
$v_z=v_z(z)$ and~$v_{\Bar{z}}=v_{\Bar{z}}(\Bar{z})$, then
Eqs.~\eqref{eq:(A31)}--\eqref{eq:(A36)} coincide with the conformal
transformation, that is an exact invariance of the massless free WZ model. As
a consequence, when $W'=0$ the conserved Noether current obtained by localizing
$v_\mu$ as~$v_\mu(x)$ must generate the conformal symmetry, i.e., it must be
related to the traceless energy--momentum tensor.

In this way, from the variation of the action
under~Eqs.~\eqref{eq:(A31)}--\eqref{eq:(A36)},
\begin{equation}
   \delta S=-\frac{1}{2\pi}\int d^2x\,
   \sum_{\mu\nu}v_\nu(x)\partial_\mu T_{\mu\nu}(x),
\label{eq:(A37)}
\end{equation}
we have
\begin{align}
   T_{\mu\nu}
   &=-2\pi\partial_\mu A^*\partial_\nu A
   -2\pi\partial_\nu A^*\partial_\mu A
\notag\\
   &\qquad{}+\pi\delta_{\mu\nu}
   \bigl[
   2\partial_\rho A^*\partial_\rho A-2F^*F-2F^*W'(A)^*-2FW'(A)
\notag\\
   &\qquad\qquad\qquad{}
   +W''(A)^*\Bar{\psi}_{\Dot{1}}\Bar{\psi}_{\Dot{2}}
   +W''(A)\psi_2\psi_1\bigr]
\notag\\
   &\qquad\qquad{}
   -\pi(\delta_{0\mu}-i\delta_{1\mu})(\delta_{0\nu}-i\delta_{1\nu})
   \left(\Bar{\psi}_{\Dot{1}}\Bar{\partial}\psi_1
   -\Bar{\partial}\Bar{\psi}_{\Dot{1}}\psi_1\right)
\notag\\
   &\qquad\qquad\qquad{}
   -\pi(\delta_{0\mu}+i\delta_{1\mu})(\delta_{0\nu}+i\delta_{1\nu})
   \left(\psi_2\partial\Bar{\psi}_{\Dot{2}}-\partial\psi_2\Bar{\psi}_{\Dot{2}}
   \right).
\label{eq:(A38)}
\end{align}
This can be written as
\begin{align}
   T(\equiv T_{zz})
   &=-4\pi\partial A^*\partial A
   -\pi\psi_2\partial\Bar{\psi}_{\Dot{2}}
   +\pi\partial\psi_2\Bar{\psi}_{\Dot{2}},
\label{eq:(A39)}
\\
   \Bar{T}(\equiv T_{\Bar{z}\Bar{z}})
   &=-4\pi\Bar{\partial}A^*\Bar{\partial}A^*
   -\pi\Bar{\psi}_{\Dot{1}}\Bar{\partial}\psi_1
   +\pi\Bar{\partial}\Bar{\psi}_{\Dot{1}}\psi_1,
\label{eq:(A40)}
\\
   T_{z\Bar{z}}=T_{\Bar{z}z}
   &=-\pi F^*F-\pi F^*W'(A)^*-\pi FW'(A)
   +\frac{\pi}{2}W''(A)^*\Bar{\psi}_{\Dot{1}}\Bar{\psi}_{\Dot{2}}
   +\frac{\pi}{2}W''(A)\psi_2\psi_1.
\label{eq:(A41)}
\end{align}
When~$W'=0$, the traceless condition of~Eq.~\eqref{eq:(A30)} is clearly
satisfied (note that $F=-W'^*$ under the equation of motion).

\subsection{$U(1)$ symmetry and the $U(1)$ current}
\label{sec:A.3}
We take the following $U(1)$ transformation ($\gamma\in\mathbb{R}$),
\begin{equation}
   \delta
   \begin{pmatrix}
   \psi_1\\\Bar{\psi}_{\Dot{2}}
   \end{pmatrix}(x)
   =i\gamma
   \begin{pmatrix}
   \psi_1\\\Bar{\psi}_{\Dot{2}}
   \end{pmatrix}(x),\qquad
   \delta
   \begin{pmatrix}
   \Bar{\psi}_{\Dot{1}}\\\psi_2
   \end{pmatrix}(x)
   =-i\gamma
   \begin{pmatrix}
   \Bar{\psi}_{\Dot{1}}\\\psi_2
   \end{pmatrix}(x),
\label{eq:(A42)}
\end{equation}
under which the WZ model is invariant; we have assigned the~$U(1)$ charge~$+1$
to~$\psi_1$ and~$\Bar{\psi}_{\Dot{2}}$, and $-1$ to~$\Bar{\psi}_{\Dot{1}}$
and~$\psi_2$. It turns out that, in the massless free WZ model, the $U(1)$
current associated with this symmetry forms the superconformal multiplet with
the supercurrent and the energy--momentum tensor.\footnote{The~$U(1)_R$
symmetry in the case of the superpotential~$W=\lambda\Phi^n/n$ would be
\begin{align}
   A(x)&\to\exp\left(i\gamma/n\right)A(x),
\label{eq:(A43)}
\\
   \psi_\alpha(x)&\to\exp[-i\gamma(n-2)/2n]\psi_\alpha(x),
\label{eq:(A44)}
\\
   \Bar{\psi}_{\Dot{\alpha}}(x)
   &\to\exp[i\gamma(n-2)/2n]\Bar{\psi}_{\Dot{\alpha}}(x),
\label{eq:(A45)}
\\
   F(x)&\to\exp[-i\gamma(n-1)/n]F(x).
\label{eq:(A46)}
\end{align}
The associated Noether current, however, can be neither holomorphic nor
anti-holomorphic even in the free-field limit~$\lambda\to0$; thus it cannot be
a member of the superconformal multiplet.} Localizing the parameter~$\gamma$
as~$\gamma(x)$, the associated Noether current can be obtained as
\begin{equation}
   \delta S=\frac{1}{2\pi}\int d^2x\,2i\gamma(x)
   \left[\partial J_z(x)+\Bar{\partial}J_{\Bar{z}}(x)\right].
\label{eq:(A47)}
\end{equation}
The explicit form is given by
\begin{align}
   J\equiv J_z&=2\pi\Bar{\psi}_{\Dot{2}}\psi_2,
\label{eq:(A48)}
\\
   \Bar{J}\equiv J_{\Bar{z}}&=2\pi\psi_1\Bar{\psi}_{\Dot{1}}.
\label{eq:(A49)}
\end{align}

It can be confirmed that the supercurrent, the energy--momentum tensor, and
the $U(1)$ current in the above form are related by the SUSY transformation in
a very simple way. This fact provides more support for the above explicit forms
of currents.

\subsection{Massless free WZ model}
We summarize explicit expressions for the above Noether currents in the
massless free WZ model, a free $\mathcal{N}=2$ SCFT, and confirm that they
actually fulfill the $\mathcal{N}=2$ super-Virasoro algebra as expected.

The supercurrent, the energy--momentum tensor, and the~$U(1)$ current in the
holomorphic sector are
\begin{align}
   S^+(z)&=4\pi\Bar{\psi}_{\Dot{2}}(z)\partial A(z),
\\
   S^-(z)&=-4\pi\psi_2(z)\partial A^*(z),
\\
   T(z)&=-4\pi\partial A^*(z)\partial A(z)
   -\pi\psi_2(z)\partial\Bar{\psi}_{\Dot{2}}(z)
   +\pi\partial\psi_2(z)\Bar{\psi}_{\Dot{2}}(z),
\\
   J(z)&=2\pi\Bar{\psi}_{\Dot{2}}(z)\psi_2(z),
\end{align}
and in the anti-holomorphic sector,
\begin{align}
   \Bar{S}^+(\Bar{z})
   &=-4\pi\psi_1(\Bar{z})\Bar{\partial}A^*(\Bar{z}),
\\
   \Bar{S}^-(\Bar{z})
   &=4\pi\Bar{\psi}_{\Dot{1}}(\Bar{z})\Bar{\partial}A(\Bar{z}),
\\
   \Bar{T}(\Bar{z})
   &=-4\pi\Bar{\partial}A^*(\Bar{z})\Bar{\partial}A(\Bar{z})
   -\pi\Bar{\psi}_{\Dot{1}}(\Bar{z})\Bar{\partial}\psi_1(\Bar{z})
   +\pi\Bar{\partial}\Bar{\psi}_{\Dot{1}}(\Bar{z})\psi_1(\Bar{z}),
\\
   \Bar{J}(\Bar{z})
   &=2\pi\psi_1(\Bar{z})\Bar{\psi}_{\Dot{1}}(\Bar{z}).
\end{align}

The OPEs between the component fields are given by
\begin{align}
   A(z,\Bar{z})A^*(0,0)
   &\sim-\frac{1}{4\pi}\ln|z|^2,
\\
   \psi_1(\Bar{z})\Bar{\psi}_{\Dot{1}}(0)
   &\sim\frac{1}{2\pi}\frac{1}{\Bar{z}},
\\
   \Bar{\psi}_{\Dot{2}}(z)\psi_2(0)
   &\sim\frac{1}{2\pi}\frac{1}{z},
\\
   \text{(otherwise)}
   &\sim0,
\end{align}
where ``$\sim$'' implies ``$=$'' up to non-singular terms. Using these, we find
that the above Noether currents in the holomorphic part satisfy the OPEs of the
$\mathcal{N}=2$ super-Virasoro algebra,
\begin{align}
   T(z)T(0)
   &\sim\frac{c}{2z^4}+\frac{2}{z^2}T(0)+\frac{1}{z}\partial T(0),
\\
   T(z)S^\pm(0)
   &\sim\frac{3}{2z^2}S^\pm(0)+\frac{1}{z}\partial S^\pm(0),
\\
   T(z)J(0)
   &\sim\frac{1}{z^2}J(0)+\frac{1}{z}\partial J(0),
\\
   S^\pm(z)S^\pm(0)
   &\sim0,
\\
   S^+(z)S^-(0)
   &\sim\frac{2c}{3z^3}+\frac{2}{z^2}J(0)
   +\frac{2}{z}T(0)+\frac{1}{z}\partial J(0),
\\
   J(z)S^\pm(0)
   &\sim\pm\frac{1}{z}S^\pm(0),
\\
   J(z)J(0)
   &\sim\frac{c}{3z^2},
\end{align}
where the central charge is~$c=3$ corresponding to a free $\mathcal{N}=2$ SCFT.

\section{A fast algorithm for the Jacobian computation}
\label{sec:B}
We can accelerate the computation
of~$\sign\det\frac{\partial(N,N^*)}{\partial(A,A^*)}$ by effectively halving
the size of the matrix,
\begin{equation}
   \frac{\partial(N,N^*)}{\partial(A,A^*)}
   =
   \begin{pmatrix}
   2ip_z&W''(A)^**\\W''(A)*&2ip_{\Bar{z}}\\
   \end{pmatrix},
\label{eq:(B1)}
\end{equation}
whose $p,q$ element is
\begin{align}
   \left[\frac{\partial(N,N^*)}{\partial(A,A^*)}\right]_{p,q}
   &=
   \begin{pmatrix}
   2ip_z\delta_{p,q}&\frac{1}{L_0L_1}W''(A)(q-p)^*\\
   \frac{1}{L_0L_1}W''(A)(p-q)&2ip_{\Bar{z}}\delta_{p,q}\\
   \end{pmatrix}
\notag\\
   &=
   \begin{pmatrix}
   2ip_z\delta_{p,q}&\frac{1}{L_0L_1}W''(A)(p-q)^\dagger\\
   \frac{1}{L_0L_1}W''(A)(p-q)&2ip_{\Bar{z}}\delta_{p,q}\\
   \end{pmatrix}.
\label{eq:(B2)}
\end{align}
Note that Eq.~\eqref{eq:(B2)} is a $[2(L_0+1)(L_1+1)]\times[2(L_0+1)(L_1+1)]$
matrix when the momentum takes the values
\begin{equation}
   p_\mu=\frac{2\pi}{L_\mu}n_\mu,\qquad
   n_\mu=0,\pm1,\dotsc,\pm\frac{L_\mu}{2},
\label{eq:(B3)}
\end{equation}
where we have assumed that both integers $L_0$ and~$L_1$ are even,

We write the matrix in~Eq.~\eqref{eq:(B2)} as
\begin{equation}
   \frac{\partial(N,N^*)}{\partial(A,A^*)}
   \equiv 
   \begin{pmatrix}
   iP&W^\dagger\\W&iP^\dagger\\
   \end{pmatrix}.
\label{eq:(B4)}
\end{equation}
It should be noted that the diagonal matrix $P$, whose $p,q$ element
is~$2p_z\delta_{p,q}$, does not have an inverse because it has zero at~$p=0$;
what we want to do is to remove this zero.

Considering the case that $P$ and~$W$ are $3\times3$ matrices for simplicity,
we can confirm that the determinant of the matrix in~Eq.~\eqref{eq:(B4)} can
be deformed as
\begin{equation}
   \det
   \begin{pmatrix}
   \lambda_1 &   &           & &           & \\
             & 0 &           & & W^\dagger & \\
             &   & \lambda_2 & &           & \\
   W_{11} & W_{12} & W_{13} & \lambda_3 &   &           \\
   W_{21} & W_{22} & W_{23} &           & 0 &           \\
   W_{31} & W_{32} & W_{33} &           &   & \lambda_4 \\
   \end{pmatrix}
   =-|W_{22}|^2\det
   \begin{pmatrix}
   \lambda_1 &           & \widetilde{W}^*_{11} & \widetilde{W}^*_{31} \\
             & \lambda_2 & \widetilde{W}^*_{13} & \widetilde{W}^*_{33} \\
   \widetilde{W}_{11} & \widetilde{W}_{13} & \lambda_3 &           \\
   \widetilde{W}_{31} & \widetilde{W}_{33} &           & \lambda_4 \\
   \end{pmatrix},
\label{eq:(B5)}
\end{equation}
where
\begin{equation}
   \widetilde{W}_{ij}\equiv\frac{1}{W_{22}}\det
   \begin{pmatrix}
   W_{ij}&W_{i2}\\W_{2j}&W_{22}\\
   \end{pmatrix}.
\label{eq:(B6)}
\end{equation}

In an analogous way, we can write, for the general case, 
\begin{equation}
   \det
   \begin{pmatrix}
   iP&W^\dagger\\W&iP^\dagger\\
   \end{pmatrix}
   =-|W_{0,0}|^2\det\nolimits'
   \begin{pmatrix}
   iP&\widetilde{W}^\dagger\\\widetilde{W}&iP^\dagger\\
   \end{pmatrix},
\label{eq:(B7)}
\end{equation}
where $W_{0,0}$ is the component at~$(p,q)=(0,0)$, $\det'$ is the determinant
in the subspace in which the components with $p=0$ or~$q=0$ are omitted, and
\begin{equation}
   \widetilde{W}_{p,q}=\frac{1}{W_{0,0}}\det
   \begin{pmatrix}
   W_{p,q}&W_{p,0}\\W_{0,q}&W_{0,0}\\
   \end{pmatrix}.
\label{eq:(B8)}
\end{equation}
Note that this is simply the determinant of a $2\times2$ matrix.

Since the right-hand side of~Eq.~\eqref{eq:(B7)} refers to the subspace in
which $P$ has an inverse, the Jacobian can be expressed as
\begin{align}
   \det
   \begin{pmatrix}
   iP&W^\dagger\\W&iP^\dagger\\
   \end{pmatrix}
   &=-|W_{0,0}|^2\det\nolimits'
   \begin{pmatrix}
   iP&0\\\widetilde{W}&I\\
   \end{pmatrix}
   \det\nolimits'
   \begin{pmatrix}
   I&(-i)P^{-1}\widetilde{W}^\dagger\\
   0&iP^\dagger-\widetilde{W}(-i)P^{-1}\widetilde{W}^\dagger\\
  \end{pmatrix}
\label{eq:(B9)}
\\
   &=-|W_{0,0}|^2\det\nolimits'
   \left(-PP^\dagger
   -P\widetilde{W}P^{-1}\widetilde{W}^\dagger\right).
\label{eq:(B10)}
\end{align}
Here, the inverse of $P$ is given by
\begin{equation}
   \left(P^{-1}\right)_{p,q}
   =\frac{1}{2p_z}\delta_{p,q}
   =\frac{p_{\Bar{z}}}{2|p_z|^2}\delta_{p,q}.
\label{eq:(B11)}
\end{equation}

Thus, substituting the matrix elements in~Eq.~\eqref{eq:(B2)}, we have
\begin{align}
   \det
   \begin{pmatrix}
   iP&W^\dagger\\W&iP^\dagger\\
   \end{pmatrix}
   &=-\det\nolimits'(-1)
   \left|\frac{1}{L_0L_1}W''(A)(0)\right|^2
\notag\\
   &\qquad{}
   \times\det\nolimits'
   \left[4|p_z|^2\delta_{p,q}
   +\left(\frac{1}{L_0L_1}\right)^2
   \sum_{l\neq0}\frac{p_z}{l_z}\widetilde{W}''(A)(p-l)
   \widetilde{W}''(A)(l-q)^\dagger\right],
\label{eq:(B12)}
\end{align}
where for $p\neq0$,
\begin{align}
   \widetilde{W}''(A)(p-l) 
   &\equiv\frac{1}{W''(A)(0)}
   \det
   \begin{pmatrix}
   W''(A)(p-l)&W''(A)(p-0)\\
   W''(A)(0-l)&W''(A)(0-0)\\
   \end{pmatrix}
\notag \\
   &=\frac{1}{W''(A)(0)}
   \left[W''(A)(p-l)W''(A)(0)-W''(A)(p W''(A)(-l)\right].
\label{eq:(B13)}
\end{align}
Here, the factor $\det'(-1)$ is
\begin{align}
   \det\nolimits'(-1)=(-1)^{(L_0+1)(L_1+1)-1}=+1,
\label{eq:(B14)}
\end{align}
for $L_0$ and $L_1$ are even.

Thus, finally, the sign of the Jacobian is given by the sign of the determinant
of a matrix with smaller
dimensions~$[(L_0+1)(L_1+1)-1]\times[(L_0+1)(L_1+1)-1]$, as
\begin{align}
   &\sign\det
   \begin{pmatrix}
   iP&W^\dagger\\W&iP^\dagger\\
   \end{pmatrix}
\notag\\
   &=-\det\nolimits'(-1)\sign\det\nolimits'
   \left[
   4|p_z|^2\delta_{p,q}
   +\left(\frac{1}{L_0L_1}\right)^2
   \sum_{l\neq0}\frac{p_z}{l_z}\widetilde{W}''(A)(p-l)
   \widetilde{W}''(A)(q-l)^*\right].
\label{eq:(B15)}
\end{align}
Since the computational cost required for the matrix determinant is~$O(N^3)$
for a matrix of size~$N$, this representation reduces the cost by~$\sim1/8$.

It turns out that the above sign is mainly negative for most of configurations
of~$A(p)$. Since the overall sign
of~$\sign\det\frac{\partial(N,N^*)}{\partial(A,A^*)}$ is irrelevant in the
expectation value of~Eq.~\eqref{eq:(2.24)}, we regard Eq.~\eqref{eq:(B15)}
as\footnote{Or we may simply say that the partition function~\eqref{eq:(2.19)}
is defined with another negative sign.}
\begin{align}
   -\sign\det\frac{\partial(N,N^*)}{\partial(A,A^*)}.
\label{eq:(B16)}
\end{align}


\begin{thebibliography}{00}

\bibitem{Zamolodchikov:1986db}
  A.~B.~Zamolodchikov,
  Sov.\ J.\ Nucl.\ Phys.\  {\bf 44} (1986) 529
   [Yad.\ Fiz.\  {\bf 44} (1986) 821].

\bibitem{Wess:1974tw}
  J.~Wess and B.~Zumino,
  Nucl.\ Phys.\ B {\bf 70} (1974) 39.
  doi:10.1016/0550-3213(74)90355-1

\bibitem{DiVecchia:1985ief}
  P.~Di Vecchia, J.~L.~Petersen and H.~B.~Zheng,
  Phys.\ Lett.\  {\bf 162B} (1985) 327.
  doi:10.1016/0370-2693(85)90932-3

\bibitem{DiVecchia:1986cdz}
  P.~Di Vecchia, J.~L.~Petersen and M.~Yu,
  Phys.\ Lett.\ B {\bf 172} (1986) 211.
  doi:10.1016/0370-2693(86)90837-3

\bibitem{DiVecchia:1986fwg}
  P.~Di Vecchia, J.~L.~Petersen, M.~Yu and H.~B.~Zheng,
  Phys.\ Lett.\ B {\bf 174} (1986) 280.
  doi:10.1016/0370-2693(86)91099-3

\bibitem{Boucher:1986bh}
  W.~Boucher, D.~Friedan and A.~Kent,
  Phys.\ Lett.\ B {\bf 172} (1986) 316.
  doi:10.1016/0370-2693(86)90260-1

\bibitem{Gepner:1986ip}
  D.~Gepner,
  Nucl.\ Phys.\ B {\bf 287} (1987) 111.
  doi:10.1016/0550-3213(87)90098-8

\bibitem{Cappelli:1986hf}
  A.~Cappelli, C.~Itzykson and J.~B.~Zuber,
  Nucl.\ Phys.\ B {\bf 280} (1987) 445.
  doi:10.1016/0550-3213(87)90155-6

\bibitem{Cappelli:1986ed}
  A.~Cappelli,
  Phys.\ Lett.\ B {\bf 185} (1987) 82.
  doi:10.1016/0370-2693(87)91532-2

\bibitem{Gepner:1986hr}
  D.~Gepner and Z.~a.~Qiu,
  Nucl.\ Phys.\ B {\bf 285} (1987) 423.
  doi:10.1016/0550-3213(87)90348-8

\bibitem{Gepner:1987qi}
  D.~Gepner,
  Nucl.\ Phys.\ B {\bf 296} (1988) 757.
  doi:10.1016/0550-3213(88)90397-5

\bibitem{Cappelli:1987xt}
  A.~Cappelli, C.~Itzykson and J.~B.~Zuber,
  Commun.\ Math.\ Phys.\  {\bf 113} (1987) 1.
  doi:10.1007/BF01221394

\bibitem{Kato:1987td}
  A.~Kato,
  Mod.\ Phys.\ Lett.\ A {\bf 2} (1987) 585.
  doi:10.1142/S0217732387000732

\bibitem{Gepner:1987vz}
  D.~Gepner,
  Phys.\ Lett.\ B {\bf 199} (1987) 380.
  doi:10.1016/0370-2693(87)90938-5

\bibitem{Kastor:1988ef}
  D.~A.~Kastor, E.~J.~Martinec and S.~H.~Shenker,
  Nucl.\ Phys.\ B {\bf 316} (1989) 590.
  doi:10.1016/0550-3213(89)90060-6

\bibitem{Vafa:1988uu}
  C.~Vafa and N.~P.~Warner,
  Phys.\ Lett.\ B {\bf 218} (1989) 51.
  doi:10.1016/0370-2693(89)90473-5

\bibitem{Martinec:1988zu}
  E.~J.~Martinec,
  Phys.\ Lett.\ B {\bf 217} (1989) 431.
  doi:10.1016/0370-2693(89)90074-9

\bibitem{Lerche:1989uy}
  W.~Lerche, C.~Vafa and N.~P.~Warner,
  Nucl.\ Phys.\ B {\bf 324} (1989) 427.
  doi:10.1016/0550-3213(89)90474-4

\bibitem{Howe:1989qr}
  P.~S.~Howe and P.~C.~West,
  Phys.\ Lett.\ B {\bf 223} (1989) 377.
  doi:10.1016/0370-2693(89)91619-5

\bibitem{Cecotti:1989jc}
  S.~Cecotti, L.~Girardello and A.~Pasquinucci,
  Nucl.\ Phys.\ B {\bf 328} (1989) 701.
  doi:10.1016/0550-3213(89)90226-5

\bibitem{Howe:1989az}
  P.~S.~Howe and P.~C.~West,
  Phys.\ Lett.\ B {\bf 227} (1989) 397.
  doi:10.1016/0370-2693(89)90950-7

\bibitem{Cecotti:1989gv}
  S.~Cecotti, L.~Girardello and A.~Pasquinucci,
  Int.\ J.\ Mod.\ Phys.\ A {\bf 6} (1991) 2427.
  doi:10.1142/S0217751X91001192

\bibitem{Cecotti:1990kz}
  S.~Cecotti,
  Int.\ J.\ Mod.\ Phys.\ A {\bf 6} (1991) 1749.
  doi:10.1142/S0217751X91000939

\bibitem{Witten:1993jg}
  E.~Witten,
  Int.\ J.\ Mod.\ Phys.\ A {\bf 9} (1994) 4783
  doi:10.1142/S0217751X9400193X
  [hep-th/9304026].

\bibitem{Kawai:2010yj}
  H.~Kawai and Y.~Kikukawa,
  Phys.\ Rev.\ D {\bf 83} (2011) 074502
  doi:10.1103/PhysRevD.83.074502
  [arXiv:1005.4671 [hep-lat]].

\bibitem{Kikukawa:2002as}
  Y.~Kikukawa and Y.~Nakayama,
  Phys.\ Rev.\ D {\bf 66} (2002) 094508
  doi:10.1103/PhysRevD.66.094508
  [hep-lat/0207013].

\bibitem{Nicolai:1979nr}
  H.~Nicolai,
  Phys.\ Lett.\  {\bf 89B} (1980) 341.
  doi:10.1016/0370-2693(80)90138-0

\bibitem{Nicolai:1980jc}
  H.~Nicolai,
  Nucl.\ Phys.\ B {\bf 176} (1980) 419.
  doi:10.1016/0550-3213(80)90460-5

\bibitem{Parisi:1982ud}
  G.~Parisi and N.~Sourlas,
  Nucl.\ Phys.\ B {\bf 206} (1982) 321.
  doi:10.1016/0550-3213(82)90538-7

\bibitem{Cecotti:1983up}
  S.~Cecotti and L.~Girardello,
  Annals Phys.\  {\bf 145} (1983) 81.
  doi:10.1016/0003-4916(83)90172-0

\bibitem{Sakai:1983dg} 
  N.~Sakai and M.~Sakamoto,
  Nucl.\ Phys.\ B {\bf 229}, 173 (1983).
  doi:10.1016/0550-3213(83)90359-0

\bibitem{Giedt:2004qs} 
  J.~Giedt and E.~Poppitz,
  JHEP {\bf 0409}, 029 (2004)
  doi:10.1088/1126-6708/2004/09/029
  [hep-th/0407135].

\bibitem{Kadoh:2010ca} 
  D.~Kadoh and H.~Suzuki,
  Phys.\ Lett.\ B {\bf 696}, 163 (2011)
  doi:10.1016/j.physletb.2010.12.012
  [arXiv:1011.0788 [hep-lat]].

\bibitem{Kadoh:2016eju}
  D.~Kadoh,
  PoS LATTICE {\bf 2015} (2016) 017
  [arXiv:1607.01170 [hep-lat]].

\bibitem{Beccaria:1998vi} 
  M.~Beccaria, G.~Curci and E.~D'Ambrosio,
  Phys.\ Rev.\ D {\bf 58}, 065009 (1998)
  doi:10.1103/PhysRevD.58.065009
  [hep-lat/9804010].

\bibitem{Catterall:2001fr} 
  S.~Catterall and S.~Karamov,
  Phys.\ Rev.\ D {\bf 65}, 094501 (2002)
  doi:10.1103/PhysRevD.65.094501
  [hep-lat/0108024].

\bibitem{Giedt:2005ae} 
  J.~Giedt,
  Nucl.\ Phys.\ B {\bf 726}, 210 (2005)
  doi:10.1016/j.nuclphysb.2005.08.004
  [hep-lat/0507016].

\bibitem{Bergner:2007pu} 
  G.~Bergner, T.~Kaestner, S.~Uhlmann and A.~Wipf,
  Annals Phys.\  {\bf 323}, 946 (2008)
  doi:10.1016/j.aop.2007.06.010
  [arXiv:0705.2212 [hep-lat]].

\bibitem{Kastner:2008zc} 
  T.~Kastner, G.~Bergner, S.~Uhlmann, A.~Wipf and C.~Wozar,
  Phys.\ Rev.\ D {\bf 78}, 095001 (2008)
  doi:10.1103/PhysRevD.78.095001
  [arXiv:0807.1905 [hep-lat]].

\bibitem{Nicolis:2017lqk} 
  S.~Nicolis,
  arXiv:1712.07045 [hep-th].

\bibitem{Kamata:2011fr}
  S.~Kamata and H.~Suzuki,
  Nucl.\ Phys.\ B {\bf 854} (2012) 552
  doi:10.1016/j.nuclphysb.2011.09.007
  [arXiv:1107.1367 [hep-lat]].

\bibitem{Kadoh:2009sp}
  D.~Kadoh and H.~Suzuki,
  Phys.\ Lett.\ B {\bf 684} (2010) 167
  doi:10.1016/j.physletb.2010.01.022
  [arXiv:0909.3686 [hep-th]].

\bibitem{Bartels:1983wm}
  J.~Bartels and J.~B.~Bronzan,
  Phys.\ Rev.\ D {\bf 28} (1983) 818.
  doi:10.1103/PhysRevD.28.818

\bibitem{Drell:1976bq}
  S.~D.~Drell, M.~Weinstein and S.~Yankielowicz,
  Phys.\ Rev.\ D {\bf 14} (1976) 487.
  doi:10.1103/PhysRevD.14.487

\bibitem{Drell:1976mj}
  S.~D.~Drell, M.~Weinstein and S.~Yankielowicz,
  Phys.\ Rev.\ D {\bf 14} (1976) 1627.
  doi:10.1103/PhysRevD.14.1627

\bibitem{Bergner:2009vg} 
  G.~Bergner,
  JHEP {\bf 1001}, 024 (2010)
  doi:10.1007/JHEP01(2010)024
  [arXiv:0909.4791 [hep-lat]].

\bibitem{Dondi:1976tx}
  P.~H.~Dondi and H.~Nicolai,
  Nuovo Cim.\ A {\bf 41} (1977) 1.
  doi:10.1007/BF02730448

\bibitem{Karsten:1979wh}
  L.~H.~Karsten and J.~Smit,
  Phys.\ Lett.\  {\bf 85B} (1979) 100.
  doi:10.1016/0370-2693(79)90786-X

\bibitem{Kato:2008sp}
  M.~Kato, M.~Sakamoto and H.~So,
  JHEP {\bf 0805} (2008) 057
  doi:10.1088/1126-6708/2008/05/057
  [arXiv:0803.3121 [hep-lat]].

\bibitem{Zamolodchikov:1986gt}
  A.~B.~Zamolodchikov,
  JETP Lett.\  {\bf 43} (1986) 730
   [Pisma Zh.\ Eksp.\ Teor.\ Fiz.\  {\bf 43} (1986) 565].

\bibitem{Cappelli:1989yu}
  A.~Cappelli and J.~I.~Latorre,
  Nucl.\ Phys.\ B {\bf 340} (1990) 659.
  doi:10.1016/0550-3213(90)90463-N

\bibitem{Cecotti:1990wz}
  S.~Cecotti,
  Nucl.\ Phys.\ B {\bf 355} (1991) 755.
  doi:10.1016/0550-3213(91)90493-H

\bibitem{Greene:1988ut}
  B.~R.~Greene, C.~Vafa and N.~P.~Warner,
  Nucl.\ Phys.\ B {\bf 324} (1989) 371.
  doi:10.1016/0550-3213(89)90471-9

\bibitem{Witten:1993yc}
  E.~Witten,
  Nucl.\ Phys.\ B {\bf 403} (1993) 159
   [AMS/IP Stud.\ Adv.\ Math.\  {\bf 1} (1996) 143]
  doi:10.1016/0550-3213(93)90033-L
  [hep-th/9301042].

\bibitem{Luscher:2009eq} 
  M.~L\"uscher,
  Commun.\ Math.\ Phys.\  {\bf 293}, 899 (2010)
  doi:10.1007/s00220-009-0953-7
  [arXiv:0907.5491 [hep-lat]].

\bibitem{Witten:1982df}
  E.~Witten,
  Nucl.\ Phys.\ B {\bf 202} (1982) 253.
  doi:10.1016/0550-3213(82)90071-2

\bibitem{Cecotti:1981fu}
  S.~Cecotti and L.~Girardello,
  Phys.\ Lett.\  {\bf 110B} (1982) 39.
  doi:10.1016/0370-2693(82)90947-9

\bibitem{eigen}
\url{http://eigen.tuxfamily.org/}

\bibitem{Julia}
\url{http://julialang.org/}

\bibitem{Bezanson:xxxxa}
  J.~Bezanson, S.~Karpinski, V.~B.~Shah and A.~Edelman,
  arXiv:1209.5145 [cs.PL].

\bibitem{Bezanson:xxxxb}
  J.~Bezanson, A.~Edelman, S.~Karpinski and V.~B.~Shah,
  arXiv:1411.1607 [cs.MS].

\bibitem{Polchinski:1998rq}
  J.~Polchinski,
  ``String theory. Vol. 1: An introduction to the bosonic string,''
  Cambridge University Press,
  1998, 402~p.

\bibitem{Polchinski:1998rr}
  J.~Polchinski,
  ``String theory. Vol. 2: Superstring theory and beyond,''
  Cambridge University Press,
  1998, 531~p.

\end{thebibliography}
\end{document}